\documentclass[12pt,letterpaper]{article}
\usepackage[a4paper, total={7in, 10in}]{geometry}

\usepackage[T1]{fontenc}
\usepackage{graphicx}
\usepackage{helvet}
\usepackage{authblk}
\usepackage{amsmath}
\usepackage{amssymb}
\usepackage{booktabs,longtable}
\usepackage{environ,adjustbox}
\usepackage{microtype}
\usepackage{placeins}
\let\ArxivFloatBarrier\FloatBarrier
\usepackage{hyperref}
\usepackage[super,comma,sort&compress]{natbib}
\let\citesupp\cite
\AddToHook{env/longtable/begin}{\setlength{\LTleft}{0pt}\setlength{\LTright}{\fill}}
\let\CellFigure\figure
\let\CellEndFigure\endfigure
\RenewEnviron{figure}[1][]{%
  \CellFigure[p]\centering
  \begin{adjustbox}{max totalheight=.94\textheight,max width=\textwidth}%
    \begin{minipage}{\textwidth}\BODY\end{minipage}%
  \end{adjustbox}%
  \CellEndFigure}
\makeatletter
\renewcommand{\maketitle}{\bgroup\setlength{\parindent}{0pt}
\begin{flushleft}\textbf{\@title}\par\medskip\@author\end{flushleft}\egroup}
\makeatother
\let\CellSection\section
\let\CellSubsection\subsection
\let\CellSubsubsection\subsubsection
\newcommand{\OriginalHeadings}{%
  \renewcommand{\section}[1]{\CellSection*{\MakeUppercase{##1}}\refstepcounter{section}}%
  \renewcommand{\subsection}[1]{\CellSubsection*{##1}\refstepcounter{subsection}}}
\newcommand{\OriginalMethodHeadings}{%
  \renewcommand{\section}[1]{\CellSubsubsection*{##1}\refstepcounter{section}}%
  \renewcommand{\subsection}[1]{\CellSubsubsection*{##1}\refstepcounter{subsection}}}
\title{
Panoramic modelling of single-cell transcriptomes at scale with CellNet 
}
\date{}
\author[1,\textdagger]{Xiaoqing Lian}
\author[1,\textdagger]{Pei Liu}
\author[1]{Jiahui Tan}
\author[1]{Luoda Tan}
\author[1]{Chaoyi Li}
\author[1]{Zhonghao Ren}
\author[1]{Hongxin Xiang}
\author[2]{Tianxu Lv}
\author[3]{He Wang}
\author[3]{Xiang Pan}
\author[4]{Quan Zou}
\author[5,*]{Chen Lin}
\author[6,7,*]{Xiangxiang Zeng}
\affil[1]{College of Computer Science and Electronic Engineering, Hunan University, Changsha, 410082, China}
\affil[2]{AI for Life Sciences Lab, Tencent, Shenzhen, 518000, China}
\affil[3]{School of Artificial Intelligence and Computer Science, Jiangnan University, Wuxi, 214122, China}
\affil[4]{University of Electronic Science and Technology of China, Chengdu, 610054 China}
\affil[5]{School of Informatics, National Institute for Data Science in Health and Medicine, Xiamen University, Xiamen, 361000, China}
\affil[6]{State Key Laboratory of Chemo and Biosensing, Institute of AI4Science, Hunan University, Changsha, 410082, China}
\affil[7]{Yuelushan Laboratory, Changsha, China}
\affil[$\dagger$]{These authors contributed equally to this work.}
\begin{document}
\maketitle
\CellSection*{SUMMARY}
Comprehensively characterizing cellular landscapes to elucidate cell diversity is a central goal in biology. Recent advances in single-cell profiling have created unprecedented opportunities for this endeavor by capturing diverse transcriptomes with remarkable breadth. However, current foundation models generally struggle to leverage such holistic and valuable information to represent a cell. Here we present CellNet, which enables panoramic modelling of single-cell transcriptomes at scale by integrating DNA-derived gene identities and measured RNA expression into a shared image representation. Pretrained on 231 million image-based cell profiles across 40 species and covering over one million genes, CellNet improves the recognition of rare and condition-associated cellular states, enables the study of genes beyond experimentally measured panels, and uncovers associations between poorly characterized non-coding RNAs and known biological programmes. CellNet paves the way for AI-enabled cell-biology discovery at the whole-transcriptome scale. The CellNet website is available at https://www.cellnet.top.

\OriginalHeadings
\let\FloatBarrier\relax
\section{Introduction}\label{sec1}

Elucidating cellular diversity is essential for understanding tissue development, identifying disease-associated cell populations, and investigating gene programmes linked to specialized functions~\cite{bunne2024build,shapiro2013singlecell,stegle2015challenges,stuart2019integrative}. Single-cell transcriptomic atlases have enabled this by characterizing cell identities and states across tissues and species~\cite{tabulasapiens,salmen2022,tarashansky2021mapping}. However, the molecular view provided by these atlases depends on which transcripts are measured. Conventional poly(A)-based profiling underrepresents many RNA classes, hindering the study of their expression and relationships with other genes. Advanced technologies, \textit{e.g.}, VASA-seq and TotalX, overcome this limitation by providing a more comprehensive view of the transcriptome, spanning both coding and non-coding RNAs~\cite{salmen2022,totalx}. This creates unprecedented opportunities to uncover new cell-type markers, developmental expression patterns, and regulatory modules.

Single-cell foundation models have made considerable progress in learning transferable representations from transcriptomic data, enabling a wide range of applications including cell annotation, gene-network analysis, and cross-species comparison~\cite{boiarsky2024deeper,jiang2026landscape}. Geneformer~\cite{geneformer2023} demonstrated how large-scale pretraining enables predictions in network biology, while scGPT~\cite{scgpt} extended generative modelling to single-cell multi-omics. UCE~\cite{uce} and TranscriptFormer~\cite{transcriptformer} learned universal cell representations across datasets and species. However, the molecular information these models leverage to represent a cell remains constrained, owing to the longstanding trade-off between gene coverage and computational efficiency, although advanced sequencing assays have captured diverse transcriptomes with remarkable breadth. Protein-derived identities cannot directly describe genes without protein products, whereas predefined gene vocabularies are ill-suited to open-set cellular representation. Moreover, gene-as-token paradigms make broader coverage more computationally demanding; as a compromise, UCE, for instance, samples 1,024 expressed genes per cell~\cite{uce}. Consequently, critical transcripts measured by advanced assays may fall outside the molecular context available to models, leaving a substantial information gap that constrains the model's capacity in key tasks such as cell-state recognition, expression prediction, and gene-association discovery.

A representation that keeps pace with these measurements must accommodate diverse gene identities while not making cellular context prohibitively expensive to models. Genomic sequence supplies a unified identity for both coding and non-coding genes, independent of any protein product~\cite{evo2,dallatorre2025nt,avsec2021enformer,dealmeida2025segmentnt}. An image-based layout offer a feasible approach to organizing numerous genes into a structured cellular representation~\cite{sharma2019deepinsight,lian2026understandinggeneexpressiontoimagelearning}. In addition, the inherent sparsity and noise of single-cell measurements motivates a learning strategy that not only captures stable cellular information but also preserves underlying gene-level expression variations~\cite{luecken2019single,2026Empowering}. These principles together provide a route towards representing the full transcriptomic space across biological scales, spanning cells, tissues, and species.

Here we present CellNet, a single-cell foundation model that enables panoramic modelling of single-cell transcriptomes at scale by integrating DNA-derived gene identities and cellular expression into a common image representation. DNA sequence-derived gene features determine gene layouts in the image, while a multiview joint-embedding predictive architecture promotes consistent representations under noisy readouts. Gene-conditioned expression prediction is progressively introduced in self-supervised pretraining to encourage the retention of gene-level information in cellular representations. Alongside the foundation model, we publicly release a collection of 231 million transcriptomic images from 40 species, providing a large-scale, reusable resource for image-based single-cell studies. Experimental results show that CellNet improves identification of rare cell populations and condition-associated states and remains competitive on other representation-based tasks. Moreover, it recovers withheld coding-gene expression and selected non-coding transcripts in total-RNA data, with extensions to spatial measurements and targets without expression-pretraining exposure. Finally, its gene representations retrieve known molecular modules and developmental or neural programmes, including from sparsely detected non-coding RNA queries. These observations suggest that CellNet provides a foundation for cell-biology discoveries at the whole-transcriptome scale through its panoramic modelling of single-cell transcriptomes.

\begin{figure}[!htbp]
\centering
\includegraphics[width=\textwidth]{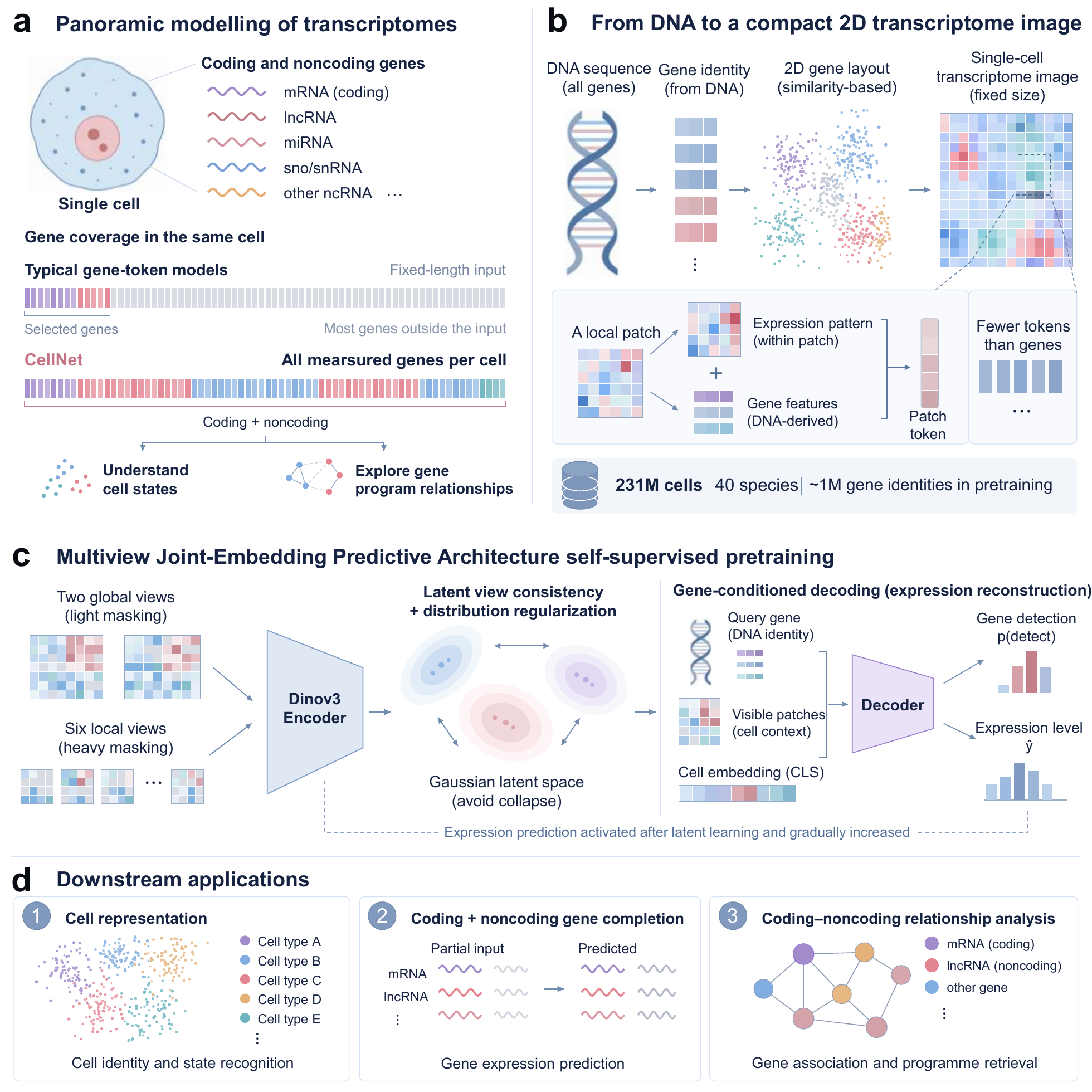}
\caption{\textbf{Panoramic modelling of single-cell transcriptomes with CellNet.}
\textbf{a}, Conventional gene-token inputs represent selected genes within a limited context, whereas CellNet organizes a broader range of coding and non-coding genes for cellular modelling.
\textbf{b}, DNA-derived gene identities define species-specific two-dimensional layouts onto which expression measurements are mapped. Local patches combine expression information with aggregated gene identities to form contextual tokens.
\textbf{c}, Multiview latent learning aligns partially masked views of the same cell, with Gaussian distribution regularization. Gene-conditioned decoding is progressively introduced to predict gene detection and expression magnitude from cellular context.
\textbf{d}, The pretrained representations support cell identity and state recognition, coding and non-coding gene-expression prediction, and the retrieval of molecular associations and biological programmes.}
\label{fig:CellNet}
\end{figure}
\section{Results}\label{sec2}
\subsection{Panoramic transcriptome modelling with DNA-informed images}

To move beyond the restricted gene coverage of conventional gene-token representations, we developed CellNet, which organizes single-cell transcriptomes into DNA-informed images (Fig.~\ref{fig:CellNet}a,b). Gene identities were derived from Evo 2 nucleotide sequence representations, providing a common basis for modelling protein-coding and non-coding genes~\cite{evo2}. For each gene, nucleotide features were pooled into a 1,920-dimensional vector and reduced to 256 dimensions by principal-component projection. These features were further used to construct species-specific two-dimensional gene layouts through dimensionality reduction and balanced spatial assignment\cite{sharma2019deepinsight}. Each gene occupied a distinct pixel on a $384 \times 384$ canvas, with positions fixed across cells of the same species. Expression measurements were mapped to these positions alongside assay-specific masks distinguishing measured zeros from unmeasured genes.

Each transcriptome image was divided into non-overlapping patches of $16 \times 16$ pixels, yielding up to 576 patch tokens (Fig.~\ref{fig:CellNet}b). Within each patch, aggregated DNA-derived gene identities were combined with local expression information before entering the DINOv3 backbone~\cite{dinov3}. This image-based organization allows many genes to contribute to a shared cellular context without allocating a separate global token to every gene. The encoder learns a cell-level representation while retaining contextual features that can be queried for individual gene predictions.

To learn cellular representations from sparse, noisy measurements, we trained CellNet using multiview latent-space learning (Fig.~\ref{fig:CellNet}c). Two lightly masked global views and six more extensively masked local views were generated from each cell. Their latent representations were encouraged to agree, while Gaussian distribution regularization mitigated representation collapse~\cite{lejepa}.

To retain gene-level expression information, we paired latent-space learning with a gene-conditioned expression decoder (Fig.~\ref{fig:CellNet}c). Given the DNA-derived identity of a target gene, the decoder uses the cell representation and visible patch context to predict whether the gene is detected and its expression magnitude. The expression-prediction objective was introduced after an initial warm-up period, with its weight gradually increased during pretraining. This design trains the shared encoder for both cellular representation learning and gene-level prediction.

We pretrained CellNet on approximately 231 million cell profiles spanning 40 species and publicly released the corresponding transcriptomic image collection. For pretraining, we constructed DNA-derived identity representations for over one million distinct genes across these species, including both protein-coding and non-coding genes. The main pretraining run took approximately two weeks on eight NVIDIA A800 GPUs. In comparison, UCE was trained for 40 days on 24 A100 GPUs~\cite{uce}, while scLong reported 35 GPU days on a cluster of 96 A100 GPUs~\cite{sclong}. TranscriptFormer reported training on a CZI computing cluster equipped with 1,000 H100 GPUs~\cite{transcriptformer}.  Broad molecular coverage, efficient image-based encoding and complementary pretraining objectives make CellNet a practical approach to panoramic transcriptome modelling at scale, enabling researchers to investigate rare cell states, predict expression for open-set genes, and uncover gene associations and coordinated gene programmes across a wider transcriptomic landscape (Fig.~\ref{fig:CellNet}d).

\begin{figure}[!htbp]
\centering
\includegraphics[width=\textwidth]{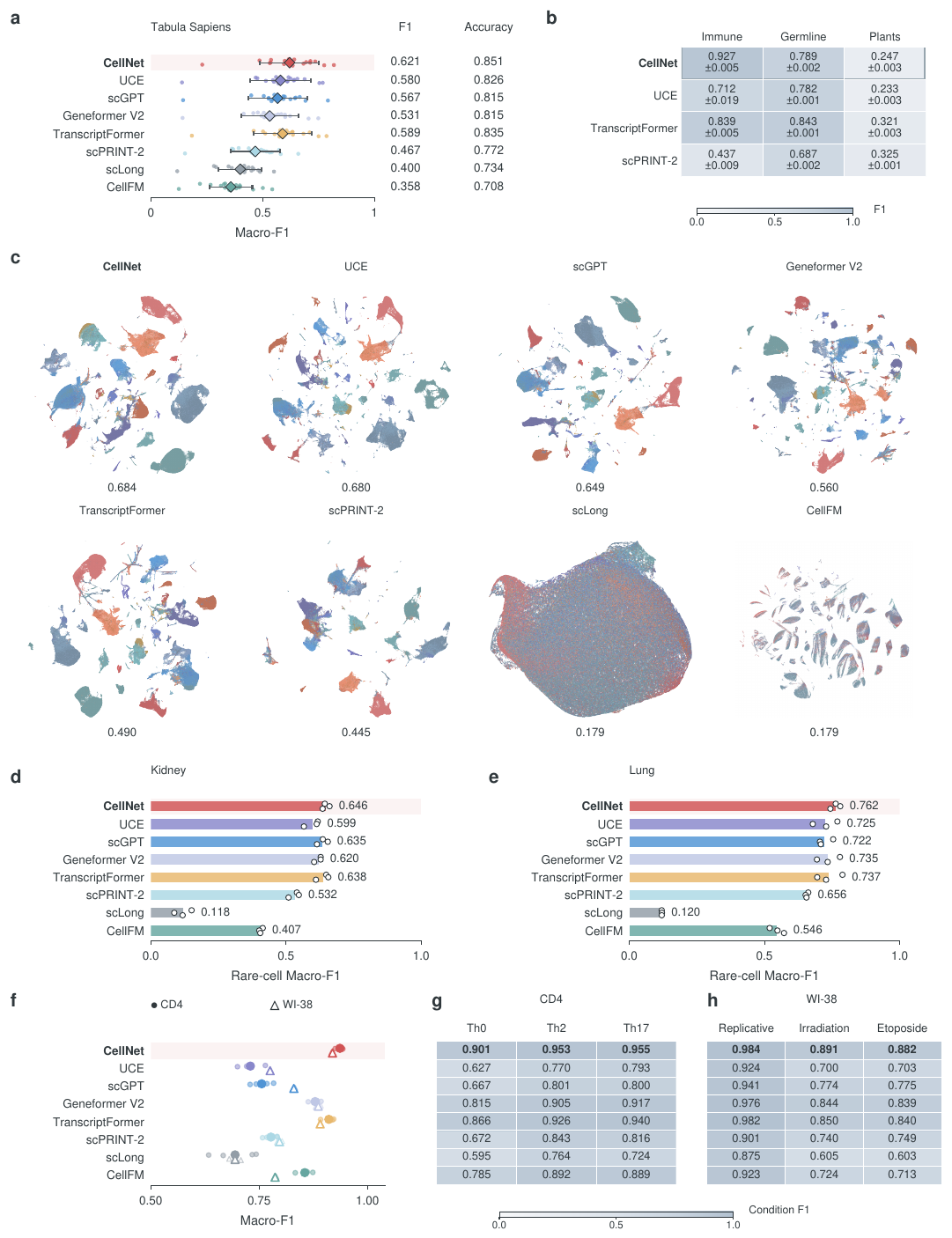}
\caption{Cell representations across tissues, species and conditions. a, Cell-type classification across 18 Tabula Sapiens v2 tissues. b, Within-species fine-label classification across immune, germline and plant references. c, Cell-type embeddings and AvgBIO across 26 tissues. AvgBIO denotes the equal-weight mean of normalized mutual information (NMI), adjusted Rand index (ARI) and cell-type silhouette rescaled as $(s+1)/2$. d,e, Rare-cell classification in kidney and lung. f, Classification of CD4 T-cell polarization and WI-38 senescence conditions. g,h, Condition-specific performance in the experiments shown in f.}
\label{fig:main2}
\end{figure}
\subsection{Cellular representations across tissues and species}

A broader view of the transcriptome is most useful when molecular information can be interpreted in the context of distinct cellular identities. We therefore evaluated whether CellNet retained these identities across tissues and donors. Using 549,018 annotated cells from 18 tissues in Tabula Sapiens v2\cite{tabulasapiensv2}, we compared seven pipelines based on frozen foundation-model representations under donor-held-out evaluation\cite{geneformer2023,scgpt,sclong,scprint2,transcriptformer,uce}. CellNet achieved a tissue-equally weighted Macro-F1 of 0.621, exceeding the strongest comparison pipeline, TranscriptFormer (0.589; Fig.~\ref{fig:main2}a). Improvements were observed across multiple organs, including bone marrow, liver and subcutaneous adipose tissue (Supplementary Fig.~\ref{fig:fig2s1}a).

Reliable cellular representations must also preserve biological differences across experiments while accommodating donor-associated variation. We assessed this balance across 26 multi-donor tissues comprising 682,741 cells using the scIB framework\cite{scib}. CellNet achieved the highest biological-conservation score (AvgBIO = 0.684, compared with 0.680 for UCE), while maintaining a batch-mixing score of 0.886 (Fig.~\ref{fig:main2}c). Tissue-level results across all 26 tissues are shown in Supplementary Fig.~\ref{fig:fig2s1}b,c.

We further examined whether CellNet could represent cellular identities across species with different gene repertoires. Evaluations covered immune, germline and plant references, including within-species identification and directional cross-species label transfer\cite{tarashansky2021mapping,uce,transcriptformer} (Fig.~\ref{fig:main2}b; Supplementary Fig.~\ref{fig:fig2s2}). Within-species recognition was particularly strong across seven immune species, with a mean Macro-F1 of 0.871 versus 0.809 for TranscriptFormer. Cross-species transfer showed task-dependent performance: CellNet led the evaluated plant settings, whereas UCE and TranscriptFormer performed better in several animal tasks. Across the evaluated datasets and organisms, CellNet retained useful cellular identity information while extending the molecular scope available for modelling.

\subsection{Resolving rare cell populations and condition-associated states}

Cellular diversity extends beyond abundant cell types. Rare populations may have specialized functions, while cells sharing the same broad identity can exhibit distinct transcriptional states. Recognizing these differences is important for identifying the cellular populations and molecular signatures associated with tissue function and disease. We therefore examined whether CellNet could resolve such fine-grained diversity.

We first evaluated rare-cell recognition in independently collected human kidney and lung atlases\cite{lake2023kidney,travaglini2020}, comprising 107,344 and 69,650 cells, respectively. Under donor-disjoint evaluation, CellNet achieved the highest rare-class Macro-F1 among seven comparison pipelines in both tissues: 0.646 in kidney and 0.762 in lung, compared with 0.638 and 0.737 for TranscriptFormer (Fig.~\ref{fig:main2}d,e). Class-level analyses showed recognition of low-abundance populations, including kidney mast cells, parietal epithelial cells and lung myofibroblasts (Supplementary Fig.~\ref{fig:fig2s4}a). Precision–recall analyses further characterized rare-population recovery (Supplementary Fig.~\ref{fig:fig2s3}b,c).

We next examined transcriptional differences within shared cellular identities using CD4 T-cell polarization\cite{canogamez2020} and WI-38 fibroblast senescence-induction experiments\cite{wechter2023}. In donor-held-out CD4 T cells, CellNet distinguished Th0, Th2 and Th17 conditions with a Macro-F1 of 0.936, compared with 0.911 for TranscriptFormer (Fig.~\ref{fig:main2}f,g). Complete-library holdouts in WI-38 cells yielded a Macro-F1 of 0.919 versus 0.891 across three induction conditions (Fig.~\ref{fig:main2}h). These advantages were observed across individual conditions and biological replicates (Supplementary Fig.~\ref{fig:fig2s5}a--c). The non-coding RNA \textit{NEAT1} also exhibited consistent condition-associated expression differences across CD4 donors, illustrating a molecular feature of these states (Supplementary Fig.~\ref{fig:fig2s5}d).

CellNet preserves both broad cellular identities and finer distinctions among rare populations and condition-associated states, providing a cellular representation for subsequent analyses of gene expression and molecular relationships.
\FloatBarrier

\begin{figure}[!htbp]
\centering
\includegraphics[width=0.9\textwidth]{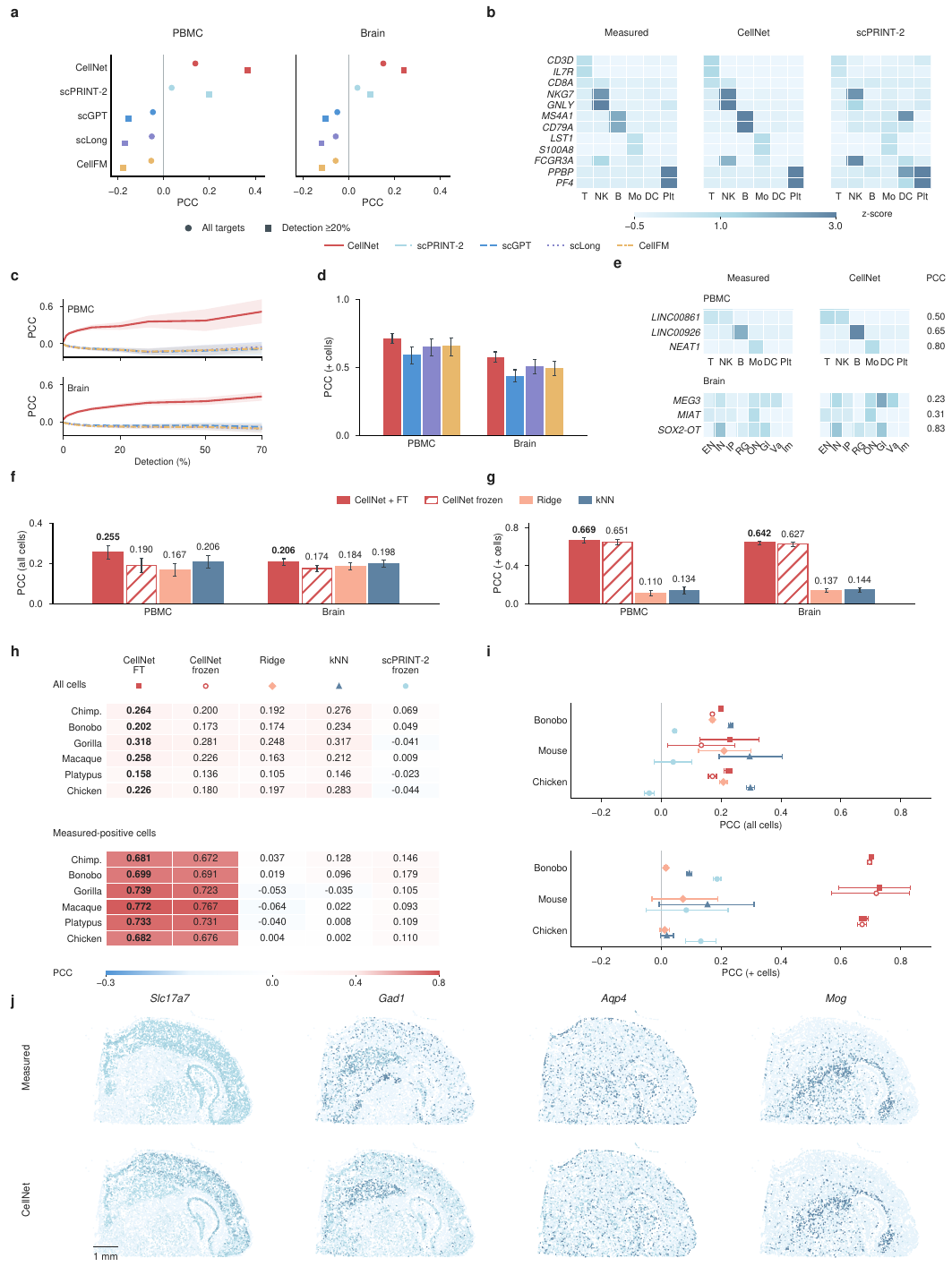}
\caption{Gene-level expression prediction across genes, species and assays. \textbf{a,b}, Frozen coding-gene recovery and measured/predicted immune-marker profiles. \textbf{c--e}, ncRNA prediction from coding-only inputs, including threshold performance and population profiles. \textbf{f,g}, ncRNA prediction after adaptation with 500 reference cells, compared with frozen CellNet and matched-reference Ridge/kNN. \textbf{h,i}, Prediction across six species, including targets unseen during CellNet expression pretraining; upper and lower panels show all-cell and measured-positive intensity PCC, respectively. \textbf{j}, Measured and predicted maps for four jointly withheld mouse-brain markers across 12,000 cells. PCCs are gene-wise means; error bars show paired 95\% bootstrap intervals (2,000 resamples). Maps use independent gene/method z-scores without smoothing; scale bar, 1\,mm. Supplementary Figs.~\ref{fig:ed1}--\ref{fig:ed6} provide complementary analyses.}
\label{fig:main}
\end{figure}
\subsection{Expression recovery across coding and non-coding genes}
Single-cell assays capture only part of the transcriptome, leaving many genes of biological interest unmeasured, including non-coding RNAs that may inform cellular identity and state. We asked whether CellNet could estimate the expression of such genes from the measured transcriptome, allowing researchers to investigate transcripts beyond the original measurement panels. We evaluated this capability by withholding target genes from TotalX profiles of human peripheral blood mononuclear cells (PBMCs) and developing brain~\cite{totalx}, using 5,000 query cells from each dataset.

We first assessed protein-coding gene recovery using fivefold gene-wise withholding (Fig.~\ref{fig:main}a,b). Among genes detected in at least 20\% of cells, frozen CellNet achieved mean gene-wise Pearson correlations (PCCs) of 0.365 in PBMCs and 0.241 in developing brain, outperforming all four evaluated pretrained comparators. The corresponding PCCs for scPRINT-2~\cite{scprint2} were 0.198 and 0.094 (Fig.~\ref{fig:main}a). Beyond these aggregate measurements, predicted immune-marker expression retained differences among lymphoid, myeloid and platelet populations, closely resembling the measured profiles (Fig.~\ref{fig:main}b). Detection-stratified analyses and additional lineage markers further characterized prediction accuracy across genes with different expression prevalence (Supplementary Fig.~\ref{fig:ed1}).

We next examined non-coding RNAs, many of which are poorly represented in conventional single-cell measurements and existing foundation models. With all non-coding targets withheld simultaneously, CellNet inferred their expression using only the observed coding transcriptome as cellular context (Fig.~\ref{fig:main}c). For non-coding targets detected in at least 20\% of cells, mean all-cell PCCs reached 0.285 in PBMCs and 0.265 in developing brain. Among cells with detected target expression, conditional intensity PCCs reached 0.716 and 0.577, respectively, exceeding scGPT, scLong and CellFM~\cite{cellfm} on the corresponding targets (Fig.~\ref{fig:main}d). Recovered expression profiles retained population-associated patterns for selected non-coding transcripts, including LINC00861, LINC00926 and NEAT1 in PBMCs, and MEG3, MIAT and SOX2-OT in developing brain (Fig.~\ref{fig:main}e). Beyond the shared accuracy benchmarks, DNA-derived identities enabled predictions for 18,502 PBMC and 26,113 brain non-coding RNA targets, substantially extending the molecular scope of the decoder (Supplementary Fig.~\ref{fig:ed2}).

We further investigated whether limited reference measurements could refine these predictions. In a separate reference-adaptation evaluation using fixed sets of 108 PBMC and 323 brain non-coding targets, adapting the native decoder with 500 reference cells increased mean all-cell PCC from 0.190 to 0.255 and from 0.174 to 0.206, respectively (Fig.~\ref{fig:main}f). Conditional intensity prediction also exceeded matched-reference Ridge regression and nearest-neighbour baselines, with PCCs of 0.669 and 0.642 (Fig.~\ref{fig:main}g). Increasing reference size yielded further improvements (Supplementary Fig.~\ref{fig:ed3}). CellNet can use observed cellular expression to predict coding and non-coding targets, with limited reference data further refining gene-level inference.

\subsection{Extending expression inference across species and assays}

The genes available for molecular analysis vary across organisms~\cite{gallego2012comparative}, while experimental technologies measure different portions of the transcriptome~\cite{moro2026targeted}. We therefore investigated whether CellNet could extend gene-level prediction across species, including genes absent from expression pretraining, and recover expression patterns under distinct measurement conditions.

Across six nonhuman species from the spermatogenesis atlas~\cite{spermatogenesis2023}, frozen CellNet achieved measured-positive expression-intensity PCCs of 0.672–0.767, compared with 0.093–0.179 for scPRINT-2 (Fig.~\ref{fig:main}h).
Adaptation with 500 within-species reference cells yielded conditional intensity PCCs of 0.681–0.772. We further evaluated genes without expression-pretraining exposure in bonobo, mouse and chicken, obtaining frozen conditional intensity PCCs of 0.696, 0.718 and 0.671, respectively (Fig.~\ref{fig:main}i). These targets retained DNA-derived identities despite lacking expression-pretraining exposure, allowing CellNet to generate predictions beyond the gene targets used in its expression training. Complete-expression correlations, target coverage and exposure-stratified results were evaluated separately (Supplementary Fig.~\ref{fig:ed4}).

Targeted spatial transcriptomics offers an additional test of expression prediction~\cite{tian2023expanding}. Because these assays use limited gene panels~\cite{zhang2024gpsfish}, accurate inference of genes of interest is particularly valuable~\cite{sun2024tissue}. We examined whether CellNet could recover such molecular information using a Xenium mouse-brain section~\cite{xenium2024mousebrain}, jointly withholding 20 predefined genes from 12,000 cells while retaining the remaining measured genes as input. Without spatial coordinates, frozen CellNet reproduced characteristic spatial expression patterns for Slc17a7, Gad1, Aqp4 and Mog, corresponding to major neuronal and glial populations (Fig.~\ref{fig:main}j). Across all 20 targets, mean all-cell PCC reached 0.478 and increased to 0.665 following same-section reference adaptation (Supplementary Fig.~\ref{fig:ed5}). Even when the observed panel was reduced to 1,000 genes, adapted predictions retained a mean PCC of 0.606 (Supplementary Fig.~\ref{fig:ed6}a). Measured-positive comparisons and additional marker maps supported recovery across panel sizes and neuronal, glial, myeloid and vascular populations (Supplementary Fig.~\ref{fig:ed6}b,c).

These evaluations show that CellNet can infer gene expression from incomplete molecular observations across diverse gene repertoires and assays. By extending prediction beyond the transcripts supplied as input, panoramic modelling offers a means to investigate additional molecular features in existing single-cell datasets.

\FloatBarrier

\begin{figure}[!htbp]
\centering
\includegraphics[width=\textwidth]{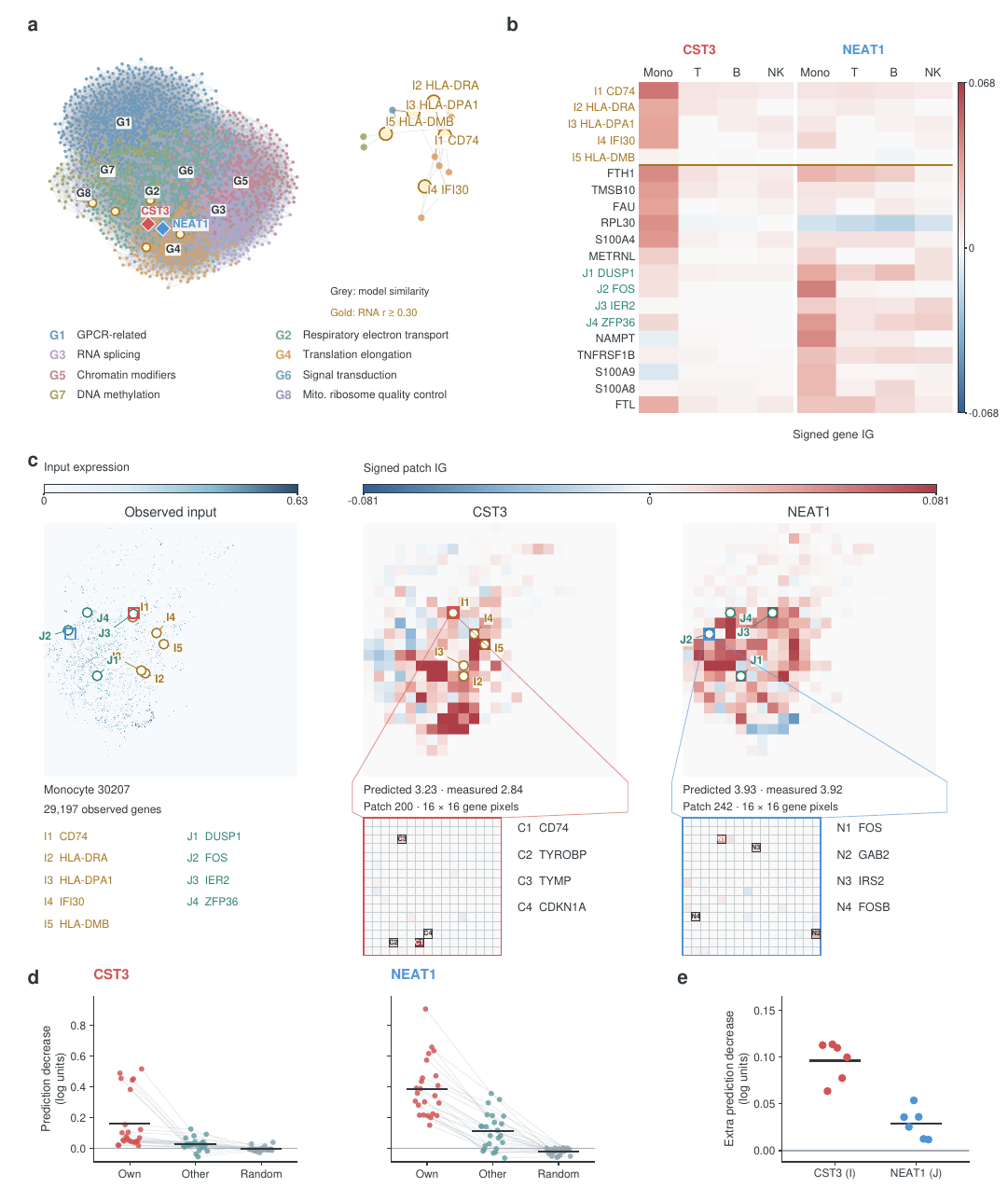}
\caption{\textbf{Distinct measured RNA inputs support coding and non-coding gene predictions.}
\textbf{a}, Network of gene representations from 126 monocytes. The 3,279 genes were selected by detection frequency, platelet expression and prespecified references (Section~\ref{note:network}).
\textbf{b}, Mean signed integrated-gradient (IG) contributions to predictions of withheld \textit{CST3} and \textit{NEAT1}, using six cells from each of four populations and a zero-expression baseline.
\textbf{c}, A shared monocyte expression image and target-specific contribution maps, with enlarged views of the strongest positive contributors (C1--C4 and N1--N4).
\textbf{d}, Prediction decreases after zeroing the 32 strongest positive-contribution RNAs selected for the same target (Own), the alternative target (Other) or eight expression-matched random sets (Random).
\textbf{e}, Additional prediction decreases relative to matched random removal for the selected antigen-presentation and response-associated gene sets in six monocytes. Other input information and normalization remain fixed. Points represent cells and bars indicate means; random trials are averaged within cells. Further details are provided in Section~\ref{note:inputs}.}
\label{fig:4}
\end{figure}
\subsection{Distinct RNA inputs inform coding and non-coding gene predictions}

Predicting the expression of unmeasured genes allows researchers to investigate transcripts beyond those captured by an assay. Understanding which measured RNAs contribute to these predictions may also help identify the cellular programmes associated with genes of interest~\cite{aibar2017scenic,pratapa2020benchmark}. We therefore examined the relationships learned by CellNet between observed RNA expression and predictions for coding and non-coding genes.

We first examined the biological organization of the learned gene representations. A network constructed from 3,279 genes in 126 monocytes contained eight communities enriched for processes including translation, RNA splicing and chromatin modification (Fig.~\ref{fig:4}a and Supplementary Fig.~\ref{ed:network}). The network also linked the MHC-II genes \textit{HLA-DRA} and \textit{HLA-DPA1}, whose measured expression was correlated ($r=0.552$). We then traced predictions of the protein-coding gene \textit{CST3} and long non-coding RNA \textit{NEAT1}, withholding both targets in 24 cells spanning monocyte, T-cell, B-cell and natural killer-cell populations. Integrated-gradient attribution~\cite{sundararajan2017} mapped their predictions to measured RNA inputs (Fig.~\ref{fig:4}b,c). Even within the same monocyte image, \textit{CST3} predictions emphasized \textit{CD74}, \textit{HLA-DRA}, \textit{HLA-DPA1}, \textit{IFI30} and \textit{HLA-DMB}, whereas \textit{NEAT1} predictions emphasized \textit{DUSP1}, \textit{FOS}, \textit{IER2} and \textit{ZFP36}. These patterns are consistent with reported links between cystatin C and MHC-II processing\cite{kitamura2005}, as well as the involvement of \textit{NEAT1} and the identified response-associated genes in stress-related programmes\cite{lellahi2018,canzoneri2020}. Contributions to \textit{NEAT1} predictions also varied across cell populations (Supplementary Fig.~\ref{ed:context}).

To test whether these attributed RNAs contributed to the predictions, we selectively removed their expression from the input while preserving other information. Removing each target's 32 strongest positive contributors reduced its predicted expression more than removing the other target's contributors or expression-matched random inputs (Fig.~\ref{fig:4}d). Relative to random removal, the additional decreases averaged 0.165 log-expression units for \textit{CST3} and 0.406 for \textit{NEAT1}, with positive contrasts across all 24 cells. Removing the identified antigen-presentation and response-associated gene sets similarly produced target-specific decreases of 0.096 and 0.029 log units beyond matched random removal in six monocytes, with negligible cross-target effects (Fig.~\ref{fig:4}e and Supplementary Fig.~\ref{ed:named}). Additional input-removal controls are presented in Supplementary Fig.~\ref{ed:controls} and Section~\ref{note:inputs}. Related analyses linked CD4 T-cell polarization scores~\cite{canogamez2020} and cellular representation features to established expression programmes~\cite{kotliar2019programs} (Supplementary Figs.~\ref{ed:state} and \ref{ed:features}).

These analyses show that CellNet organizes gene representations into biologically coherent modules, uses distinct measured-RNA inputs to predict coding and non-coding targets, and captures reproducible cell-state programmes in its cellular representations.

\FloatBarrier

\begin{figure}[!htbp]
\centering
\includegraphics[width=\textwidth]{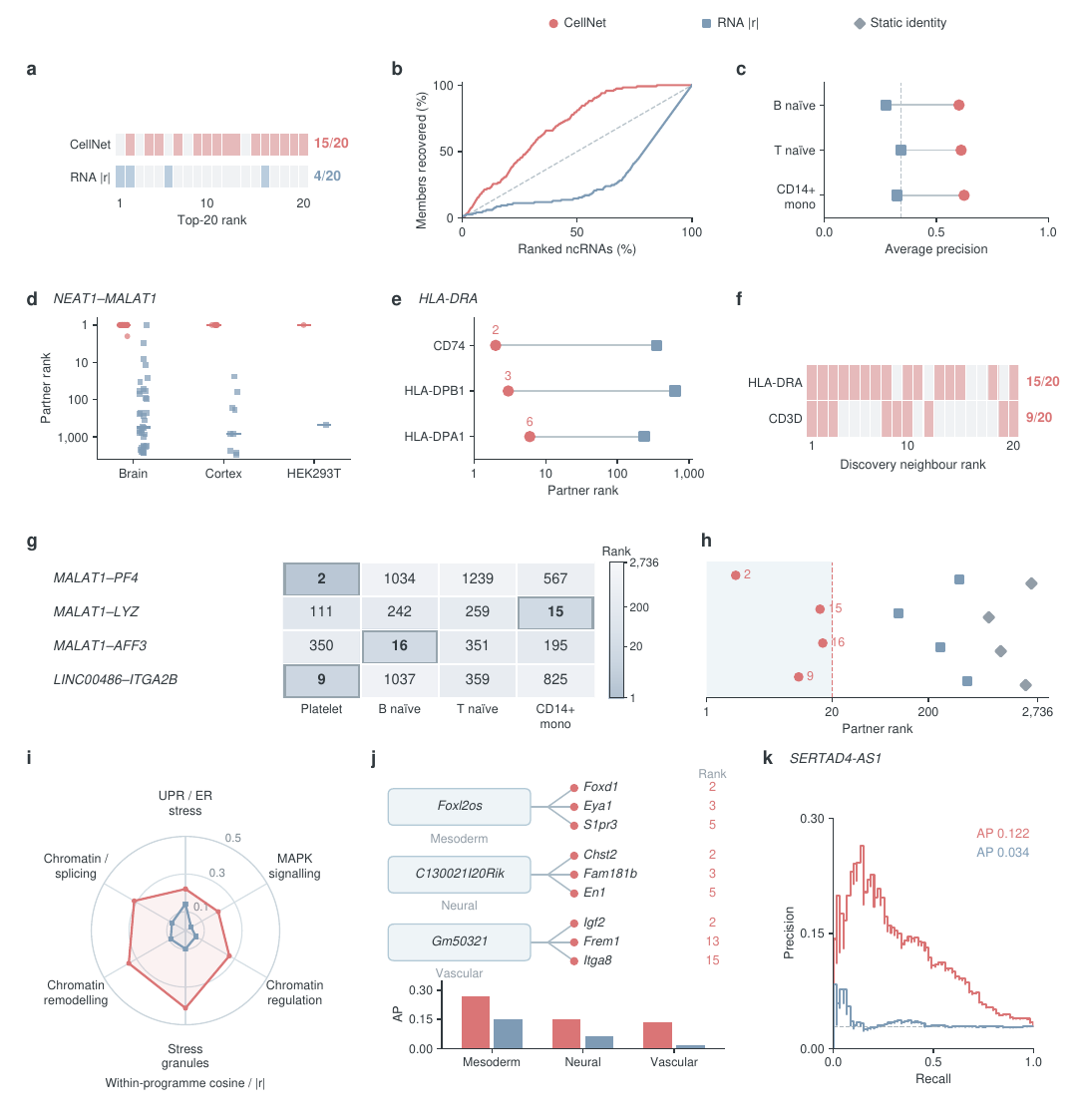}
\caption{\textbf{Recovery of molecular associations and cellular programmes across RNA classes.}
Gene-representation similarity and absolute RNA Pearson correlation were compared on matched candidates.
\textbf{a--c}, A published non-coding RNA (ncRNA) module (30 reference genes; 343 candidates, including 117 known members): top-20 ranking in naive B cells (a), cumulative recovery (b) and average precision (AP) across naive B, naive T and CD14-positive monocytes (c); tiles mark known members.
\textbf{d}, \textit{NEAT1}-guided ranking of \textit{MALAT1} across 45 supported cell populations. \textbf{e}, \textit{HLA-DRA} partner retrieval in 100 memory B cells. \textbf{f}, Recurrence of fixed top-20 neighbours in an external PBMC dataset.
\textbf{g,h}, Rankings of selected ncRNA--coding-gene pairs in protocol-matched populations (g), compared with RNA correlation and static gene-identity similarity (h).
\textbf{i}, Within-programme representation similarity for six host-response programmes in 500 DENV-exposed Huh7 cells versus size-matched random sets.
\textbf{j}, Recovery of developmental coding-gene programmes from three ncRNA queries in VASA-seq mouse embryos, with representative partners and full-ranking AP.
\textbf{k}, Neural stem-cell programme retrieval using \textit{SERTAD4-AS1} in 3,423 non-targeting human neural-induction controls.
See Sections~\ref{note:retrieval} and \ref{note:programmes} for reference definitions and evaluation protocols.}
\label{fig:5}
\end{figure}
\subsection{Panoramic gene representations recover relationships across RNA classes}

Cellular specialization involves coordinated molecular programmes that can span both coding and non-coding RNAs~\cite{salmen2022,totalx}. Some single-cell foundation models represent genes using protein-sequence-derived identities and therefore lack direct gene representations for transcripts without protein products~\cite{uce}. By representing genes through DNA-derived identities, CellNet brings both RNA classes into a shared, cell-contextual gene space. We investigated whether this expanded representation could recover established molecular modules and gene associations.

Using 30 reference RNAs from a published TotalX non-coding coexpression module\cite{totalx}, CellNet recovered 15 known members among the top 20 of 343 candidates in naive B cells, compared with four using RNA correlation (Fig.~\ref{fig:5}a,b). Mean average precision (AP) across naive B cells, naive T cells and CD14-positive monocytes was 0.612 versus 0.314 (Fig.~\ref{fig:5}c). Individual queries also retrieved known partners across RNA classes. \textit{NEAT1} ranked \textit{MALAT1} highly across supported cell populations, whereas \textit{HLA-DRA} retrieved the antigen-presentation genes \textit{CD74}, \textit{HLA-DPB1} and \textit{HLA-DPA1} at ranks 2, 3 and 6 (Fig.~\ref{fig:5}d,e). Prior studies provide context for these retrievals: NEAT1 and MALAT1 associate with active chromatin, CD74 interacts with HLA-DR, and HLA-DPA1/DPB1 form the HLA-DP heterodimer~\cite{west2014neatmalat,karakikes2012,hollenbach2012}. In an external PBMC dataset~\cite{zheng2017}, 15 and nine of the fixed top-20 neighbours of \textit{HLA-DRA} and \textit{CD3D}, respectively, were recovered (Fig.~\ref{fig:5}f).

We next examined how cellular context influenced these relationships. \textit{MALAT1} retrieved \textit{PF4} in platelets, \textit{LYZ} in monocytes and \textit{AFF3} in naive B cells, whereas \textit{LINC00486} retrieved \textit{ITGA2B} in platelets (Fig.~\ref{fig:5}g,h). Prior work reports MALAT1 activity in platelets, AFF3 function in B-cell class switching, and PF4/ITGA2B expression in megakaryocytes~\cite{sun2023malatplatelet,tsukumo2022aff3,liu2021megakaryocyte}. These selected pairs ranked above their positions under RNA correlation and static gene-identity similarity. All four associations remained among the top 20 in both cell halves, while shuffling input expression reduced their similarities (Supplementary Fig.~\ref{ed:associations}e,f). At the programme level, six host-response programmes in DENV-exposed Huh7 cells showed greater within-programme similarity than size-matched random gene sets (Fig.~\ref{fig:5}i). CellNet's shared gene representation supports retrieval of established molecular relationships across RNA classes and cellular contexts.

\subsection{Non-coding RNA queries retrieve developmental and neural programmes}

Many long non-coding RNAs (lncRNAs) show cell-type-specific expression and are detected in only a subset of cells, potentially complicating direct expression-correlation analysis~\cite{liu2016lncrna}. Because CellNet defines gene identities independently of protein products, these RNAs can serve as queries within the same representation used for coding genes. We investigated whether this capability could recover coding-gene programmes associated with cellular development and specialization.

In VASA-seq mouse embryos\cite{salmen2022}, the non-coding RNA \textit{Foxl2os} retrieved mesenchymal-programme genes including \textit{Foxd1} and \textit{Eya1}; \textit{C130021I20Rik} retrieved neural/ectodermal genes including \textit{En1}; and \textit{Gm50321} retrieved early vascular-lineage genes including \textit{Igf2}, \textit{Frem1} and \textit{Itga8} (Fig.~\ref{fig:5}j). Full-ranking AP exceeded RNA correlation for all three queries, reaching 0.132 versus 0.018 for the vascular programme (Supplementary Fig.~\ref{ed:programmes}c,d).

We further examined \textit{SERTAD4-AS1} in human neural induction\cite{wu2022}. Although detected in only 13 of 3,423 non-targeting control cells, CellNet retrieved the published neural stem-cell programme genes \textit{WNT5B}, \textit{DLK1} and \textit{SERPINF1} among its top 20 neighbours (Fig.~\ref{fig:5}k). Using all control cells as expression context, CellNet achieved an AP of 0.122, compared with 0.034 for RNA correlation. Panoramic modelling makes sparsely measured non-coding RNAs accessible to gene-programme analysis, extending biological interpretation beyond the molecules routinely represented by protein-centred models.

\FloatBarrier
\section{Discussion}

CellNet combines DNA-derived gene identities with image-based transcriptome representations to place diverse RNA molecules in a shared cellular context while retaining gene-level representations. It improves recognition of rare populations and condition-associated states and performs competitively in dataset integration and cross-species evaluations. These results show that broad molecular coverage can coexist with useful cellular representations across tissues and biological conditions.

A key application of panoramic transcriptome modelling is predicting genes omitted from individual experiments. CellNet predicts withheld expression from partial transcriptomes, including selected non-coding RNAs and targets without expression-pretraining exposure, and recovers spatial expression patterns for genes outside targeted measurement panels. Its gene representations also recover established molecular modules and retrieve coding-gene programmes using non-coding RNA queries. For example, the sparsely detected \textit{SERTAD4-AS1} retrieved genes from a published neural stem-cell programme. These results make underrepresented RNAs more accessible to biological analysis and provide hypotheses about their expression and cellular associations.

The shared representation also provides opportunities to investigate the molecular information underlying model predictions. Input-attribution and targeted-removal analyses identified distinct sets of measured RNAs contributing to predictions of different coding and non-coding targets. Cell-contextual gene representations further recovered known partners and coordinated programmes across diverse cellular backgrounds. 

Computational tractability is essential for extending molecular coverage at scale. By organizing multiple genes within image patches, CellNet avoids assigning a separate global token to every gene. Multiview latent learning captures shared information across incomplete observations, while gene-conditioned expression prediction retains gene-level information. This design supported pretraining on 231 million cell profiles spanning 40 species using eight NVIDIA A800 GPUs over approximately two weeks. 
\CellSubsection*{Limitations of the study}
Prediction accuracy nevertheless varies across molecular classes, particularly for sparsely detected non-coding transcripts, and inferred associations require independent functional validation. As single-cell profiling continues to expand, more comprehensive RNA measurements and perturbation datasets will offer opportunities to refine these predictions and investigate their regulatory significance. CellNet provides a scalable framework for leveraging this growing molecular information to characterize cellular diversity and explore the gene programmes associated with specialized functions.

\CellSection*{RESOURCE AVAILABILITY}
% The source contains no Lead contact or Materials availability statements.
% These required declarations are recorded in the accompanying requirements report.
\CellSubsection*{Data and code availability}
\subsection{Data availability}

All datasets analysed in this study are publicly available from the original studies or official repositories; source links and accession details will be provided on the CellNet website (\mbox{\url{https://cellnet.top/}}). The pretraining compendium comprised 231,190,557 profiles from GEO, ArrayExpress, BioProject-associated and CELLxGENE sources. 

\subsection{Code availability}

The code used to construct transcriptomic images and pretrain CellNet is publicly available through the CellNet website: \mbox{\url{https://cellnet.top/}}.

\CellSection*{ACKNOWLEDGMENTS}
This work was supported by the National Natural Science Foundation of China (grant nos. U22A2037, 62425204, 62122025, 6245002, 62432011, 62602189).

% Author contributions and Declaration of interests are absent
% from the source. No example declarations or factual statements are inserted.
\CellSection*{SUPPLEMENTAL INFORMATION INDEX}
\noindent\hyperref[arxiv:supplement]{Supplementary Information (appendix at the end of this article)}
\CellSection*{STAR METHODS}
% A compliant Key resources table must be completed by the authors; the source
% dataset and software descriptions are retained verbatim below.
% This computational study uses public data; no new experimental-model claim is added.
\CellSubsection*{Method details}
\OriginalMethodHeadings
\setcounter{section}{3}\refstepcounter{section}\setcounter{subsection}{0}
\label{sec:methods}

\subsection{Overview and pretraining collection}

CellNet combines DNA-derived gene identities with image-based expression encoding to learn cellular representations and gene-level predictions within a shared model. Each transcriptome is represented by an expression image and a measurement mask. A transformer encodes these inputs into cellular context, which a gene-conditioned decoder uses to predict specified transcripts and generate contextual gene representations.

The main pretraining collection comprised 231,190,557 cell profiles from 26,467 datasets spanning 40 species. Genes were indexed using species-aware identifiers, and the model's molecular representation contained 1,015,163 genes. Expression processing, sampling and optimization are described in Supplementary Methods~\ref{sm:expression}-\ref{sm:architecture}.

\subsection{DNA-informed transcriptome images}

We represented gene identity using nucleotide sequence features from Evo 2~\cite{evo2}. For gene $g$, sequence features were averaged over nucleotide positions, yielding a 1,920-dimensional vector $\mathbf e_g$. A frozen principal-component transform provided the 256-dimensional representation $\mathbf a_g$ used for trainable adaptation:

\begin{equation}
\mathbf e_g = \operatorname{Pool}\!\left(f_{\mathrm{DNA}}(s_g)\right),
\qquad
\mathbf a_g = \mathbf P^{\mathsf T}(\mathbf e_g-\boldsymbol\mu).
\label{eq:main-dna-identity}
\end{equation}

Here, $s_g$ denotes the gene-associated nucleotide sequence, $f_{\mathrm{DNA}}$ the pretrained sequence encoder, and $\mathbf P$ and $\boldsymbol\mu$ the fixed projection basis and centring vector. The sequence features remained frozen during transcriptome pretraining, while downstream identity projections were learned jointly with the cellular encoder.

A fixed, species-specific layout was derived from the DNA sequence features by principal-component projection followed by t-SNE. The resulting coordinates were scaled to the $384\times384$ canvas and assigned to distinct pixels using balanced greedy k-d tree assignment~\cite{lian2026understandinggeneexpressiontoimagelearning}. The assignment distributed genes across $16\times16$ patches while retaining one gene per pixel. For cell $c$, normalized expression $x_{cg}$ was placed at the pixel assigned to gene $g$. A corresponding indicator $m_{cg}$ recorded whether the gene was measured by the assay. The same layout was used across cells, retaining measured zero-expression genes and accommodating protein-coding and non-coding identities through the sequence-based interface.

The image was divided into non-overlapping $16\times16$ patches. Image organization allowed multiple genes to contribute to each contextual token, providing a bounded patch grid as molecular coverage increased.

\subsection{Joint encoding of expression and gene identity}

Each patch combined expression, molecular composition and expression-modulated molecular identity. Let $v_{cg}$ indicate that gene $g$ is visible in the current view, and let $\mathcal G_p$ denote genes assigned to patch $p$. Two independent trainable projections produced $\mathbf b_g=T_{\mathrm{mean}}(\mathbf a_g)$ and the unit-normalized identity $\mathbf r_g=\operatorname{Norm}_2(T_{\mathrm{mod}}(\mathbf a_g))$. Patch tokens were constructed as

\begin{equation}
\begin{aligned}
\mathbf t^{\mathrm{expr}}_{cp}
&= \operatorname{LN}\!\left(
  C_{\mathrm{expr}}\!\left([\mathbf v_c\odot\mathbf x_c,\mathbf v_c]_p\right)
\right),\\
\mathbf t^{\mathrm{id}}_{cp}
&= \operatorname{LN}\!\left(
  W_{\mathrm{id}}
  \frac{\sum_{g\in\mathcal G_p}v_{cg}\mathbf b_g}
       {\max(1,\sum_{g\in\mathcal G_p}v_{cg})}
\right),\\
\mathbf t^{\mathrm{mod}}_{cp}
&= \operatorname{LN}\!\left(
  C_{\mathrm{mod}}\!\left(
    [v_{cg}(x_{cg}+\delta)\mathbf r_g]_{g\in\mathcal G_p}
  \right)
\right),\\
\mathbf t_{cp}
&= \frac{
  \mathbf t^{\mathrm{expr}}_{cp}
  +\mathbf t^{\mathrm{id}}_{cp}
  +\mathbf t^{\mathrm{mod}}_{cp}}
  {\sqrt{3}}.
\end{aligned}
\label{eq:main-patch-token}
\end{equation}

$C_{\mathrm{expr}}$ and $C_{\mathrm{mod}}$ are convolutional patch projections, $W_{\mathrm{id}}$ is an affine projection, and $\operatorname{LN}$ denotes layer normalization~\cite{ba2016layernorm}. The identity offset $\delta$ preserves a molecular contribution from visible genes with zero expression. Each branch produces a 768-dimensional vector.

A DINOv3 base model encoded the patch tokens into a cellular summary and contextual patch representations~\cite{dinov3}. Empty patches were omitted, and absolute positional embeddings were disabled. During training, local gene permutations varied within-patch ordering while preserving the correspondence between expression and identity. The resulting encoder supplied a common cellular context for representation learning and expression prediction.

\subsection{Multi-view JEPA representation learning}

We trained cellular representations using two global views and six more strongly masked local views of each transcriptome. All views retained the image extent and varied through patch masking and expression perturbations. Hidden genes were removed from the expression and visibility inputs.

Following the joint-embedding objective of LeJEPA~\cite{lejepa}, the encoder output was mapped to a 256-dimensional projection space. Let $\mathbf z_c^{(v)}$ denote the projected representation of cell $c$ in view $v$. Representations were aligned to the mean of the two global views:

\begin{equation}
\overline{\mathbf z}_c
=\frac{1}{2}\sum_{v\in\mathcal V_{\mathrm{global}}}\mathbf z_c^{(v)},
\qquad
\mathcal L_{\mathrm{align}}
=\frac{1}{BVD_z}
\sum_{c=1}^{B}\sum_{v=1}^{V}
\left\|\mathbf z_c^{(v)}-\overline{\mathbf z}_c\right\|_2^2.
\label{eq:main-alignment-loss}
\end{equation}

Here, $B$ denotes the batch size, $V=8$ the number of views and $D_z=256$ the projection dimension. Gradients were propagated through both the individual views and their global-view mean. Sketched isotropic Gaussian regularization constrained the projected distribution across the training batch, supporting non-collapsed representations without a separate teacher encoder. Details of the regularizer are provided in Supplementary Methods~\ref{sm:projector}.

\subsection{Gene-conditioned expression prediction}

A cross-attention decoder queried cellular context using projected DNA identities. The context combined the cellular summary with 32 spatially pooled patch tokens. For gene $g$, its query $\mathbf q_g$ interacted with this context through three decoder blocks, yielding $\mathbf u_{cg}$. The final contextual gene representation was

\begin{equation}
\mathbf q_g=T_{\mathrm{query}}(\mathbf a_g),
\qquad
\mathbf u_{cg}=D_{\mathrm{attn}}(\mathbf q_g,\mathbf C_c),
\qquad
\mathbf h_{cg}
=\operatorname{LN}\!\left(
W_{\mathrm{int}}
[\mathbf u_{cg};\mathbf u_{cg}\odot\mathbf q_g]
\right).
\label{eq:main-contextual-gene}
\end{equation}

$T_{\mathrm{query}}$ includes the learned identity projection and decoder query transformation. $\mathbf C_c$ denotes encoded cellular context, and $[\,;\,]$ denotes concatenation. This interaction representation combines gene identity with information retrieved from the observed transcriptome. Decoder queries can therefore address sequence-represented genes independently of whether their expression appears in the current input.

Expression was modelled using a compound Poisson–exponential distribution, corresponding to a Tweedie model with power parameter $1.5$~\cite{dunn2005tweedie,2025Diversity}. Two decoder outputs parameterized the event rate $\lambda_{cg}$ and positive-expression scale $\theta_{cg}$:

\begin{equation}
\begin{aligned}
N_{cg}&\sim\operatorname{Poisson}(\lambda_{cg}),\\
Y_{cg}&=\sum_{j=1}^{N_{cg}}E_{cgj},\\
E_{cgj}&\sim\operatorname{Exponential}(\text{mean}=\theta_{cg}),\\
\widehat{x}_{cg}&=\mathbb E[Y_{cg}]=\lambda_{cg}\theta_{cg}.
\end{aligned}
\label{eq:main-compound-poisson}
\end{equation}

The distribution jointly represents zero expression and continuous positive values. Supervision used measured genes hidden from each encoder view. Inverse-sampling weights accounted for separate sampling of positive and zero-expression targets. A deviation-weighted positive-expression term complemented the likelihood objective, emphasizing expression variation relative to source-specific gene statistics.

\subsection{Joint pretraining objective}

Cellular and molecular learning were coupled through the shared encoder using the joint objective

\begin{equation}
\mathcal L
=(1-\alpha)\mathcal L_{\mathrm{align}}
+\alpha\mathcal L_{\mathrm{SIGReg}}
+\lambda_{\mathrm{expr}}\mathcal L_{\mathrm{expr}}
\qquad
\mathcal L_{\mathrm{expr}}
=\mathcal L_{\mathrm{CPE}}+\beta\mathcal L_{\mathrm{dev}}.
\label{eq:main-joint-objective}
\end{equation}

We used $\alpha=0.05$, $\lambda_{\mathrm{expr}}=8$ and $\beta=0.25$. Expression gradients updated the shared encoder, linking cellular representation learning to gene-level prediction. We also used coarse-class prototype learning to support cross-species alignment (Supplementary Methods~\ref{sm:prototypes}).

\subsection{Cellular and molecular readouts}

The pretrained encoder remained frozen for cellular classification, integration and reference mapping. Molecular association analyses used the frozen gene decoder. Expression completion compared frozen predictions with separately specified reference-based decoder adaptation (Supplementary Methods~\ref{sm:expression-evaluation}). Cellular classification used layer normalization of the sum of the final eight raw cellular summary states. Shrinkage linear discriminant analysis (LDA) was the default cellular classification readout for CellNet, motivated by the isotropic Gaussian regularization used during representation learning. The six comparison models used logistic regression, a conventional linear probing approach in single-cell representation evaluation~\cite{sctab}. These heads were fitted using training labels, with feature standardization restricted to the corresponding training partition. The Gaussian regularizer acted in the pretraining projection space; downstream LDA estimated class means and a shared covariance from the encoder features.

Integration used the raw cellular summary from the sixth block counted from the encoder output. Reference mapping used the eighth block counted from the output and a shared distance-weighted nearest-neighbour classifier. These readouts and evaluation partitions are specified in Supplementary Methods~\ref{sm:cellular-evaluation}.

Expression completion queried the gene decoder using context assembled exclusively from observed genes. Withheld targets were absent from encoder expression, gene identities and measurement masks. Complete reconstruction used the expected-expression output, $p_{cg}M_{cg}$, across all query cells. Conditional intensity used $M_{cg}$ and was scored separately on measured-positive observations. Total-RNA, multispecies and spatial protocols are described in Supplementary Methods~\ref{sm:expression-evaluation}.

For molecular association, we extracted the final interaction representation $\mathbf h_{cg}$ for each queried gene. Within a cell population $\mathcal C_k$, features were averaged across cells, centred across the complete decoder gene panel $\mathcal P_k$ and normalized:

\begin{equation}
\begin{aligned}
\overline{\mathbf h}_{kg}
&=\frac{1}{|\mathcal C_k|}\sum_{c\in\mathcal C_k}\mathbf h_{cg},\\
\widetilde{\mathbf h}_{kg}
&=\overline{\mathbf h}_{kg}
-\frac{1}{|\mathcal P_k|}\sum_{j\in\mathcal P_k}\overline{\mathbf h}_{kj},\\
A_k(g,j)
&=\frac{\widetilde{\mathbf h}_{kg}^{\mathsf T}\widetilde{\mathbf h}_{kj}}
{\|\widetilde{\mathbf h}_{kg}\|_2\|\widetilde{\mathbf h}_{kj}\|_2}.
\end{aligned}
\label{eq:main-association}
\end{equation}

This centred cosine score ranked molecular neighbours within each analysed cell population. Module retrieval averaged similarities to fixed seed genes. Query-based analyses connected individual non-coding transcripts to published developmental and neural programmes. For module recovery, encoder expression and measurement masks were set to zero across the query panel. Observed-context association analyses retained measured query RNA. Gene networks connected each gene to its highest positive-cosine neighbours; weighted community detection and pathway enrichment characterized the resulting molecular organization. Association, gene-network construction, host-cell programme analysis and biological programme retrieval are described in Supplementary Methods~\ref{sm:associations}-\ref{sm:programme-retrieval}. Data sources and analysed cohorts are detailed in \ref{sm:datasets}, comparison methods in \ref{sm:comparators}, and evaluation metrics and uncertainty in \ref{sm:metrics}.

% Detailed supplementary methods remain in the original appendix, as requested.
\let\section\CellSection
\let\subsection\CellSubsection
\bibliography{sn-bibliography}
% One document and one bibliography: all references resolve within this PDF.
\clearpage
\let\FloatBarrier\ArxivFloatBarrier
% The printed S. identifiers remain as in the supplied supplement. Distinct
% hyperref anchors prevent collisions with the main article's counters.
\renewcommand{\theHsection}{S.\arabic{section}}
\renewcommand{\theHsubsection}{\theHsection.\arabic{subsection}}
\renewcommand{\theHsubsubsection}{\theHsubsection.\arabic{subsubsection}}
\renewcommand{\theHfigure}{S.\arabic{figure}}
\renewcommand{\theHtable}{S.\arabic{table}}
\renewcommand{\theHequation}{S.\arabic{equation}}
\phantomsection
\label{arxiv:supplement}
\pdfbookmark[0]{Supplementary Information}{arxiv-supplement}
\section*{Supplementary Information}
\setcounter{section}{0}
\setcounter{figure}{0}
\setcounter{table}{0}
\setcounter{equation}{0}
\renewcommand{\thesection}{S.\arabic{section}}
\renewcommand{\thesubsection}{\thesection.\arabic{subsection}}
\renewcommand{\thesubsubsection}{\thesubsection.\arabic{subsubsection}}
\renewcommand{\thefigure}{S.\arabic{figure}}
\renewcommand{\thetable}{S.\arabic{table}}
\renewcommand{\theequation}{S.\arabic{equation}}

\section{Supplementary evidence}\label{sec:supplementary-evidence}

\subsection{Tissue-resolved cellular identity and AvgBIO}
\begin{figure}[p]
\centering
\includegraphics[width=\textwidth]{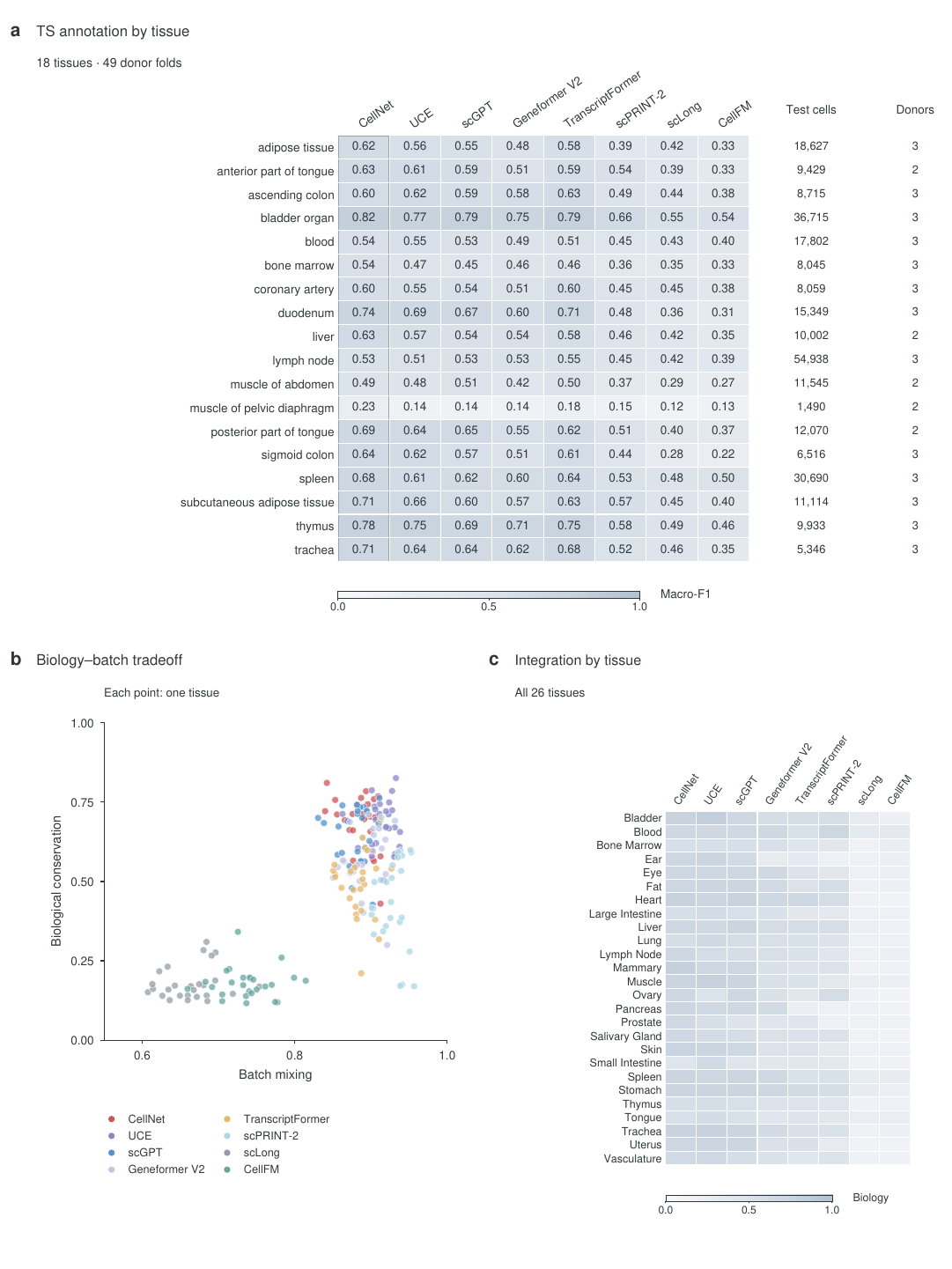}
\caption{Tissue-resolved cellular identity and AvgBIO. \textbf{a}, Annotation Macro-F1 across 18 Tabula Sapiens tissues and seven methods. Values average donor-held-out folds within each tissue; right-hand columns give test-cell and donor-fold counts. The evaluation comprises 49 folds and 276,385 test cells. \textbf{b}, AvgBIO versus batch mixing for each of 26 multi-donor tissues. Colours identify methods; points show tissue-level scores. Both axes increase with performance. \textbf{c}, Tissue-specific AvgBIO across all 26 tissues. AvgBIO is the equal-weight mean of normalized mutual information (NMI), adjusted Rand index (ARI) and cell-type silhouette rescaled as $(s+1)/2$. Batch mixing is the equal-weight mean of scaled batch silhouette and graph connectivity. Scores are calculated in the original feature space. Source data retain the individual metric components and cell supports.}
\label{fig:fig2s1}
\end{figure}

The tissue-level annotation matrix localizes the variation underlying the aggregate Tabula Sapiens results (Supplementary Fig.~\ref{fig:fig2s1}a). CellNet achieved a Macro-F1 of 0.537 in bone marrow, compared with 0.469 for UCE, and 0.634 in liver, compared with 0.575 for TranscriptFormer. AvgBIO was examined jointly with donor mixing across all 26 evaluated tissues (Supplementary Fig.~\ref{fig:fig2s1}b). Overall AvgBIO was 0.684 for CellNet and 0.680 for UCE; the corresponding batch-mixing scores were 0.886 and 0.915. Tissue-level AvgBIO and batch-mixing results across all 26 tissues are shown in Supplementary Fig.~\ref{fig:fig2s1}b,c; for example, lymph-node AvgBIO was 0.711 for CellNet and 0.606 for UCE.

\subsection{Species-resolved identity and directional reference mapping}
\begin{figure}[p]
\centering
\includegraphics[width=\textwidth]{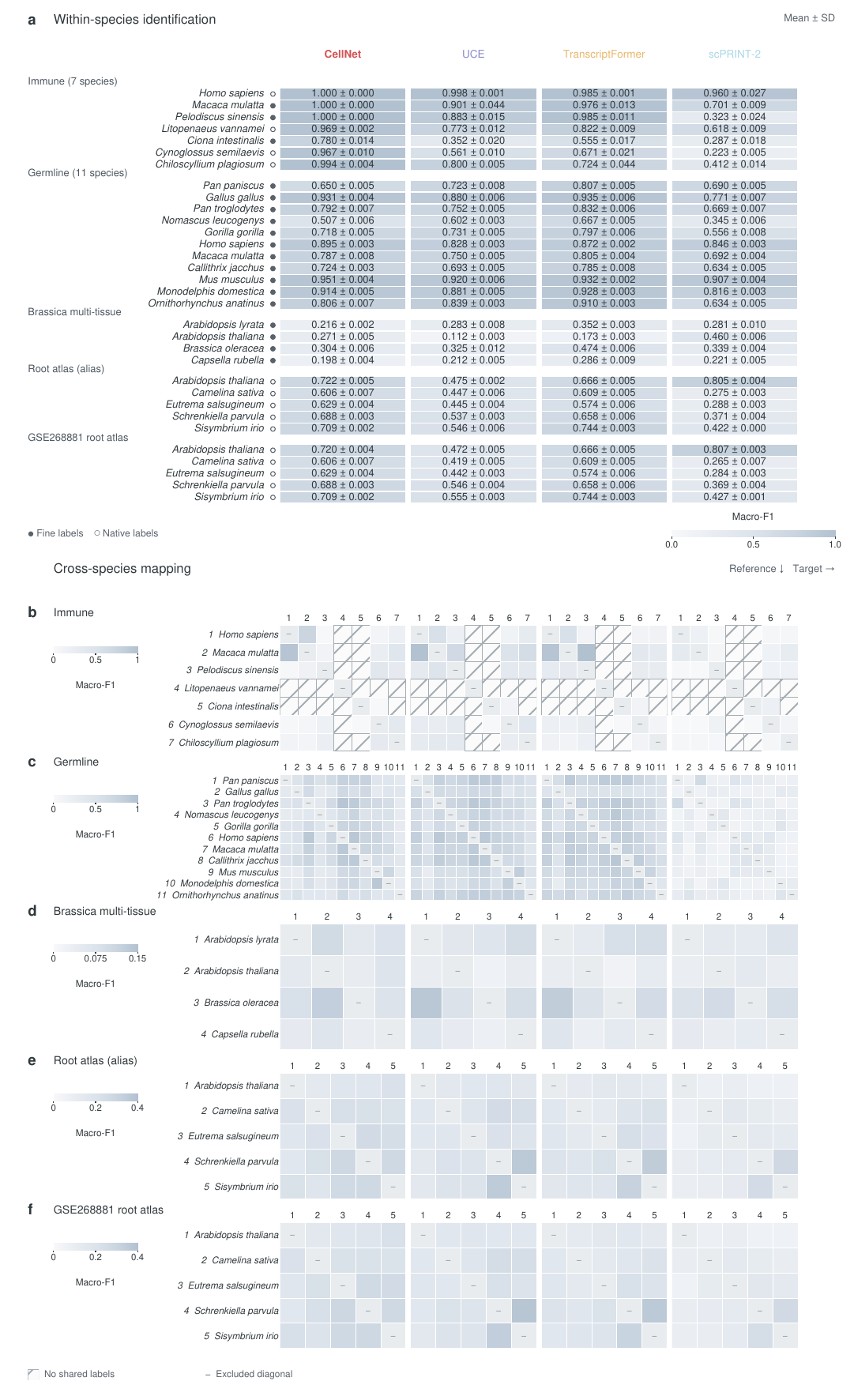}
\caption{Within-species identity and directional reference mapping. \textbf{a}, Macro-F1 for 32 task--species combinations and four methods, shown as mean $\pm$ sample s.d. across five shared, approximately 80/20 reference--query splits. Filled and open symbols distinguish strictly refined labels from native labels at coarse resolution. \textbf{b--f}, Coarse-label reference mapping for seven immune species, eleven germline species, four Brassica multi-tissue species and two five-species root-atlas variants. Rows identify reference species; columns identify target species using the numbered scientific-name keys. Colour denotes Macro-F1 on a common 0--1 scale. Hatching denotes no shared evaluable labels; grey minus signs denote omitted same-species diagonals. The root-atlas and GSE268881 variants use the same biological cells. Split s.d. describes sensitivity across overlapping cell partitions. The two evaluation protocols are specified in Supplementary Methods~\ref{sm:cellular-evaluation}. Exact scores, label counts and query supports are supplied in source data.}
\label{fig:fig2s2}
\end{figure}

Within-species identification retained all 32 task--species combinations, including 18 with strictly refined labels and 14 with native labels at coarse resolution (Supplementary Fig.~\ref{fig:fig2s2}a). CellNet had the highest mean score in all seven immune species. Fine-label F1 in \textit{Ciona intestinalis} was 0.780, compared with 0.555 for TranscriptFormer; fine-label germline recognition reached 0.895 in human and 0.951 in mouse. Brassica multi-tissue fine-label Macro-F1 averaged 0.247 for CellNet, 0.233 for UCE, 0.321 for TranscriptFormer and 0.325 for scPRINT-2. The five-species root-atlas and GSE268881 variants yielded CellNet means of 0.671 and 0.671; these variants shared biological cells and represented different task-specific feature caches.

Directional reference mapping evaluated coarse labels separately for the complete immune, germline and three plant tasks (Supplementary Fig.~\ref{fig:fig2s2}b--f). CellNet's mean F1 over evaluable directions was 0.229 for immune and 0.552 for germline transfer; UCE and TranscriptFormer scored higher in these animal tasks. Across the three plant variants, CellNet achieved the highest directional means among the four methods: 0.069 for multi-tissue Brassica, 0.215 for the root atlas and 0.220 for GSE268881. Results are shown by reference and target species. Of 42 possible immune directions, 22 had shared evaluable labels; all germline and plant directions were evaluable. Within-species and directional transfer used different split and k-nearest-neighbour protocols.

\subsection{Kidney and lung annotation and rare-population detection}
\begin{figure}[p]
\centering
\includegraphics[width=\textwidth]{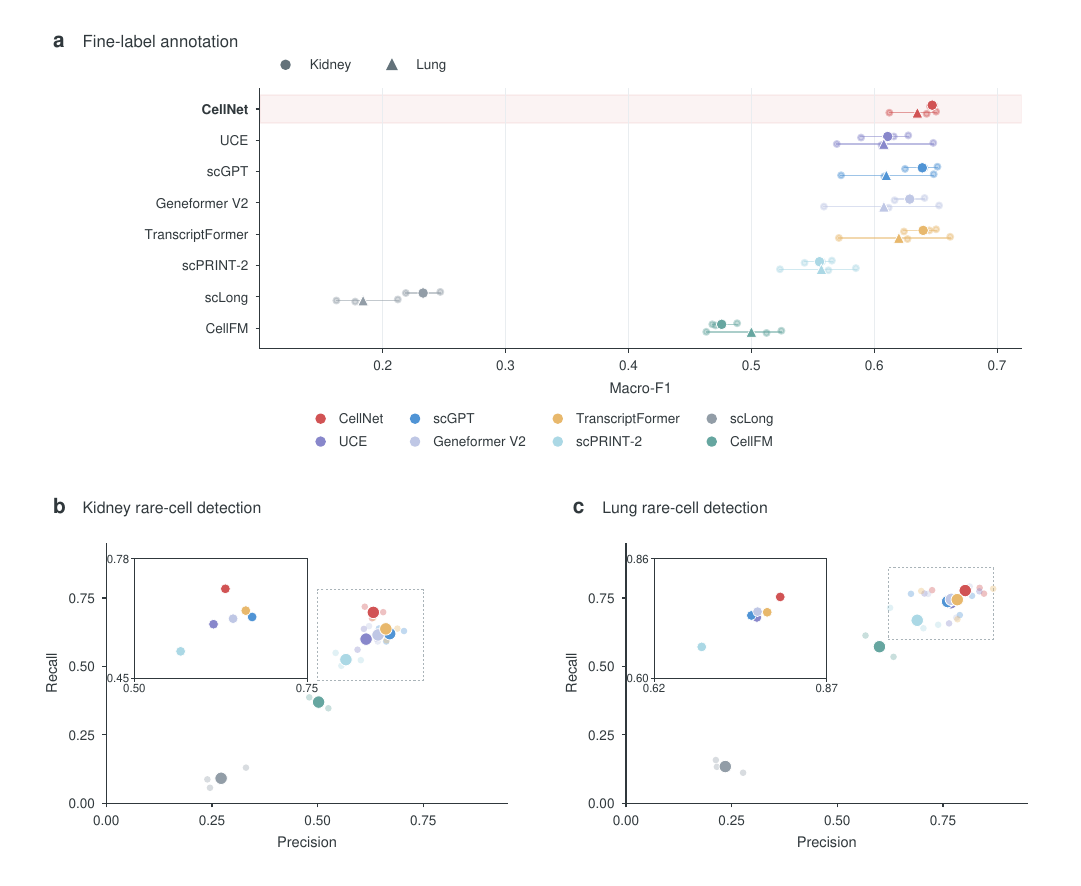}
\caption{Kidney and lung annotation and rare-population detection. \textbf{a}, Complete-label Macro-F1 in kidney and lung. Kidney comprises 107,344 cells, 45 donors and 57 labels; lung comprises 69,650 cells, three donors and 59 labels. Large circles and triangles show three-fold means for kidney and lung, respectively; small points show individual donor-disjoint test folds. Horizontal lines span the observed fold range. \textbf{b,c}, Rare-population macro precision and recall in kidney and lung. Large points show fold means; small points show individual folds, and insets enlarge the mean clusters. These are classification operating points, with higher values on both axes indicating better detection. Rare eligibility requires training prevalence $\leq 1\%$, at least 20 cohort cells represented in at least three donors, and positive training and test support. There are 34 kidney and 17 lung identities eligible in at least one fold. Metrics average eligible classes within each fold, then average the three folds. Method colours are shared across panels.}
\label{fig:fig2s3}
\end{figure}

The kidney and lung evaluations retained all 57 and 59 author-defined fine labels, respectively (Supplementary Fig.~\ref{fig:fig2s3}a). CellNet achieved complete-label Macro-F1 of 0.647 in kidney and 0.635 in lung, compared with 0.640 and 0.620 for TranscriptFormer. Rare-population detection was then resolved into precision and recall using the same held-out predictions (Supplementary Fig.~\ref{fig:fig2s3}b,c). In kidney, CellNet's macro precision and recall were 0.632 and 0.697; scGPT had higher precision at 0.670, whereas CellNet had the highest recall. In lung, CellNet led both endpoints, with precision of 0.803 and recall of 0.777. The precision--recall plots show how rare-cell recovery contributed to the F1 results, with enlarged insets distinguishing methods with similar mean values.

\subsection{Class-resolved rare-population performance and error outcomes}
\begin{figure}[p]
\centering
\includegraphics[width=\textwidth]{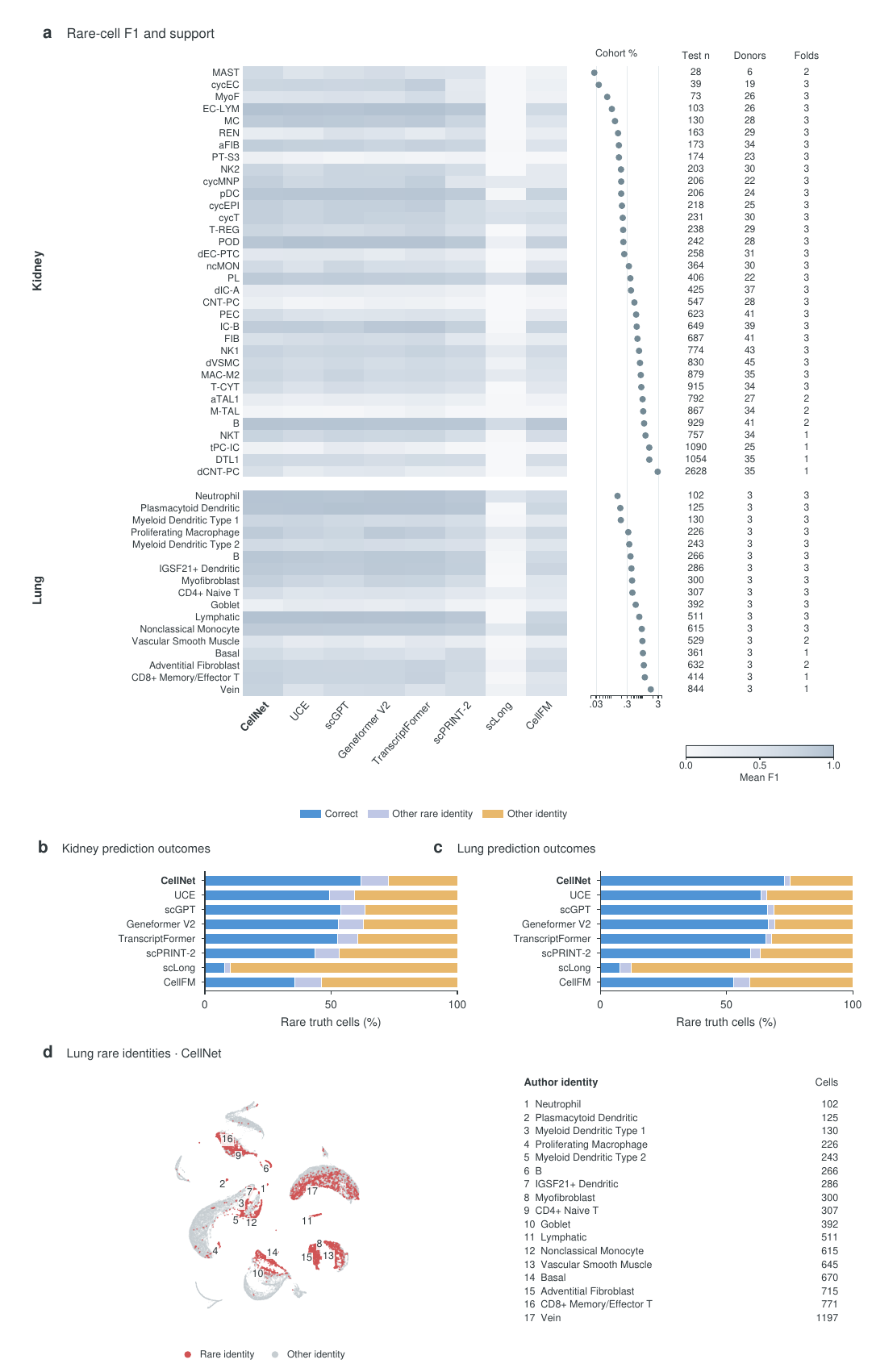}
\caption{Complete rare-population performance, support and errors. \textbf{a}, Class-level F1 for all 34 kidney and 17 lung eligible identities across seven methods, ordered by cohort abundance. Colour shows each class's mean across its eligible test folds. Side columns show cohort prevalence, eligible-fold test-cell support, donor count and eligible-fold count. Exact F1 values are supplied in source data. \textbf{b,c}, Outcomes for eligible rare truth cells in kidney and lung, pooled across the test folds. Bars partition cells into correct identity, another eligible rare identity and any other identity, summing to 100\% per method. These proportions are weighted by cell counts. \textbf{d}, Fixed UMAP of all 69,650 lung cells in the CellNet classification representation. Pink marks the 17 eligible author identities; grey marks other identities. Numbers correspond to the full identity key and whole-cohort cell counts. Whole-cohort counts in \textbf{d} differ from eligible-fold test supports in \textbf{a}. Rare eligibility and classification settings follow Fig.~\ref{fig:main2}d,e and Supplementary Fig.~\ref{fig:fig2s3}.}
\label{fig:fig2s4}
\end{figure}

Class-resolved analysis retained every eligible identity, together with cell abundance, held-out support, donor coverage and eligible-fold count (Supplementary Fig.~\ref{fig:fig2s4}a). Kidney mast cells (MAST) constituted 0.026\% of the cohort and achieved CellNet F1 of 0.658, compared with 0.609 for Geneformer V2. Kidney parietal epithelial cells (PEC) achieved F1 of 0.622, compared with 0.513 for TranscriptFormer. Lung myofibroblasts comprised 300 cells and achieved F1 of 0.806, compared with 0.769 for Geneformer V2.

Pooling eligible rare truth cells across test folds, CellNet correctly assigned 61.8\% of kidney cells and 73.1\% of lung cells (Supplementary Fig.~\ref{fig:fig2s4}b,c). The highest comparator proportions were 53.8\% for scGPT in kidney and 66.6\% for Geneformer V2 in lung. These cell-weighted proportions describe correct identities and error destinations; class-macro performance is reported in Fig.~\ref{fig:main2}d,e. A projection of all 69,650 lung cells locates the 17 eligible author identities within the frozen CellNet representation (Supplementary Fig.~\ref{fig:fig2s4}d).

\subsection{Condition recognition across donors and libraries}
\begin{figure}[p]
\centering
\includegraphics[width=\textwidth]{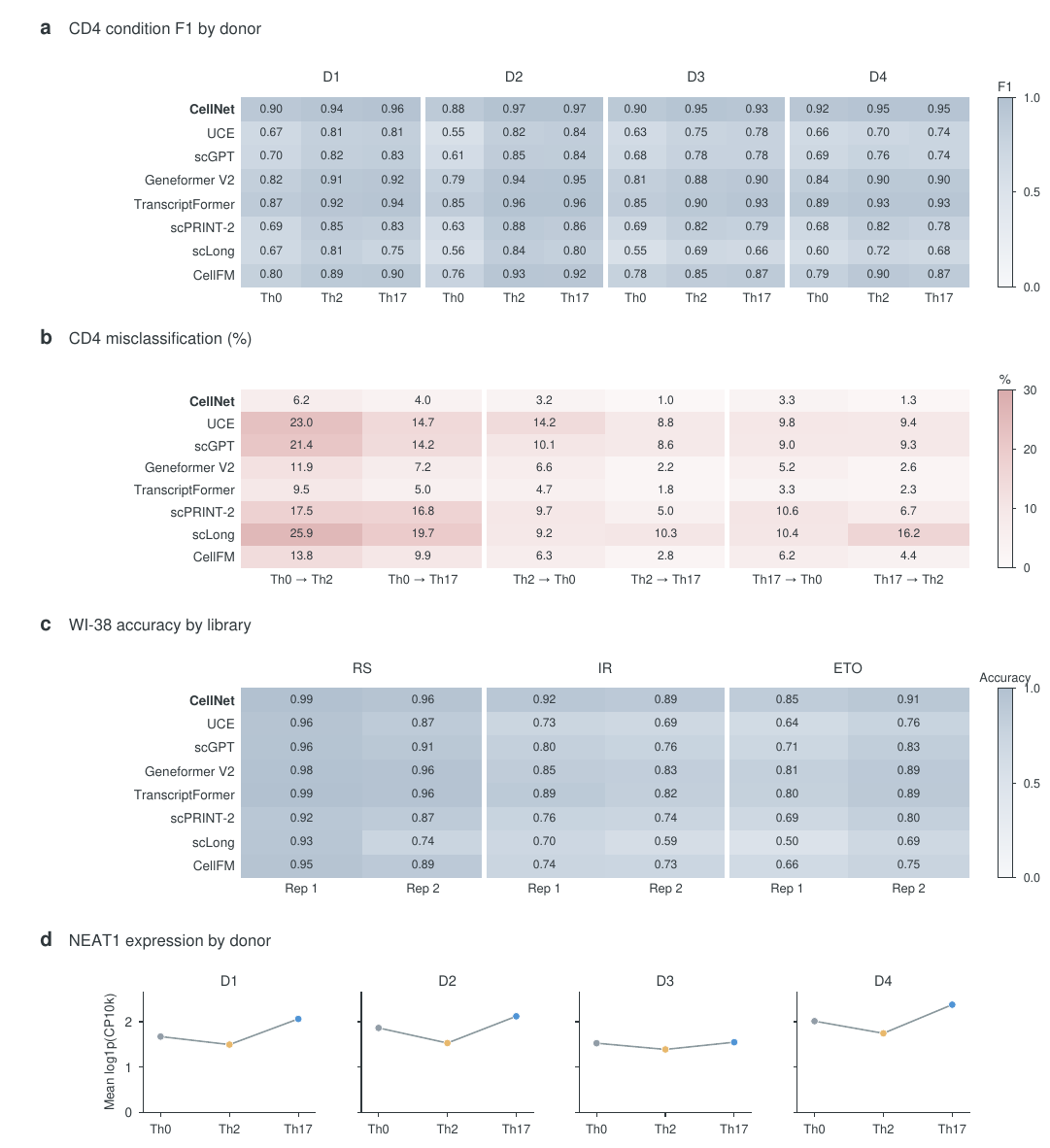}
\caption{Condition recognition across donors and libraries. \textbf{a}, Condition-specific F1 for Th0, Th2 and Th17 in each of four CD4 donor holdouts, covering 9,777 cells. \textbf{b}, Directed misclassification percentages pooled across donor holdouts. Arrows indicate true $\to$ predicted condition; denominators are 2,543 Th0, 4,040 Th2 and 3,194 Th17 cells. Lower values indicate fewer errors. \textbf{c}, WI-38 accuracy for six complete held-out libraries, covering 14,457 cells. RS, IR and ETO denote replicative, irradiation-induced and etoposide-induced senescence conditions; each has two author culture replicates. Each fold holds out one library per condition, and single-condition library performance is reported as accuracy. \textbf{d}, Measured NEAT1 expression by CD4 donor and condition. Points show cell-mean log1p(CP10k), normalized using all 20,953 supplied genes; lines join conditions from the same donor. NEAT1 supplies condition-associated expression context. Heatmaps in \textbf{a,c} use a common 0--1 score scale; \textbf{b} shows error percentages. Library accessions, cell supports and expression measurements are supplied in source data.}
\label{fig:fig2s5}
\end{figure}

Donor-resolved CD4 evaluation yielded CellNet Macro-F1 values of 0.935, 0.941, 0.926 and 0.943, leading the comparison methods in every donor holdout (Supplementary Fig.~\ref{fig:fig2s5}a). CellNet also had the lowest pooled error percentage for each directed condition pair (Supplementary Fig.~\ref{fig:fig2s5}b). For example, 6.2\% of Th0 cells were assigned to Th2, compared with 9.5\% for TranscriptFormer. WI-38 accuracy was highest for CellNet in five of six held-out libraries (Supplementary Fig.~\ref{fig:fig2s5}c), with values ranging from 0.854 to 0.992 across both author culture replicates of the three conditions.

Measured NEAT1 expression provided a non-coding-transcript correlate of the CD4 conditions (Supplementary Fig.~\ref{fig:fig2s5}d). Mean log1p-normalized expression followed Th17 $>$ Th0 $>$ Th2 in all four donors. Th17--Th0 differences ranged from 0.022 to 0.388, and Th2--Th0 differences ranged from $-0.332$ to $-0.137$. These measurements describe expression associated with the author-defined conditions; classification labels were taken from the original experimental metadata.

\clearpage

\subsection{Testable coding outputs across target prevalence}
\begin{figure}[p]
\centering
\includegraphics[width=\textwidth]{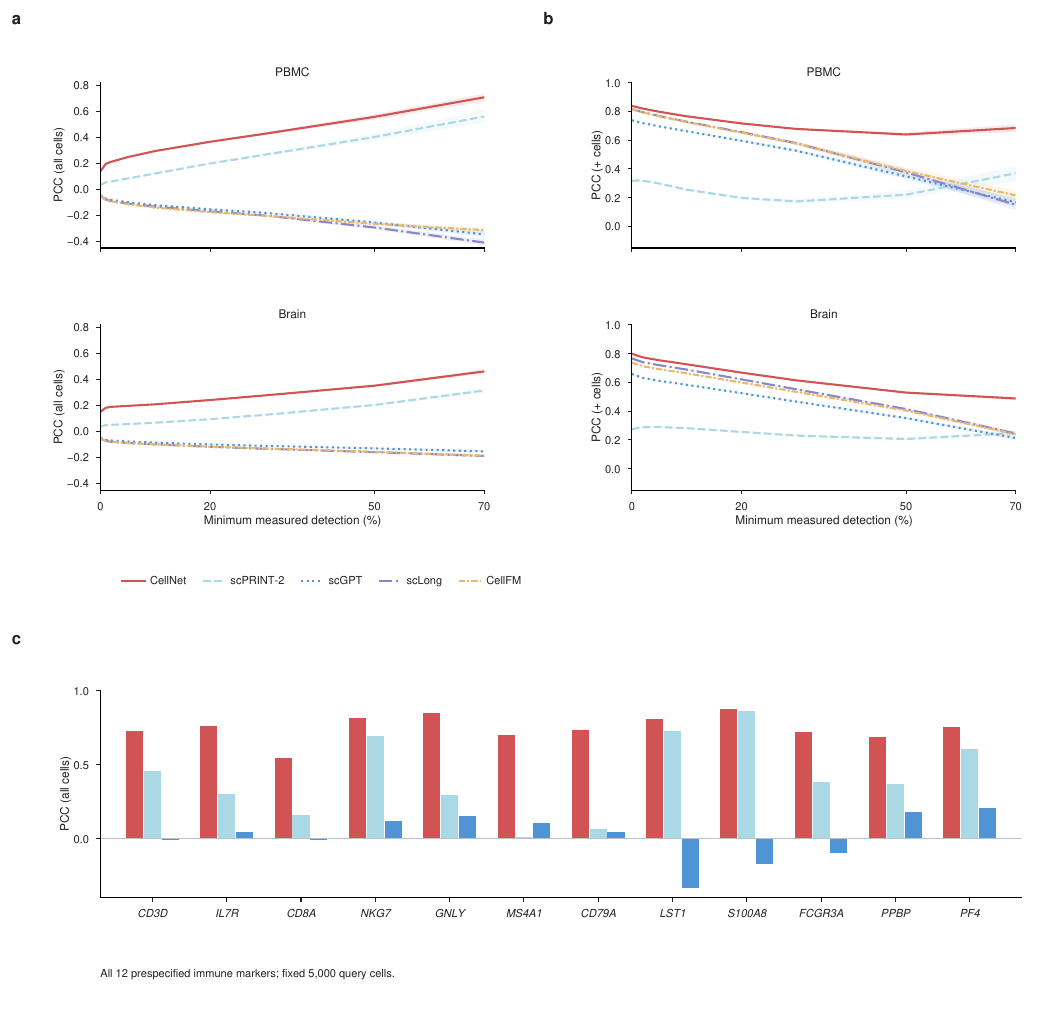}
\caption{\textbf{a,b}, All-cell PCC (\textbf{a}) and measured-positive intensity PCC (\textbf{b}) across cumulative gene-detection thresholds in TotalX PBMC and brain. Five frozen models are evaluated on comparison-specific shared targets from five complementary hidden-gene folds and 5,000 fixed query cells per tissue. Detection is the fraction of query cells with positive raw counts; it selects genes, not cells for all-cell scoring. Positive PCC uses only each gene's measured-positive cells; CellNet supplies $m$. Bands are paired 95\% target-gene bootstrap intervals (2,000 draws). \textbf{c}, All-cell PCC for all twelve markers shown in Fig.~\ref{fig:main}b, retaining their prescribed order without a detection cutoff. Bars are gene-wise scores in the fixed cohort, not estimates from independent biological replicates. Baseline expression outputs retain their native definitions.}
\label{fig:ed1}
\end{figure}

Coding threshold curves evaluated both all-cell recovery and intensity among measured-positive cells (Supplementary Fig.~\ref{fig:ed1}a,b). All twelve prescribed immune markers were retained without a detection cutoff, testing lineage-associated predictions beyond the representative heatmap. Their CellNet PCCs ranged from 0.547 to 0.880 and exceeded both recorded comparators. Measured detection ranged from 1.92\% to 53.92\% (Supplementary Fig.~\ref{fig:ed1}c).

\subsection{Non-coding identities as prediction targets}
\begin{figure}[p]
\centering
\includegraphics[width=\textwidth]{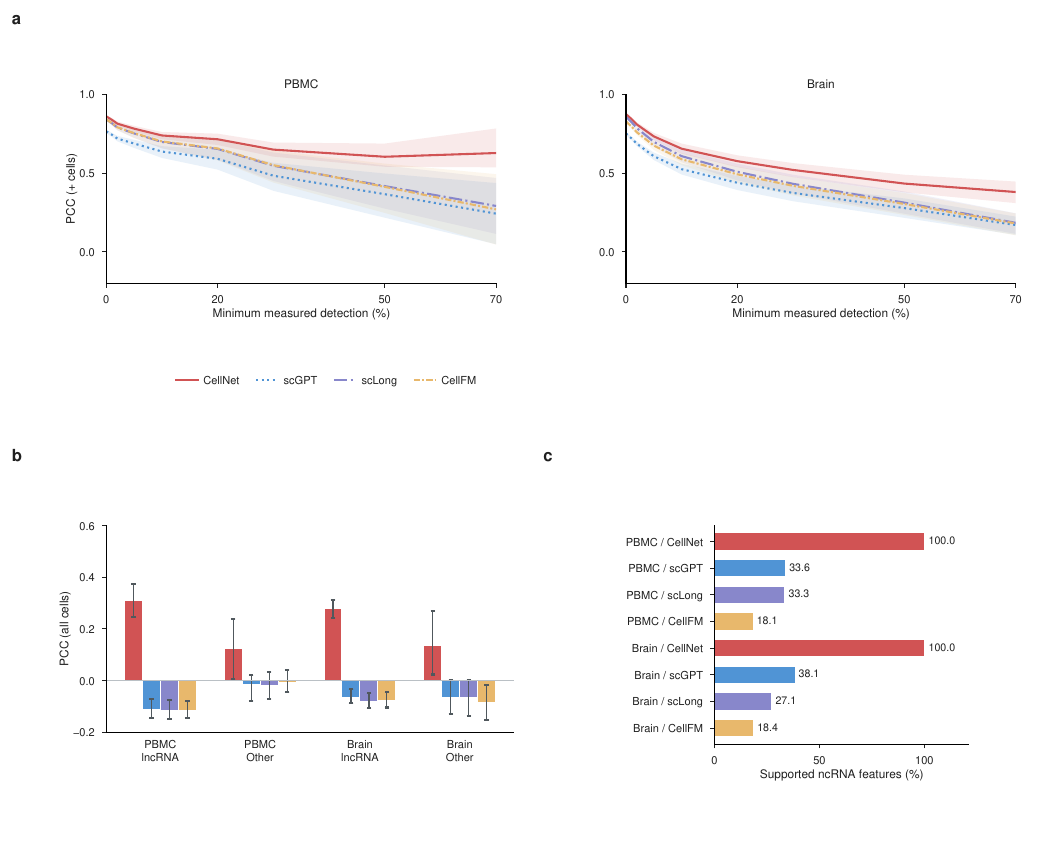}
\caption{\textbf{a}, Measured-positive ncRNA intensity PCC across cumulative gene-detection thresholds in TotalX PBMC and brain, using only coding inputs and withholding all nonprotein targets together. Each tissue contains 5,000 fixed query cells. Scoring uses each gene's raw-count-positive cells; CellNet supplies $m$ and comparators supply native expression predictions. \textbf{b}, All-cell PCC for annotated lncRNAs and other ncRNAs detected in at least 20\% of cells. The shared pools contain 28/4 targets in PBMC and 69/7 in brain, respectively. Bands and error bars are paired 95\% target-gene bootstrap intervals (2,000 draws). \textbf{c}, Coverage of 18,502 PBMC and 26,113 brain assay ncRNA identities. CellNet coverage includes actual hidden-target predictions after DNA-identity extension with the original DNA encoder, fixed identity transformation and frozen expression model. Comparator coverage reflects native vocabulary support. Coverage establishes molecular range; it is not a prediction-accuracy percentage. Accuracy pools in \textbf{a,b} and Fig.~\ref{fig:main}c--e remain unchanged.}
\label{fig:ed2}
\end{figure}

Measured-positive threshold curves and transcript-class summaries tested expression output across ncRNA prevalence and classes (Supplementary Fig.~\ref{fig:ed2}a,b). The DNA-identity extension supplied actual predictions for every annotated assay ncRNA while preserving the original accuracy pools (Supplementary Fig.~\ref{fig:ed2}c). The two analyses established predictive accuracy on shared targets and output availability across the assay's wider molecular repertoire.

\subsection{Refinement of molecular predictions with reference expression}
\begin{figure}[p]
\centering
\includegraphics[width=\textwidth]{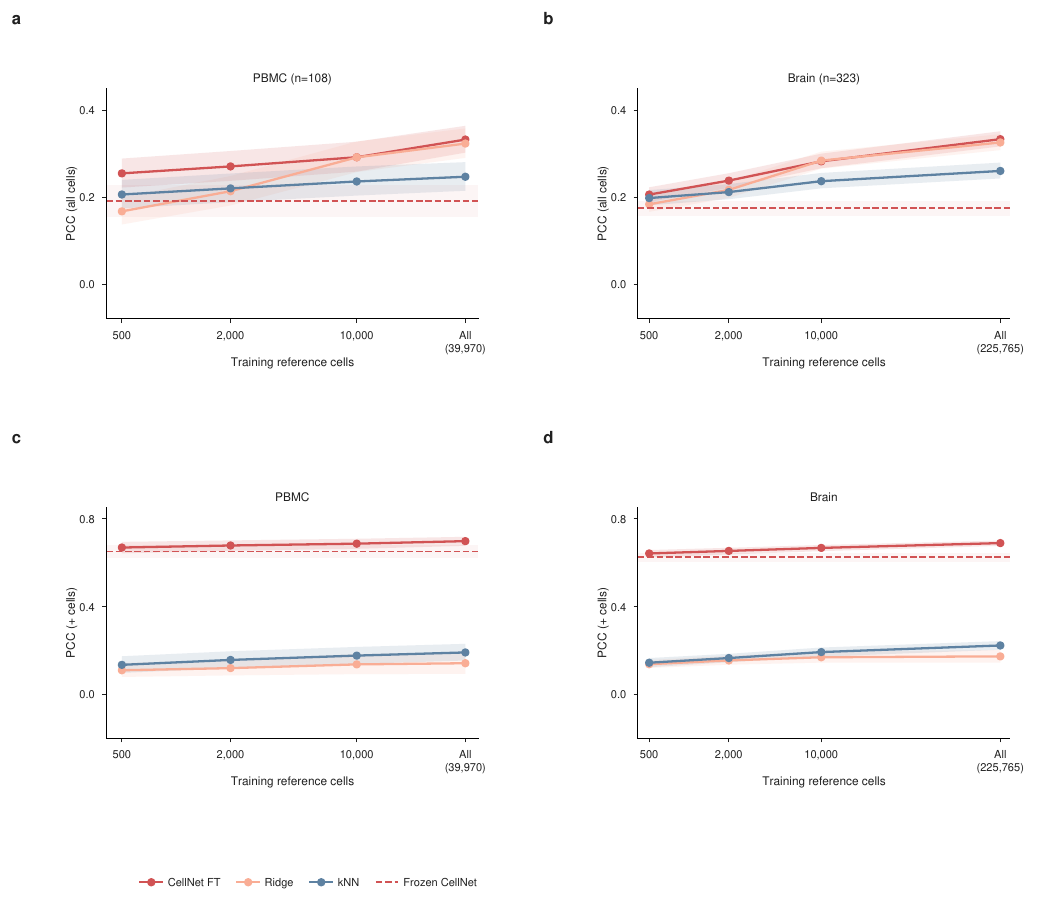}
\caption{\textbf{a,b}, All-cell PCC in PBMC (\textbf{a}) and brain (\textbf{b}). \textbf{c,d}, Corresponding intensity PCC within measured-positive cells, using CellNet $m$. Native decoder adaptation, Ridge and kNN use identical training-reference members at 500, 2,000, 10,000 and all non-test cells (39,970 PBMC; 225,765 brain). Dashed lines denote frozen CellNet; the reference-count axis is logarithmic. The fixed comparison pools comprise 108 and 323 ncRNAs detected in at least 20\% of the 5,000 query cells per tissue. Shading denotes paired 95\% target-gene bootstrap intervals (2,000 draws). Smaller cohorts use 2,000 separate validation cells for epoch selection; full-reference fits reincorporate these cells after epoch selection is fixed. Native fits use thirty epochs for smaller cohorts and 29/30 epochs for full PBMC/brain. Fig.~\ref{fig:main}f,g shows the 500-training-reference endpoint.}
\label{fig:ed3}
\end{figure}

Four matched reference sizes tested whether additional cellular examples improved ncRNA outputs on fixed target pools (Supplementary Fig.~\ref{fig:ed3}). At full reference, all-cell PCC reached 0.333 in PBMCs and 0.334 in brain. Measured-positive intensity PCC reached 0.698 and 0.689, respectively, with an advantage over Ridge and nearest-neighbour prediction across all reference sizes.

\subsection{Prediction scope across gene identities and expression exposure}
\begin{figure}[p]
\centering
\includegraphics[width=\textwidth]{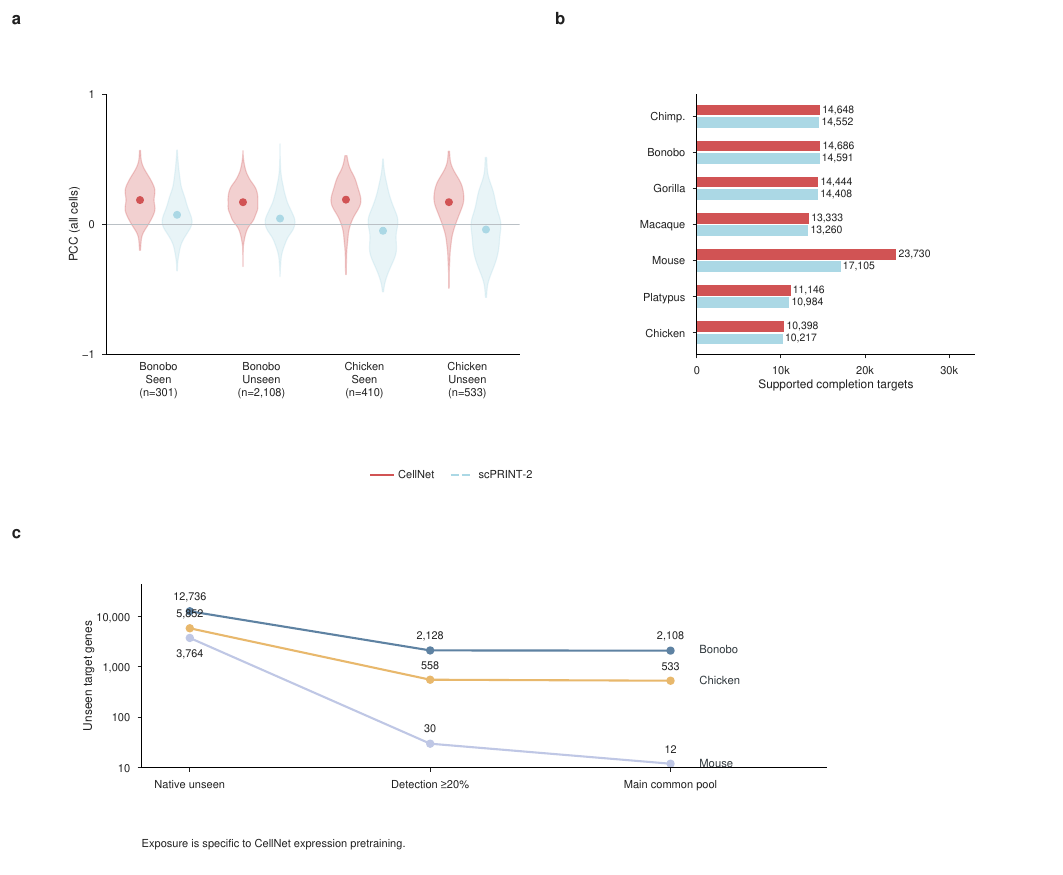}
\caption{\textbf{a}, Frozen CellNet/scPRINT-2 all-cell PCC distributions for CellNet-expression-pretraining-seen and unseen targets in bonobo (301/2,108 genes) and chicken (410/533). All genes belong to the prespecified common pools used in Fig.~\ref{fig:main}h. Violin densities have equal widths; points and intervals denote means and paired 95\% target-gene bootstrap intervals (2,000 draws). \textbf{b}, Exact vocabulary support within the original registered completion-target universes of seven species. scPRINT-2 uses checkpoint-native identities or official frozen species extensions, including gene embeddings and positions, for gorilla, platypus and legacy chicken. Support counts describe identity range, not accuracy or full-universe inference in this evaluation. \textbf{c}, Counts of original CellNet-supported unseen identities, measured-detection-qualified genes ($\geq20\%$) and current common scoring targets. The final pools are 2,108 bonobo, twelve mouse and 533 chicken genes; mouse support is twelve of thirty detection-qualified candidates. “Unseen” refers exclusively to CellNet expression pretraining; DNA identities remain available.}
\label{fig:ed4}
\end{figure}

Frozen CellNet exceeded scPRINT-2 in mean all-cell PCC for both expression-pretraining-seen and unseen bonobo and chicken targets (Supplementary Fig.~\ref{fig:ed4}a). Species support counts described the original registered target universes independently of accuracy (Supplementary Fig.~\ref{fig:ed4}b). Detection and vocabulary intersections yielded 2,108 bonobo, twelve mouse and 533 chicken unseen targets, specifying the evaluated extension beyond expression-training exposure (Supplementary Fig.~\ref{fig:ed4}c).

\subsection{Spatial recovery across the prescribed marker repertoire}
\begin{figure}[p]
\centering
\includegraphics[width=\textwidth]{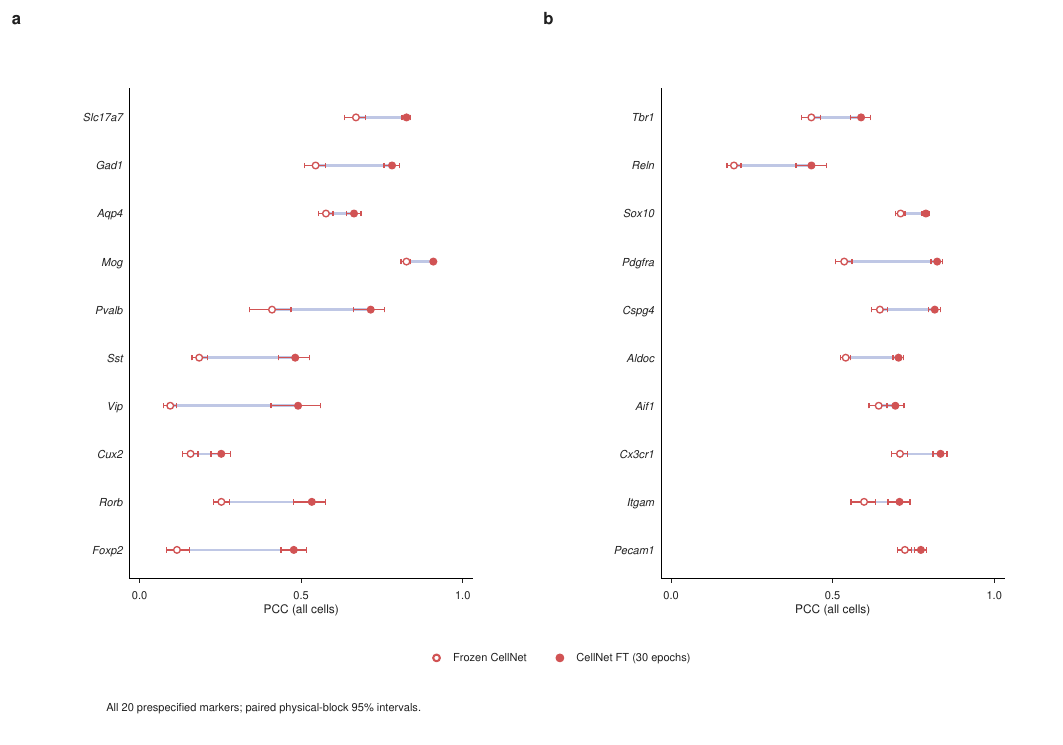}
\caption{\textbf{a,b}, All-cell PCC for all twenty prescribed Xenium mouse-brain markers, retaining their order across two columns. Markers are withheld together from 12,000 fixed query cells with 4,986 observed genes. Open/filled symbols denote frozen CellNet and its native decoder after thirty fixed adaptation epochs with 51,173 nonquery cells from the same section. Lines connect paired marker estimates. Intervals are paired 95\% percentiles from 2,000 shared resamples of 130 nonempty $500\times500\,\mu\mathrm{m}$ physical blocks, recalculating each marker's PCC in every draw. Fig.~\ref{fig:main}j shows four representative frozen maps; this figure retains quantitative results for the entire marker set. Supervised comparisons and measured-positive intensity scoring are provided in Supplementary Fig.~\ref{fig:ed6}a,b.}
\label{fig:ed5}
\end{figure}

All twenty prescribed targets were evaluated with physical-block uncertainty estimates (Supplementary Fig.~\ref{fig:ed5}). Every marker's all-cell PCC increased after same-section adaptation, extending the quantitative evidence beyond the four frozen maps in the main figure. With 4,986 observed genes, adapted CellNet achieved mean all-cell PCC of 0.665, compared with 0.620 for Ridge and 0.563 for nearest neighbours.

\subsection{Prediction under restricted spatial measurement}
\begin{figure}[p]
\centering
\includegraphics[width=\textwidth]{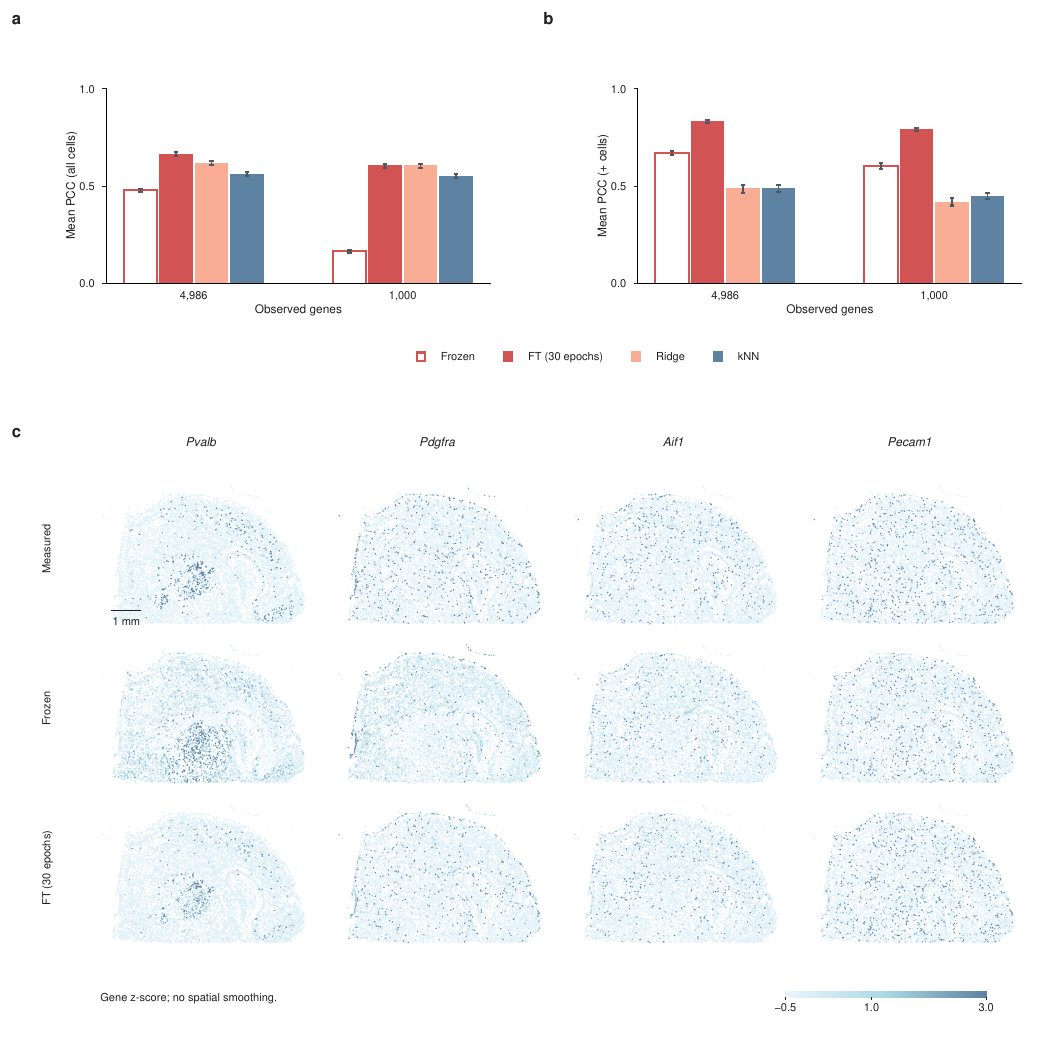}
\caption{\textbf{a,b}, Mean all-cell PCC (\textbf{a}) and measured-positive intensity PCC (\textbf{b}) across the same twenty prescribed markers with 4,986 or 1,000 observed genes. Frozen CellNet, its thirty-epoch adapted decoder, Ridge and kNN share 12,000 query cells; supervised methods share 51,173 same-section nonquery references. In \textbf{b}, each gene is scored only in its raw-count-positive query cells (300--9,015 per gene). This scorer-only mask changes neither inputs nor marker membership. CellNet supplies $m$; comparators supply native expression predictions. Error bars are paired 95\% physical-block bootstrap intervals from 2,000 shared resamples of 130 blocks ($500\times500\,\mu\mathrm{m}$), recalculating each marker's PCC before equal averaging. \textbf{c}, Measured, frozen and adapted maps of \textit{Pvalb}, \textit{Pdgfra}, \textit{Aif1} and \textit{Pecam1}, extending Fig.~\ref{fig:main}j to interneuronal, precursor, myeloid and endothelial populations. Maps retain original coordinates and independent gene/method z-scores without smoothing; scale bar, 1\,mm. Adaptation uses thirty fixed epochs, without spatial validation or test-based epoch selection.}
\label{fig:ed6}
\end{figure}

With 1,000 observed genes, adaptation increased mean all-cell PCC from 0.166 to 0.606 (Supplementary Fig.~\ref{fig:ed6}a). Measured-positive intensity PCC reached 0.831 with 4,986 observed genes and 0.792 with 1,000, exceeding both supervised comparators (Supplementary Fig.~\ref{fig:ed6}b). Maps of \textit{Pvalb}, \textit{Pdgfra}, \textit{Aif1} and \textit{Pecam1} tested recovery of additional population-associated spatial patterns (Supplementary Fig.~\ref{fig:ed6}c). These analyses assessed dependence on measured input breadth and biological coverage within the same tissue section.

\clearpage

\subsection{Biological organization of gene representations}\label{note:network}

The network used 126 CD14-positive monocytes with observed RNA available. Its 3,279 query genes were selected before network analysis. The preparation cohort contained 126 platelets and 300 cells each from naive B, naive T and monocyte populations. We combined the 2,048 protein-coding and 512 other annotated genes with the highest detection rates across these 1,026 cells, all 1,625 supported coding genes detected in at least seven platelets, and 83 prespecified platelet-reference and query genes. Removing duplicates and retaining supported, unambiguous gene mappings yielded the 3,279-gene panel. The 512-gene annotated group was not restricted to non-coding RNA. The network queries did not restrict the full measured-RNA input.

\begin{figure}[p]
\centering
\includegraphics[width=\textwidth]{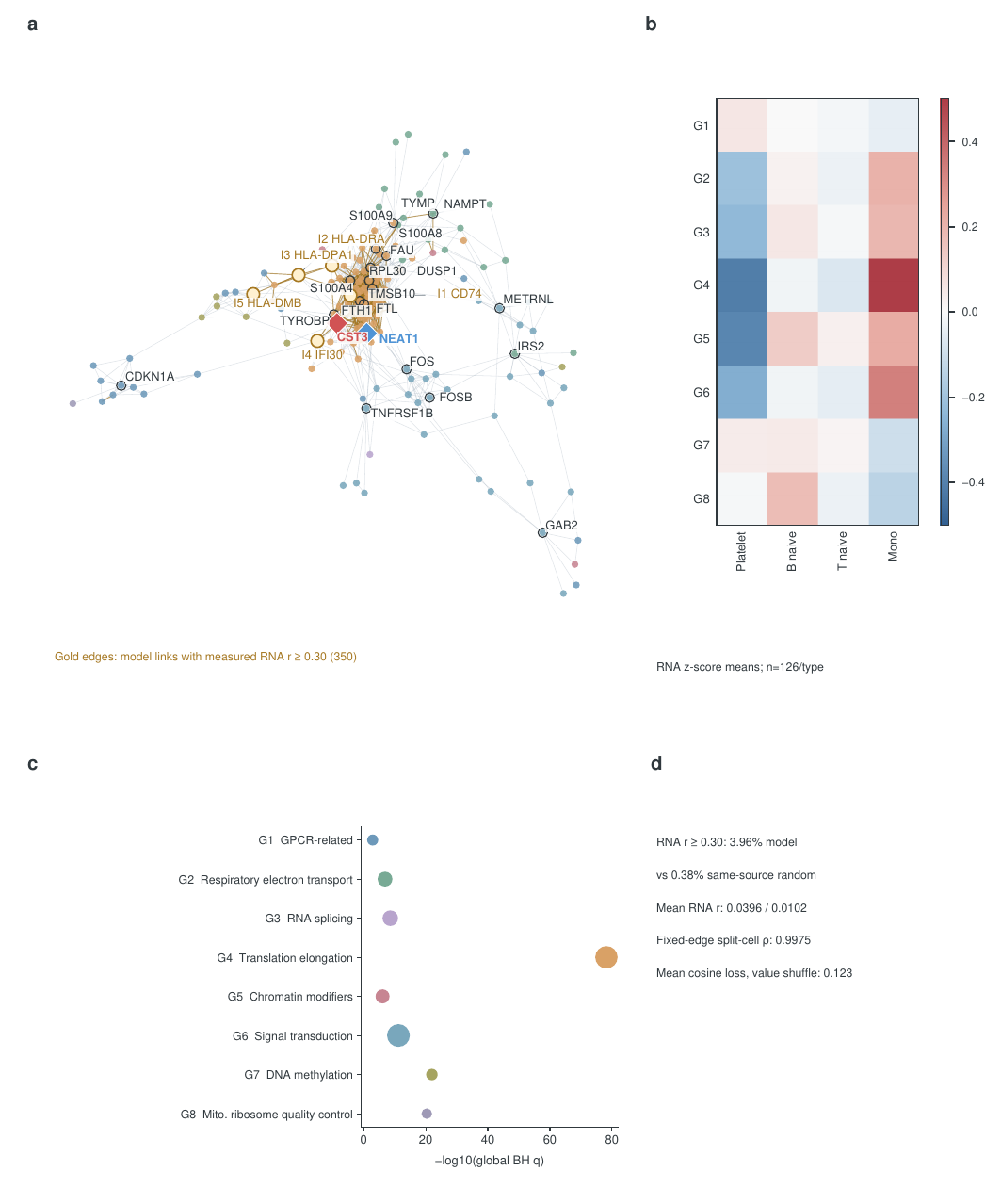}
\caption{Biological organization and measured RNA support of the gene network. The network contains 3,279 genes selected as described in Section~\ref{note:network}, using 126 monocytes with observed RNA. \textbf{a}, A neighbourhood induced by 26 displayed genes and their eight highest-similarity neighbours: 158 genes and 814 existing links. Gold identifies 350 links with measured RNA Pearson $r\geq0.30$. \textbf{b}, Mean standardized community RNA scores across 504 protocol-matched PBMCs, 126 per population. Gene expression is standardized across cells, averaged within each community, then within population. \textbf{c}, Strongest Reactome enrichment per community against the 3,279-gene background. Benjamini--Hochberg correction covers 10,720 tests; point size indicates gene overlap. \textbf{d}, RNA support among model links and random non-edge partners matched to the source gene, cosine agreement between disjoint monocyte halves, and mean cosine changes after within-cell expression shuffling. The displayed summaries characterize the existing representation network.}
\label{ed:network}
\end{figure}

Gene vectors were averaged across monocytes, centred across the query panel and normalized to unit length. Connecting each gene to its 15 highest positive-cosine neighbours, retaining a link selected in either direction, produced 36,638 links and eight weighted Leiden communities. All eight communities showed Reactome enrichment after global Benjamini--Hochberg correction over 10,720 tests against the 3,279-gene background (Supplementary Fig.~\ref{ed:network}). Representative enrichments included translation elongation ($q=6.28\times10^{-79}$) and mRNA splicing ($q=2.64\times10^{-9}$). The HLA-DRA--HLA-DPA1 link joined curated MHC-II pathway members with measured RNA $r=0.552$. Across the network, 3.96\% of links had RNA $r\geq0.30$, compared with 0.38\% of random non-edge partners matched to the source gene. These analyses characterize biological organization of the representations. Prediction attribution used the separate hidden-target design below.

\subsection{Target-specific input evidence and removal controls}\label{note:inputs}

The prediction analysis used 24 cells from one SSTX PBMC library, six each from monocyte, T-cell, B-cell and natural killer-cell groups. Within each population, cells were selected at equally spaced ranks spanning the 55th--90th percentiles of summed original \textit{CST3} and \textit{NEAT1} predictions. The entire 7,302-gene target partition, including both targets, was withheld from input expression, measurement indicators and input gene identities; 29,197 measured genes supplied the image. Expression was log-transformed after normalization to 10,000 observed-gene counts per cell. The monocyte displayed in Fig.~\ref{fig:4}c was subsequently selected for measured targets and the strongest combined two-target removal contrast among the selected monocytes. The four \textit{NEAT1} examples in Supplementary Fig.~\ref{ed:context} required detected targets and absolute prediction errors no greater than 0.4 log units, and were selected by the largest removal contrast within each group, excluding the main monocyte. All 24 cells contributed to the summaries and complete map collections.

Integrated gradients\citesupp{sundararajan2017} followed a 24-step path from zero to observed expression, with measurement indicators, gene identities and detection information fixed. Signed gene contributions were summed within $16\times16$ image patches. In Supplementary Fig.~\ref{ed:context}, full signed gene-contribution vectors were more similar within than between monocyte, B-cell and natural killer-cell groups; the T-cell contrast was approximately zero. The complete \textit{CST3} and \textit{NEAT1} collections use a common scale across all 48 cell/target maps (Supplementary Figs.~\ref{ed:cst3} and \ref{ed:neat1}).

Expression-only removal set selected input values to zero while retaining measurement information, gene identity and the original normalization. Eight random sets were averaged within each cell before comparison. Complete hiding additionally removed measurement, identity and detection information. For the 32-gene sets, target-selected removal exceeded random removal in every cell under complete hiding; patch-removal contrasts were positive in 23 or 24 of 24 cells, depending on target and removal protocol (Supplementary Fig.~\ref{ed:controls}). Summed integrated gradients approximated the prediction change from the zero-expression reference with a median relative residual of 1.18\%. The maximum residual was 98.15\%, with large relative errors in low-contrast \textit{CST3} cases. The figure retains all residuals and expression-matching diagnostics.

The complete named-set comparison gave extra decreases of 0.0963, $-0.0018$, $-0.0002$ and 0.0293 log units for the antigen-presentation set on \textit{CST3} and \textit{NEAT1}, and the response-associated set on \textit{CST3} and \textit{NEAT1}, respectively (Supplementary Fig.~\ref{ed:named}). In 2,273 reference monocytes excluding attribution cells, the response-set expression score correlated negatively with measured \textit{NEAT1} ($r=-0.138$), although removing that set lowered its prediction relative to random removal. Thus, measured expression covariation and the model's use of an input set conveyed different information in this cohort.

\subsection{Cell-state evidence across donors and expression-associated features}\label{note:states}

For the Th17--Th2 analysis, a linear discriminant classifier was trained on three donors and applied to the fourth, rotating across four donors. Attribution explained each held-out cell's Th17-minus-Th2 log-posterior score. Gradient multiplied by input expression gave signed contributions; positive values favoured Th17. The 96 attribution cells comprised 12 cells per state and donor. Within-state donor-mean gene contributions had Spearman correlations of 0.717--0.752 (Supplementary Fig.~\ref{ed:state}). The rotating classifiers shared training donors. Two displayed examples were selected retrospectively from correctly classified cells by the largest existing patch-removal contrasts.

Removing the 10\% strongest supporting patches was compared with random removal of the same patch count. Of six random sets per cell, only sets matching removed expression mass and detected-gene count within 10\% were averaged. This strict matching yielded 38 eligible cells, all Th17, and each showed a larger score decrease after targeted removal. Ribosomal genes were prominent among the classifier's contributing inputs.

Separately, six-component sparse nonnegative matrix factorization was fitted to cell representations from 2,700 cells sampled across donors D1--D3 and Th0, Th17 and Th2 conditions, then applied to all 9,777 cells. D4 was held out from fitting and gene ranking. Four selected features were associated with DNA replication, mitosis, endoplasmic-reticulum stress and interferon programmes (Supplementary Fig.~\ref{ed:features}). The eight genes per feature with the highest within-donor expression correlations in D1--D3 were drawn from genes detected in at least 488 cells. All 32 selected correlations remained positive in D4, and the training-defined high/low state polarity persisted across all four donors. Feature activations also covaried with total RNA counts, as shown for each feature.

\subsection{Module and partner retrieval across RNA classes}\label{note:retrieval}

Module recovery used a 1,020-gene query panel whose expression and measurement indicators were withheld from the input. Thirty reference genes were excluded from the 343 ranked non-coding candidates, which contained 117 additional published members and 226 candidates outside the member list. Scores averaged candidate-to-reference model cosine similarities or absolute RNA Pearson correlations. Average precision sums precision over the complete ranking, weighted by each increase in reference-member recall. Across naive B, naive T and monocyte populations, it was 0.601, 0.611 and 0.624 for CellNet, compared with 0.275, 0.341 and 0.325 for RNA correlation. Module membership was defined by the published TotalX coexpression analysis from the same resource\cite{totalx}.

Other neighbourhood and programme analyses retained measured query RNA. Population-mean gene vectors were centred across the analysed gene panel and normalized before cosine scoring; query self-matches were excluded. RNA comparators used absolute Pearson correlation of log-transformed expression in the same cells, including measured zeros. These rankings measured associations between representations. \textit{TCP1} retrieved \textit{CCT4}, \textit{CCT5}, \textit{CCT2}, \textit{CCT6A} and \textit{CCT3} at ranks 1--5 in both HEK293T and untreated Huh7, with \textit{CCT7} at rank 17 or 9, respectively (Supplementary Fig.~\ref{ed:associations}a,b). RNA correlation retrieved no CCT partners in either top 20. In the external PBMC cohort, \textit{HLA-DRA}'s partners \textit{CD74}, \textit{HLA-DPB1} and \textit{HLA-DPA1} retained ranks 4, 7 and 11. The complete discovery lists include recurrent and non-recurrent neighbours (Supplementary Fig.~\ref{ed:associations}c,d).

The four cell-context examples were selected after an exploratory screen, including comparison with baseline ranks. Each primary population contained 126 cells with matched protocol composition: eight 10X-dUTSO, 74 SSTX and 44 SSTX-pA+ cells. All used the same 2,736 coding candidates. Their observed similarities ranged from 0.211 to 0.369, compared with $-0.088$ to 0.034 after a single within-cell expression-value shuffle (Supplementary Fig.~\ref{ed:associations}e). Paired, protocol-stratified cell bootstrap intervals used 1,000 resamples of the observed and shuffled populations and describe that cohort and shuffle realization. All four partners remained within the top 20 in both disjoint 63-cell halves. Support across sequencing protocols varied: \textit{MALAT1}--\textit{AFF3} ranked 27 in SSTX-pA+, while its half-cell ranks were 17 and 15 and its expanded-population rank was 15 (Supplementary Fig.~\ref{ed:associations}f).

\subsection{Host programme organization and non-coding entry points}\label{note:programmes}

The Huh7 analysis contained 500 untreated and 500 DENV-exposed cells. Host-expression scores assigned 267 exposed cells to an active-like group and 233 to a quiescent-like group (Supplementary Fig.~\ref{ed:programmes}a,b). Scores averaged gene-wise standardized log-expression within predefined signatures; standardization used all 1,000 cells. Five scores defined the exposed-cell annotations, and the stress-granule score supplied additional expression context. Each displayed programme had a higher mean within-programme cosine than the mean of 200 size-matched random gene sets. The gene cosine and RNA Pearson scores in Fig.~\ref{fig:5}i summarize programme organization within their respective association spaces.

Developmental references came from the published VASA-seq combined-stage marker table\cite{salmen2022}. Coding markers with adjusted $P<0.01$ and log fold change greater than 1 were ranked by the published marker score, retaining up to 60 genes before measurement-support filtering. The \textit{Foxl2os}, \textit{C130021I20Rik} and \textit{Gm50321} analyses contained 60 of 2,149, 60 of 2,152 and 46 of 2,117 reference members/candidates, respectively. Coding candidates required detection in at least $\max(5,\lceil0.05N\rceil)$ of the context's $N$ cells; non-coding queries required at least five RNA-positive cells. All cells contributed, including those with zero measured query expression. VASA-seq contributed to pretraining; these analyses evaluated recovery of known programmes in the specified developmental contexts.

The complete developmental top-20 lists contained eight, seven and three reference members, compared with six, four and zero for RNA correlation (Supplementary Fig.~\ref{ed:programmes}c). Each method used identical cells and fixed full-population candidate and reference sets in the disjoint half-population comparisons. CellNet retained higher average precision in all six halves (Supplementary Fig.~\ref{ed:programmes}d).

The human neural analysis used the published neural stem-cell state-12 marker programme\cite{wu2022}. Coding markers with adjusted $P<0.01$ and average log$_2$ fold change greater than 0.25 were ranked by the published fold change, retaining the top 60 before coverage filtering. The eligible reference contained 59 genes among 2,082 coding candidates. All 3,423 verified non-targeting controls contributed to both methods' associations; 13 had measured \textit{SERTAD4-AS1}. \textit{WNT5B}, \textit{DLK1} and \textit{SERPINF1} were recovered at ranks 7, 10 and 17. The original perturbation study supplied the programme reference; this retrieval analysis used its control cells.

\clearpage

\begin{figure}[p]
\centering
\includegraphics[width=\textwidth]{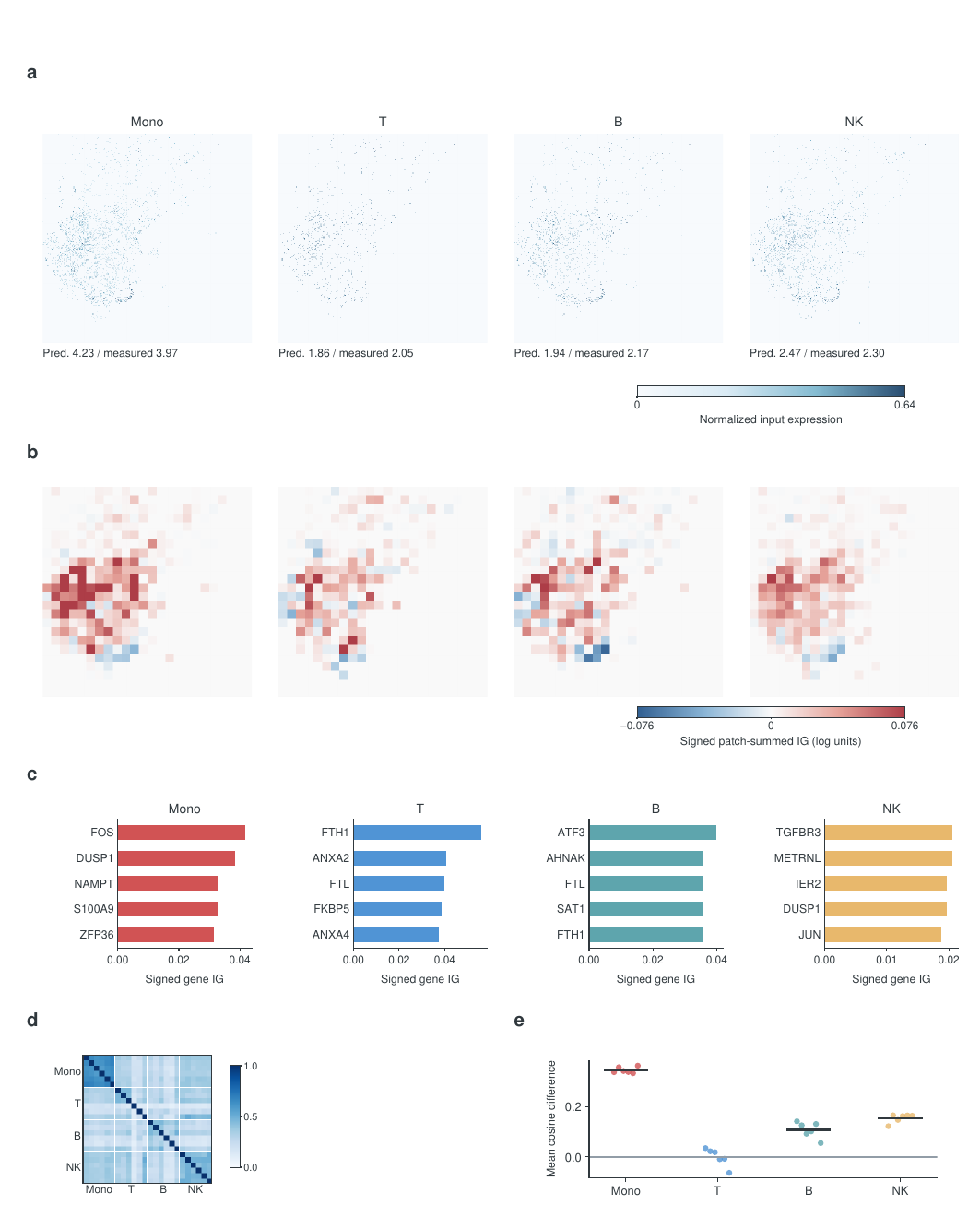}
\caption{Cell-context variation in the measured inputs supporting NEAT1 predictions. \textit{NEAT1} is withheld from the input in 24 PBMC cells, six per population; cell selection and integrated gradients are defined in Section~\ref{note:inputs}. \textbf{a}, Measured-input images for one monocyte, T, B and natural killer cell, selected retrospectively for detected targets, absolute prediction error $\leq0.4$ log units and the largest removal contrast within group. \textbf{b}, Signed contributions summed within $16\times16$ patches. All four maps share a symmetric 99th-percentile scale. \textbf{c}, Five highest positive-contribution measured genes in each displayed cell. \textbf{d}, Cosine similarity of full signed gene-contribution vectors across all 24 cells. Diagonal entries denote self-similarity. \textbf{e}, Each cell's mean similarity to its five same-group neighbours minus its mean similarity to the 18 other-group cells. Points denote cells and bars group means; pairwise entries share cells. Input-image and contribution colours have separate scales. Coordinates identify genes.}
\label{ed:context}
\end{figure}

\begin{figure}[p]
\centering
\includegraphics[width=0.9\textwidth]{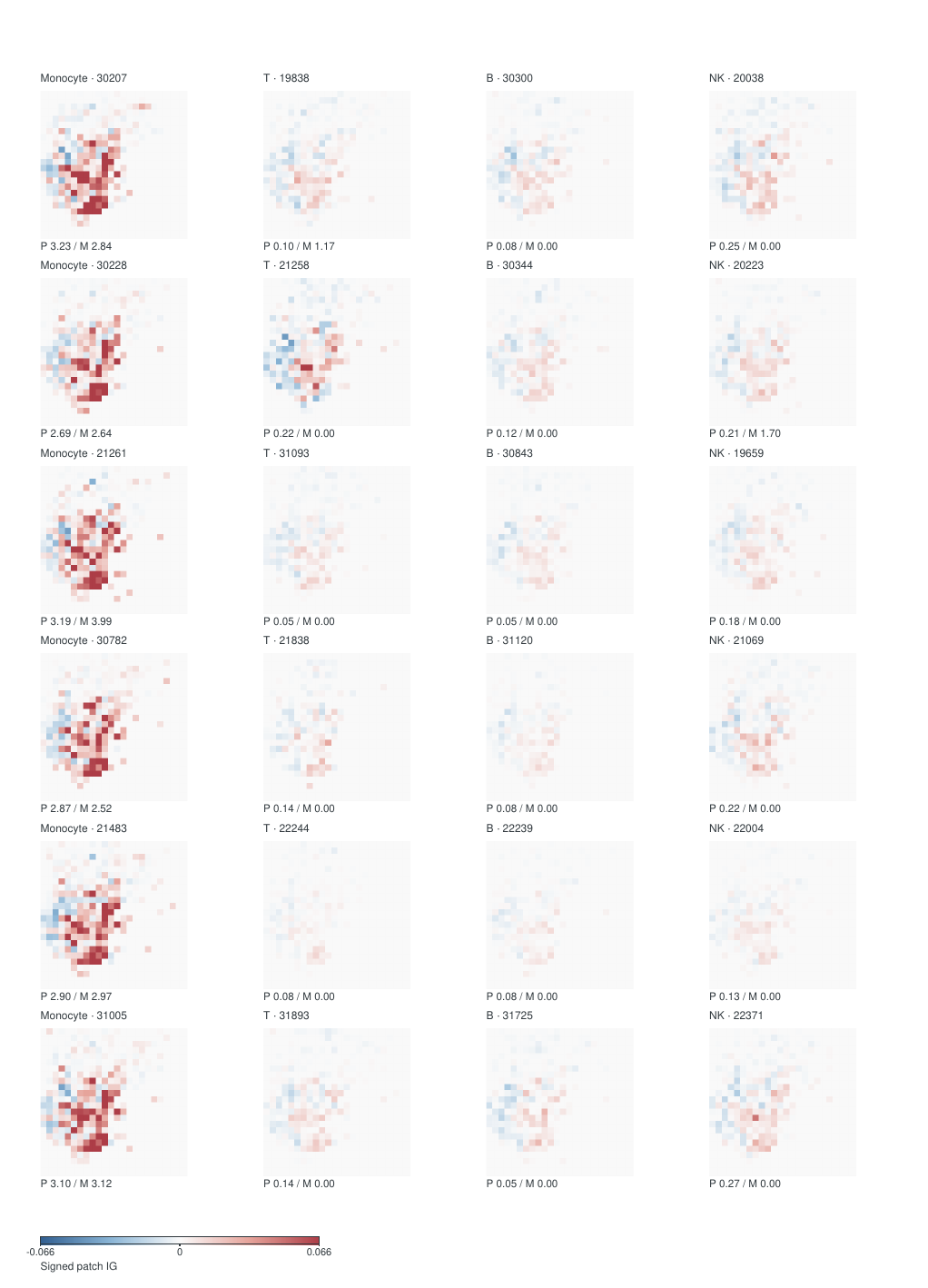}
\caption{Complete input-contribution maps for CST3 predictions. All 24 selected PBMC cells are shown, with six cells per monocyte, T, B and natural killer-cell column. Cells span the 55th--90th percentiles of summed original \textit{CST3} and \textit{NEAT1} predictions within population. The complete target partition is withheld from the input (Section~\ref{note:inputs}). Each $24\times24$ map sums signed integrated gradients within $16\times16$ patches. Positive and negative colours indicate inputs that raise or lower the \textit{CST3} prediction relative to the zero-expression reference. Cell identifiers and predicted (P) and measured (M) log-expression values accompany each map. The scale uses the absolute 99th percentile across all 48 maps in this figure and Supplementary Fig.~\ref{ed:neat1}. Coordinates identify gene positions.}
\label{ed:cst3}
\end{figure}

\begin{figure}[p]
\centering
\includegraphics[width=0.9\textwidth]{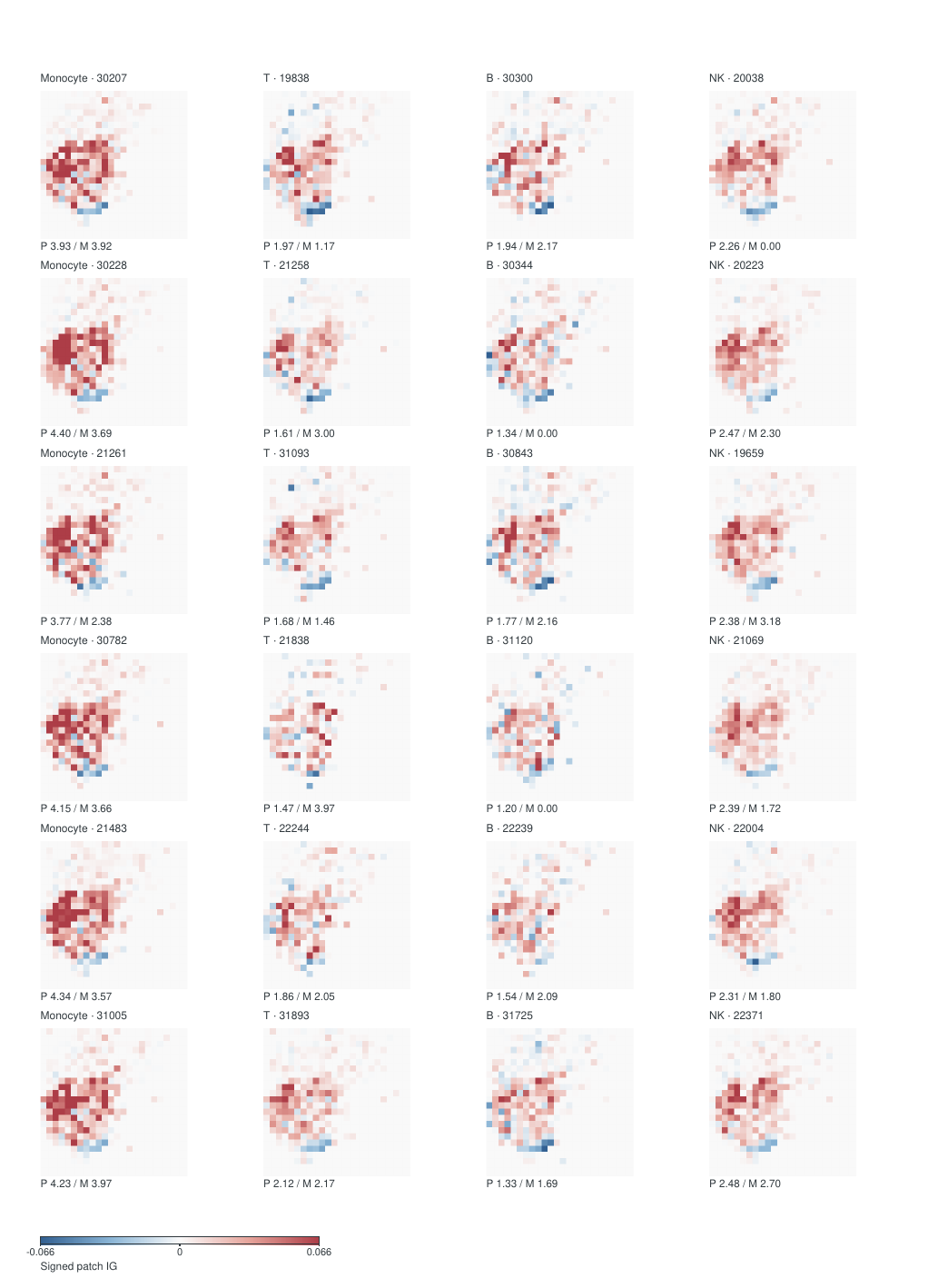}
\caption{Complete input-contribution maps for NEAT1 predictions. All 24 selected PBMC cells are shown, six per monocyte, T, B and natural killer-cell column, using the cell selection and hidden-target protocol in Section~\ref{note:inputs}. Each map sums signed integrated gradients within $16\times16$ patches; positive and negative colours indicate contributions that raise or lower the \textit{NEAT1} prediction relative to zero input expression. Cell identifiers and predicted (P) and measured (M) log-expression values accompany each map. The symmetric 99th-percentile scale is shared with Supplementary Fig.~\ref{ed:cst3}, allowing direct comparisons across the two targets and all cells. Coordinates identify gene positions.}
\label{ed:neat1}
\end{figure}

\begin{figure}[p]
\centering
\includegraphics[width=\textwidth]{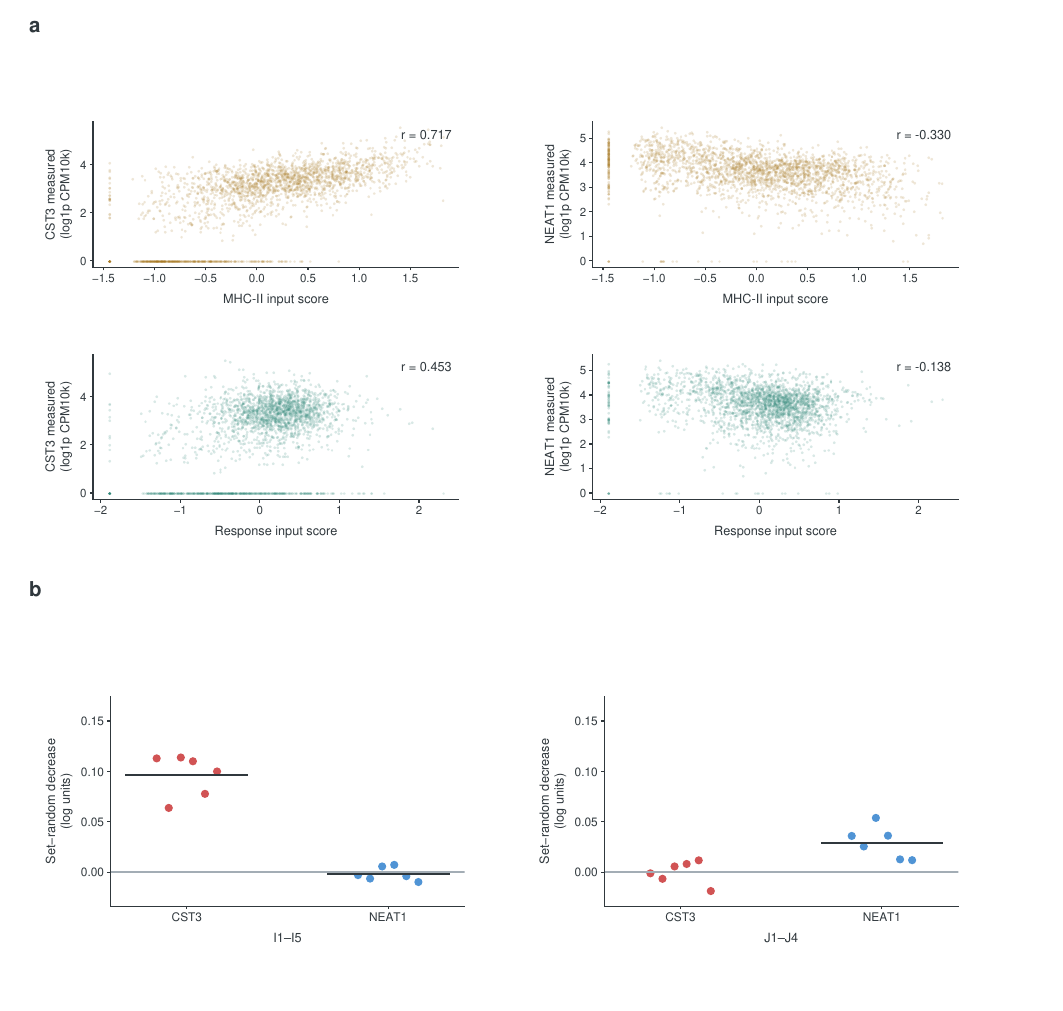}
\caption{Expression context and target-specific removal of named input sets. I1--I5 denote \textit{CD74}, \textit{HLA-DRA}, \textit{HLA-DPA1}, \textit{IFI30} and \textit{HLA-DMB}; J1--J4 denote \textit{DUSP1}, \textit{FOS}, \textit{IER2} and \textit{ZFP36}. \textbf{a}, Measured target expression against each set's mean standardized log-expression in 2,273 reference monocytes, excluding attribution cells. Points denote cells; $r$ is Pearson correlation. \textbf{b}, Joint expression-only removal of either set for both hidden targets in six selected monocytes. Eight random sets per cell match expressed-input count and expression mass; other input information and normalization remain fixed. Points show the named-set prediction decrease minus the cell's mean random decrease; bars show cell means. I--\textit{CST3}, I--\textit{NEAT1}, J--\textit{CST3} and J--\textit{NEAT1} contrasts average 0.0963, $-0.0018$, $-0.0002$ and 0.0293 log units. Expression relationships and removal effects are discussed in Section~\ref{note:inputs}.}
\label{ed:named}
\end{figure}

\begin{figure}[p]
\centering
\includegraphics[width=\textwidth]{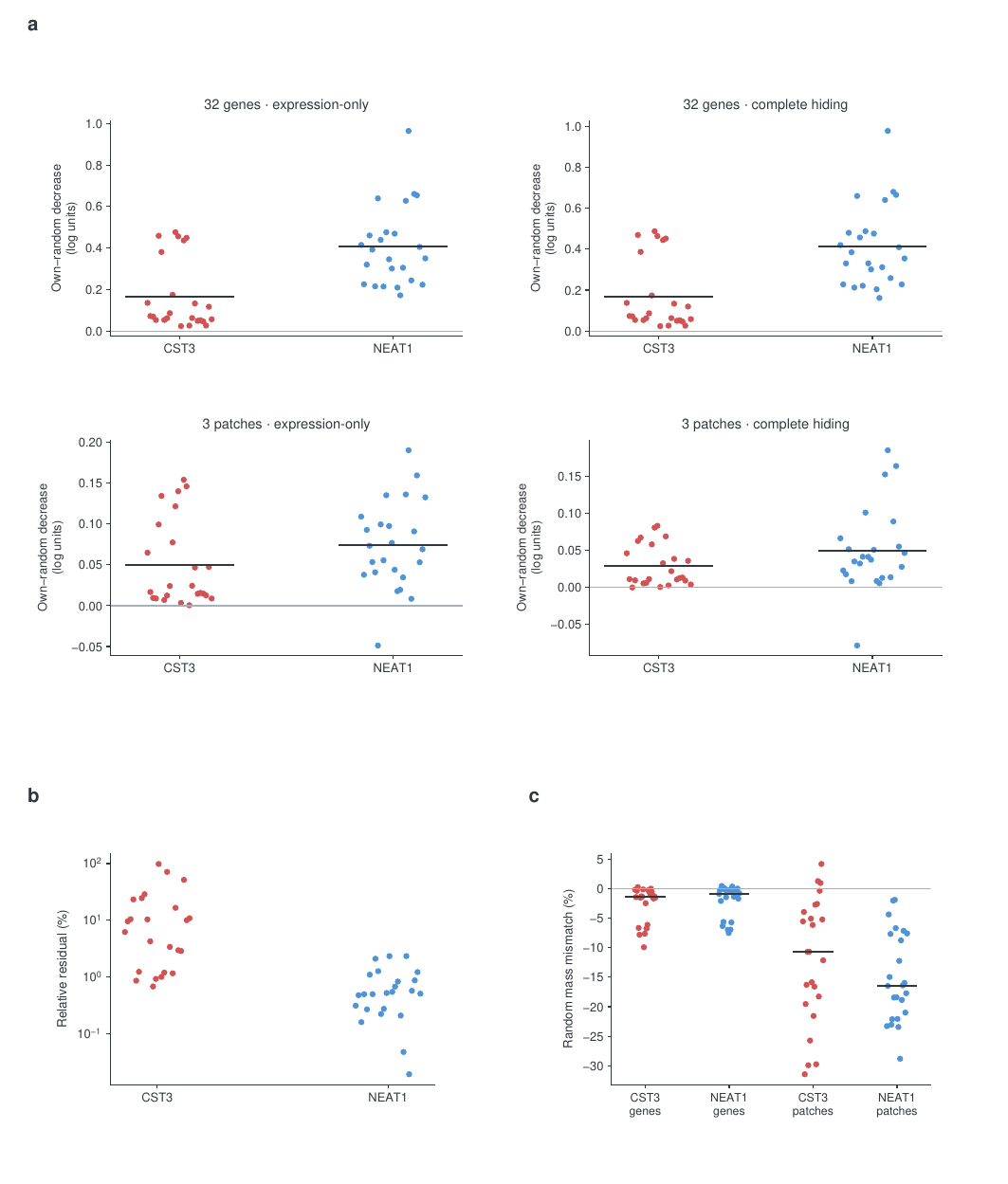}
\caption{Input-removal controls and integrated-gradient diagnostics. Both targets are hidden in the 24 selected PBMC cells described in Section~\ref{note:inputs}. \textbf{a}, Removal of each target's 32 highest positive-contribution detected genes or three highest-contribution patches. Expression-only removal sets expression to zero; complete hiding also removes measurement, identity and detection information. Original normalization is retained. Points show target-selected minus mean matched-random prediction decreases; eight random trials are averaged per cell and bars show cell means. Gene controls approximately match expression mass; patch controls also match detected-gene count. \textbf{b}, Relative discrepancy between summed integrated gradients and the prediction change from the zero-expression reference: $100\times|\sum_g\mathrm{IG}_g-\Delta\hat{x}|/|\Delta\hat{x}|$, with denominator floor $10^{-8}$. All 48 cell/target values are retained on a symmetric logarithmic scale; median and maximum residuals are 1.18\% and 98.15\%. \textbf{c}, Percentage removed-expression mismatch, $100\times(\mathrm{mean\ random\ mass}-\mathrm{selected\ mass})/\mathrm{selected\ mass}$, with the same denominator floor. Points denote cells and bars medians.}
\label{ed:controls}
\end{figure}

\begin{figure}[p]
\centering
\includegraphics[width=\textwidth]{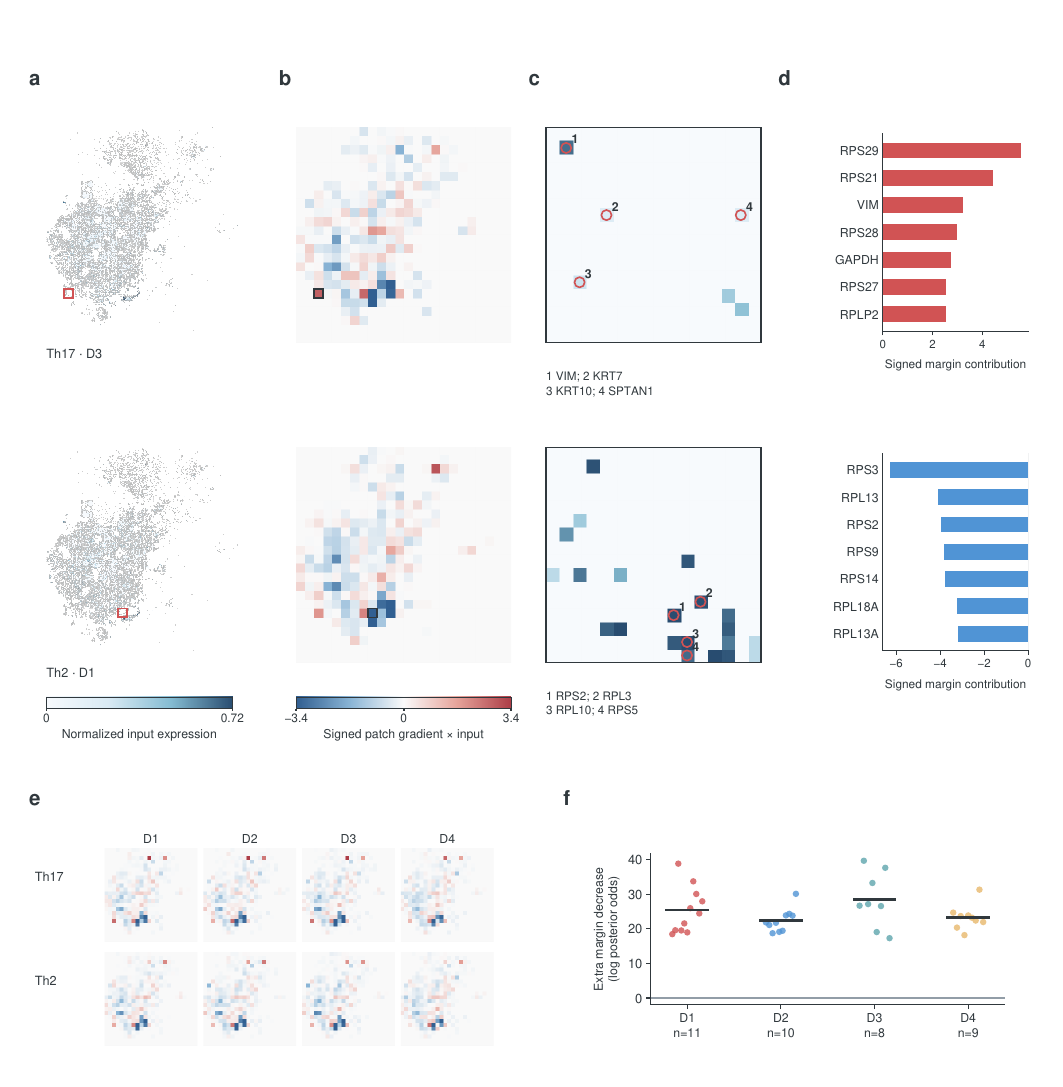}
\caption{Cell-state prediction evidence across donors. A linear discriminant classifier trained on three donors predicts Th17 and Th2 conditions in the fourth, rotating across four donors. Gradient-times-input attributions explain the Th17-minus-Th2 log-posterior score; positive contributions support Th17. \textbf{a--d}, Two correctly classified examples selected retrospectively by the largest 10\% patch-removal contrasts: input images (a), signed patch contributions (b), enlarged strongest supporting regions (c), and seven strongest class-supporting gene contributions (d). Numbers in c identify the four strongest contributors within each region. \textbf{e}, Donor/state mean maps from 96 cells, 12 per state and donor, on a common scale. Within-state donor-mean gene contributions have Spearman correlations of 0.717--0.752. \textbf{f}, Removal of the 10\% strongest supporting patches against random sets of the same patch count. Eligible random sets match expression mass and detected-gene count within 10\%. All 38 qualifying cells are Th17. Points show each cell's extra score decrease relative to random removal; bars show donor means. Sampling and classifier relationships are given in Section~\ref{note:states}.}
\label{ed:state}
\end{figure}

\begin{figure}[p]
\centering
\includegraphics[width=\textwidth]{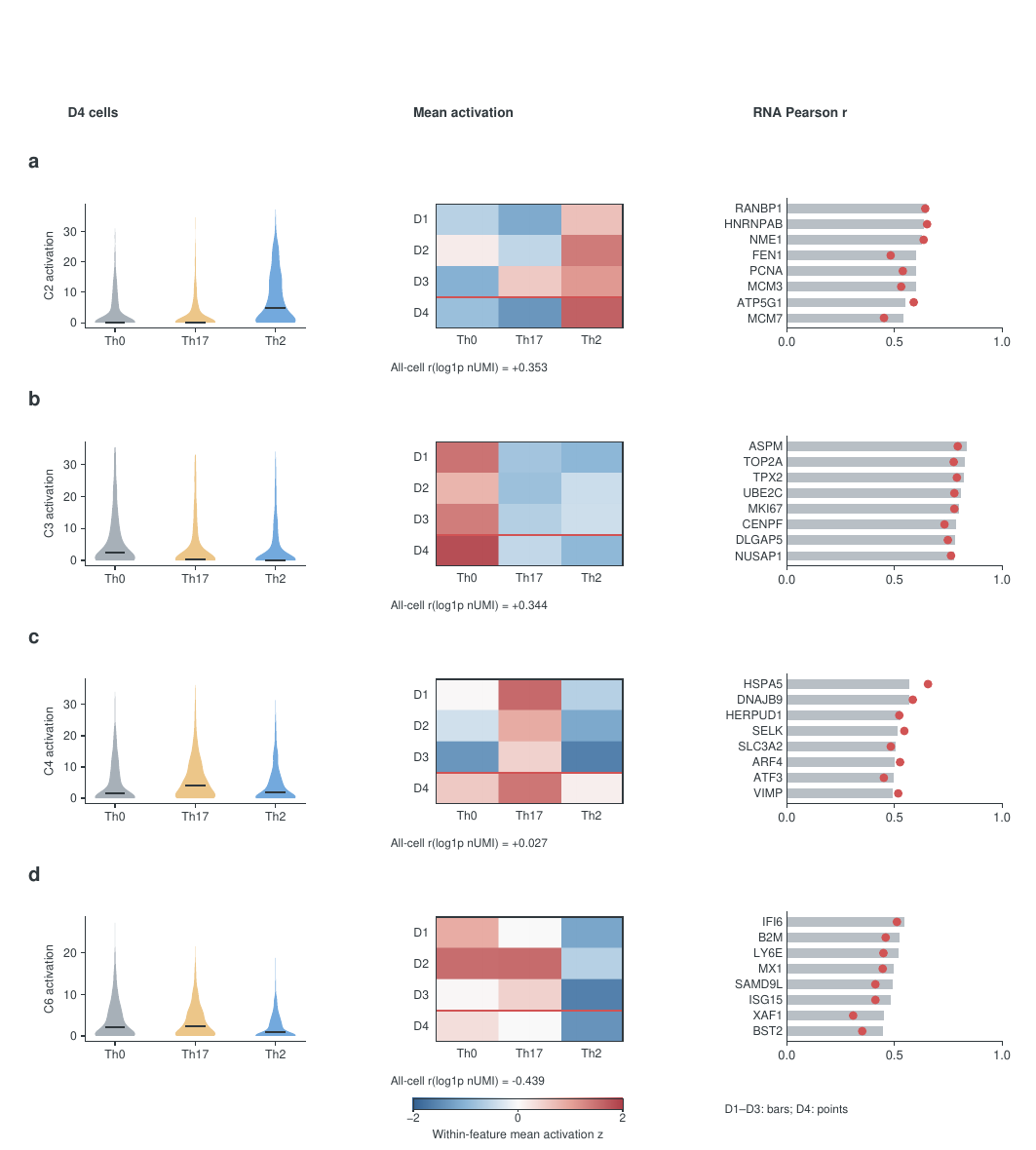}
\caption{Expression-associated features of cell representations. Six-component sparse nonnegative matrix factorization was fitted using 2,700 cells from donors D1--D3 and applied to all 9,777 CD4 cells; D4 was held out from fitting and gene ranking (Section~\ref{note:states}). \textbf{a--d}, Selected features associated with DNA replication, mitosis, endoplasmic-reticulum stress and interferon programmes. Left, D4 activation distributions by state; violins show distributions and horizontal marks medians (Th0, 712; Th17, 655; Th2, 1,012 cells). Centre, donor/state mean activations standardized within feature; frames identify D4. Correlations with log-transformed total RNA counts describe depth dependence. Right, eight genes per feature with the highest within-donor expression correlations in D1--D3. Grey bars show these training correlations and coral points the corresponding D4 correlations. All 32 selected associations remain positive in D4. Gene expression is log-transformed counts per 10,000, with donor means removed before correlation.}
\label{ed:features}
\end{figure}

\begin{figure}[p]
\centering
\includegraphics[width=0.8\textwidth]{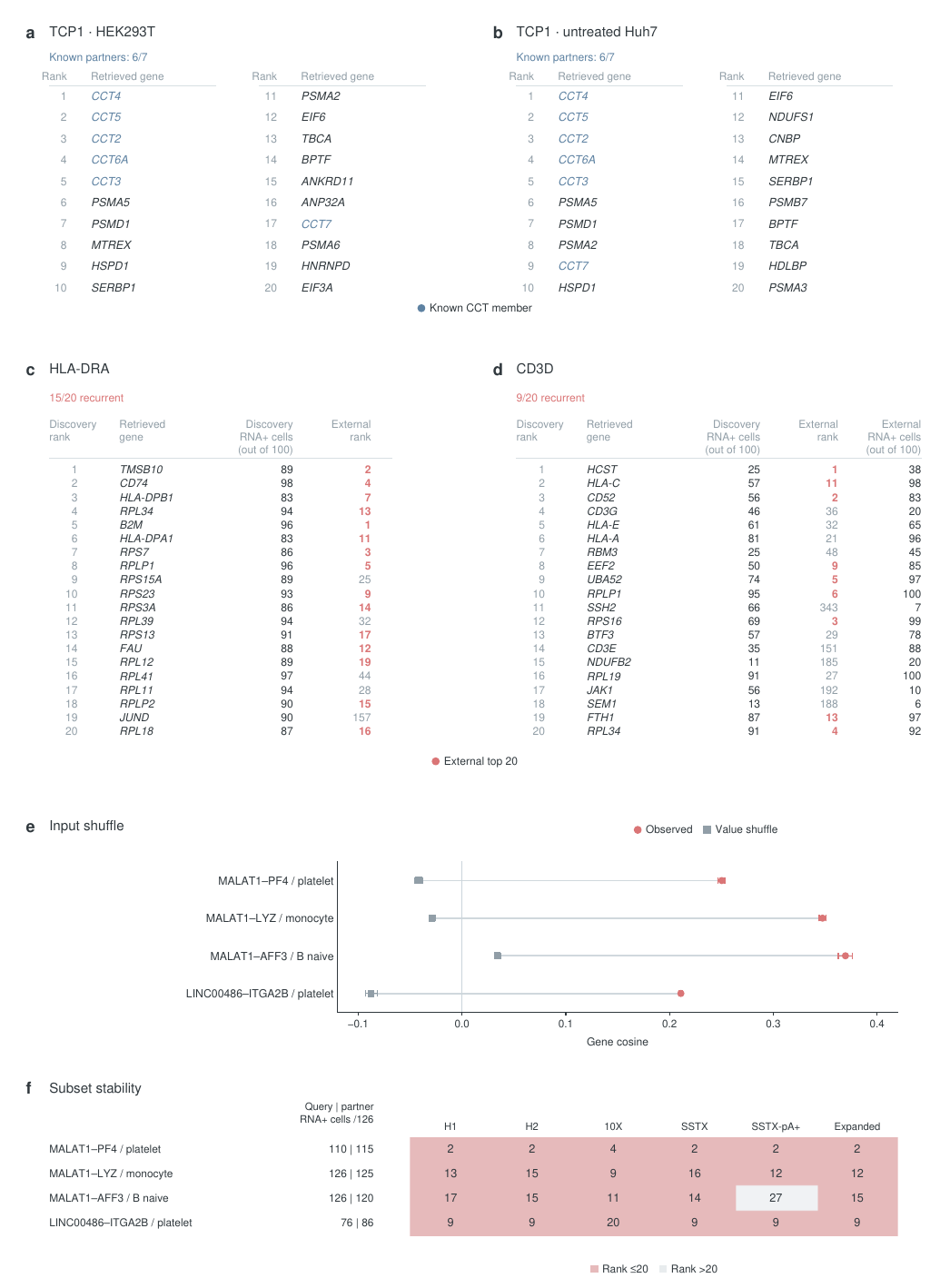}
\caption{Partner identities, cross-dataset recurrence and cell-context support. \textbf{a,b}, \textit{TCP1} top-20 neighbours in HEK293T and untreated Huh7, 500 cells per population. Blocks continue ranks 1--10 and 11--20; blue names identify six of seven eligible CCT partners. \textbf{c,d}, Fixed discovery lists for \textit{HLA-DRA} and \textit{CD3D}. Columns report discovery/external ranks and available RNA-positive cell counts for each candidate, out of 100; coral marks external ranks $\leq20$. \textbf{e}, Observed (circles) and within-cell expression-shuffled (squares) similarities for the four cases in Fig.~\ref{fig:5}g,h. Horizontal intervals are paired percentile 95\% cell-bootstrap intervals from 1,000 protocol-stratified resamples of the same cohort and shuffle realization. \textbf{f}, Complete-pool ranks in disjoint 63-cell halves (H1, H2), individual protocols (10X-dUTSO, $n=8$; SSTX, $n=74$; SSTX-pA+, $n=44$), and expanded populations (126 platelets or 300 B, T or monocyte cells). Blue cells indicate ranks $\leq20$. Query and partner RNA-positive counts each use 126 cells.}
\label{ed:associations}
\end{figure}

\begin{figure}[p]
\centering
\includegraphics[width=\textwidth]{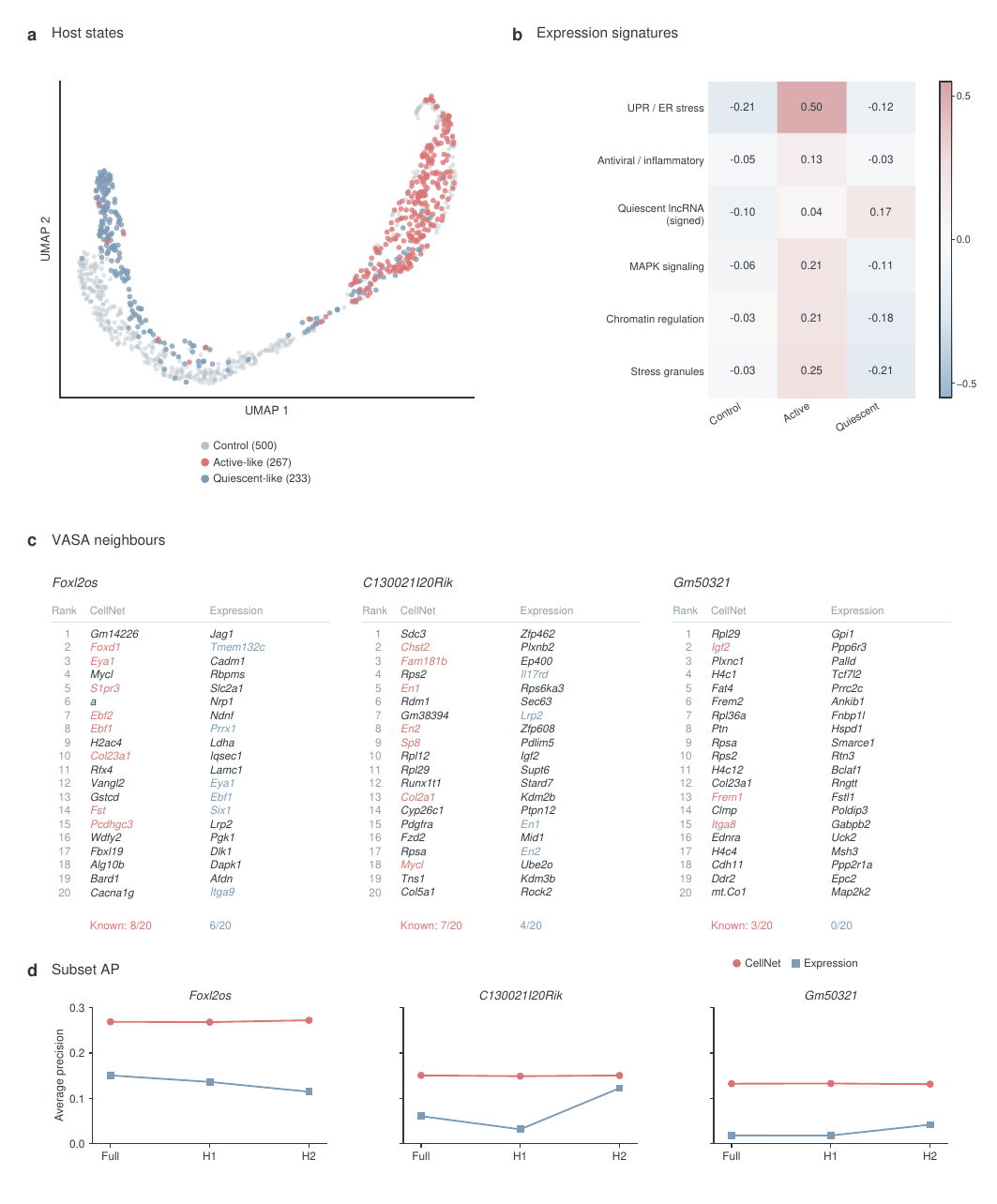}
\caption{Host-expression context and developmental programme support. \textbf{a}, Huh7 cell embeddings: 500 untreated controls and 500 DENV-exposed cells, with the latter annotated as active-like (267) or quiescent-like (233) from host expression. Each point denotes a cell. \textbf{b}, Mean standardized expression-programme scores in the three groups. Red and blue denote higher and lower scores; five scores define the exposed-cell annotations and stress granules supply additional expression context. \textbf{c}, Complete top-20 lists for each developmental query in Fig.~\ref{fig:5}j. Independently ranked model and RNA candidates are shown side by side; coloured names denote published programme members. Model member counts are 8, 7 and 3, compared with 6, 4 and 0 for RNA correlation. \textbf{d}, Complete-ranking average precision in full populations and disjoint matched halves (50/50, 50/50 and 61/61 cells). Both methods use the same cells and fixed full-population candidate and reference sets. Model half-population AP values are 0.268/0.272, 0.149/0.150 and 0.133/0.131. Programme references, host annotations and evaluation conditions are specified in Section~\ref{note:programmes}.}
\label{ed:programmes}
\end{figure}

\clearpage
\section{Supplementary Methods}\label{sec:supplementary-methods}

\subsection{Datasets}\label{sm:datasets}

We used a public pretraining collection and fifteen public datasets or dataset collections spanning human tissues, cell lines, vertebrate and invertebrate cell atlases, plant tissues and spatial transcriptomics. Table~\ref{tab:methods-s2} summarizes the datasets and adopted cohort sizes; their sources, biological material and measurement details are described below. Species-level composition of the multispecies atlases is provided in Table~\ref{tab:methods-s3}. Original donor, sample, cell-type and experimental-condition annotations were retained. Data partitioning, analysis-specific sampling and evaluation procedures are described in \ref{sm:cellular-evaluation}–\ref{sm:programme-retrieval} and \ref{sm:metrics}.

\textbf{Main pretraining collection.} The main collection contained 231,190,557 cell or nucleus profiles from 26,467 constituent expression-matrix entries, representing 40 source species designations, 1,015,163 species-qualified gene identities and 2,233 harmonized cell-type categories. Data were obtained from GEO, ArrayExpress, BioProject-associated repositories and CELLxGENE Discover, a curated and interoperable single-cell data platform~\citesupp{cellxgene2025,sccompass2025}. Constituent entries denote expression matrices rather than independent publications or biological replicates. Source quality control and measured-gene annotations were retained; expression normalization and species-aware gene harmonization are described in \ref{sm:expression}-\ref{sm:projector}.

\textbf{Tabula Sapiens 2.0.} We used the \href{https://www.biorxiv.org/content/10.1101/2024.12.03.626516v2}{Tabula Sapiens 2.0 human multi-organ atlas}, retaining source tissue, donor and cell-type annotations. The dataset is available from \href{https://figshare.com/articles/dataset/Tabula_Sapiens_v2/27921984}{Figshare}. The adopted collections comprised a 692,102-profile matrix and a 549,018-profile labelled collection from nine donors; nine profiles without valid labels were excluded from the latter. These numbers describe the supplied subsets used here rather than the complete atlas. Eligibility of tissues and labelled cells is specified in \ref{sm:cellular-evaluation}.

\textbf{Human kidney atlas.} The \href{https://www.nature.com/articles/s41586-023-05769-3}{atlas of healthy and injured human kidney} was obtained from the \href{https://cellxgene.cziscience.com/collections/bcb61471-2a44-4d00-a0af-ff085512674c}{corresponding CELLxGENE collection}. The adopted cohort contained 107,344 profiles from 45 donors and 57 author-provided cell labels. Donor identities and injury annotations were retained. Proximal-tubule subsets were drawn from this same atlas.

\textbf{Human lung atlas.} We used the \href{https://www.nature.com/articles/s41586-020-2922-4}{Travaglini et al. human lung atlas}, available from its \href{https://cellxgene.cziscience.com/collections/5d445965-6f1a-4b68-ba3a-b8f765155d3a}{CELLxGENE collection}, combining its 10x and Smart-seq2 lung transcriptomes. After excluding blood samples, the adopted cohort comprised 69,650 cells from three donors with 59 author-provided labels. The original donor identities and cell annotations were retained.

\textbf{Cano-Gamez CD4 T-cell dataset.} The \href{https://www.nature.com/articles/s41467-020-15543-y}{Cano-Gamez et al. CD4 T-cell dataset} was obtained from \href{https://www.ebi.ac.uk/biostudies/studies/S-BSST2978}{BioStudies accession S-BSST2978}. We retained the author-defined Naive starting population in 10x technical batch 1 under Th0, Th2 and Th17 conditions. The resulting collection contained 9,777 cells from four donors and 20,953 measured genes. Condition labels were taken from the experimental metadata.

\textbf{WI-38 senescence dataset.} The \href{https://www.aging-us.com/article/204666/text}{WI-38 fibroblast transcriptomes} were obtained from \href{https://www.ncbi.nlm.nih.gov/geo/query/acc.cgi?acc=GSE226225}{GSE226225}. Six 10x libraries, GSM7068355–GSM7068360, represented replicative senescence, irradiation-associated senescence and etoposide-associated senescence, with two libraries per condition. The adopted matrix contained 14,457 cells and 36,601 measured genes after the quality-control criteria described in \ref{sm:cellular-evaluation}. Library replicates were derived from the same human cell line.

\textbf{Circulating immune-cell atlas.} Author-annotated circulating immune-cell matrices were obtained from the \href{https://doi.org/10.6084/m9.figshare.28008611}{seven-species Figshare collection 28008611}. Species comprised human, rhesus macaque, Chinese softshell turtle, Pacific white shrimp, Ciona intestinalis, half-smooth tongue sole and whitespotted bamboo shark. The adopted labelled cohort contained 19,727 profiles in total (Table~\ref{tab:methods-s3}). The human matrix was supplied as counts per million without raw counts and retained its source expression convention.

\textbf{Spermatogenesis atlas.} We used the single-nucleus transcriptomes from \citesupp{spermatogenesis2023}, \href{https://www.nature.com/articles/s41586-022-05547-7}{The molecular evolution of spermatogenesis across mammals}, with data available through the \href{https://apps.kaessmannlab.org/SpermEvol/}{author portal}. The adopted atlas encompassed human, chimpanzee, bonobo, gorilla, gibbon, rhesus macaque, common marmoset, mouse, opossum, platypus and chicken, with 96,464 labelled nuclei in total (Table~\ref{tab:methods-s3}). Original cell annotations were retained. Gibbon is reported using the established common species designation for the analysed collection.

\textbf{Brassicaceae multi-tissue atlas.} The adopted author-annotated matrices, deposited under \href{https://www.ncbi.nlm.nih.gov/geo/query/acc.cgi?acc=GSE297576}{GEO accession GSE297576}, contained 86,015 profiles from Arabidopsis thaliana, Arabidopsis lyrata, Brassica oleracea and Capsella rubella (Table~\ref{tab:methods-s3}). Source tissue and sample annotations were retained, together with the species-specific cell labels and shared coarse labels supplied with the matrices.

\textbf{Brassicaceae root atlas.} Root transcriptomes were obtained from \href{https://www.ncbi.nlm.nih.gov/geo/query/acc.cgi?acc=GSE268881}{GSE268881}. The adopted collection contained 184,951 profiles from Arabidopsis thaliana, Camelina sativa, Eutrema salsugineum, Schrenkiella parvula and Sisymbrium irio, spanning control and author-defined stress conditions. Source condition, replicate and cell-type annotations were retained. The root atlas and its accession refer to the same underlying collection.

\textbf{TotalX dataset collection.} The \href{https://www.nature.com/articles/s41587-026-03068-6}{TotalX study} provided human PBMC, developing-brain, HEK293T and Huh7 total-RNA transcriptomes. Processed matrices and annotations were obtained from \href{https://doi.org/10.6084/m9.figshare.29803628}{Figshare 29803628}. Raw brain and PBMC sequencing data were deposited under Synapse accession syn72382564, and cell-line sequencing data under \href{https://www.ncbi.nlm.nih.gov/geo/query/acc.cgi?acc=GSE315939}{GSE315939}. The PBMC and brain source matrices contained 58,467 and 301,515 profiles, respectively. PBMC assay groups 10X-dUTSO, SSTX and SSTX-pA+ contained 19,431, 18,497 and 20,539 profiles. These are technical capture groups rather than distinct biological conditions. Cell-line collections included untreated and author-defined DENV-exposed Huh7 profiles. Original cell-type, sample and assay annotations were retained.

\textbf{Xenium Prime 5K mouse-brain dataset.} We used the public \href{https://www.10xgenomics.com/datasets/xenium-prime-fresh-frozen-mouse-brain}{10x Xenium Prime fresh-frozen mouse-brain dataset}~\citesupp{xenium2024mousebrain}, comprising one coronal section with 63,173 cells and 5,006 assayed genes. Direct gene measurements, cell annotations and original micrometre coordinates were retained. Query and reference subsets from this single section are specified in \ref{sm:expression-evaluation}.

\textbf{Purified PBMC dataset.} The \href{https://www.nature.com/articles/ncomms14049}{Zheng et al. purified human PBMC collections}~\cite{zheng2017} were obtained in the processed form distributed through \href{https://docs.scvi-tools.org/en/1.5.0/api/reference/scvi.data.purified_pbmc_dataset.html}{scvi-tools}. The supplied matrix contained 105,868 cells, 21,932 measured genes and ten annotated populations. Original population and sample annotations were retained.

\textbf{Perirhinal cortex dataset.} Human perirhinal-cortex nuclear transcriptomes were obtained from the \href{https://cellxgene.cziscience.com/collections/283d65eb-dd53-496d-adb7-7570c7caa443}{Human Brain Cell Atlas v1.0 collection}, using the processed release distributed with the \href{https://github.com/bowang-lab/scGPT/blob/main/data/README.md}{scGPT datasets}. The source matrix represented the rostral subdivision of perirhinal area 35 and contained 17,535 nuclei, 59,357 measured genes, ten annotated cell types and two sample groups.

\textbf{VASA-seq developmental dataset.} Mouse developmental total-RNA transcriptomes were obtained from the \href{https://www.nature.com/articles/s41587-022-01361-8}{VASA-seq study}, accession \href{https://www.ncbi.nlm.nih.gov/geo/query/acc.cgi?acc=GSE176588}{GSE176588}. The adopted profiles spanned embryonic stages E6.5, E7.5 and E8.5, retaining author-provided lineage and developmental-stage annotations. Published coding-gene programmes were obtained from the original Supplementary Table 5. This source contributed to pretraining; the developmental programme analyses are described in \ref{sm:programme-retrieval}.

\textbf{Human neural-induction dataset.} We used the \href{https://doi.org/10.1016/j.xgen.2022.100177}{Wu et al. human neural-induction study}, accession \href{https://www.ncbi.nlm.nih.gov/geo/query/acc.cgi?acc=GSE150062}{GSE150062}. The adopted transcriptomes comprised all 3,423 author-verified non-targeting control cells. Source log-transformed expression and published neural-state marker annotations were retained. The non-targeting controls supplied observed transcriptomes; the original study supplied the programme reference.

\small
\setlength{\LTleft}{0.05\textwidth}
\setlength{\LTright}{0pt}
\begin{longtable}{@{}p{0.30\textwidth}p{0.65\textwidth}@{}}
\caption{Summary of datasets and adopted cohorts.}\label{tab:methods-s2}\\
\toprule
Dataset or collection & Adopted cohort used in this study \\
\midrule
\endfirsthead
\toprule
Dataset or collection & Adopted cohort used in this study \\
\midrule
\endhead
Main pretraining collection & 231,190,557 cell or nucleus profiles; 26,467 expression matrices; 40 source species \\
Tabula Sapiens 2.0 & Two adopted collections: 692,102-profile matrix and 549,018 labelled profiles \\
Human kidney atlas & 107,344 profiles; 45 donors; 57 labels \\
Travaglini lung atlas & 69,650 cells; three donors; 59 labels \\
Cano-Gamez CD4 T-cell dataset & 9,777 cells; four donors; Th0, Th2 and Th17 conditions \\
WI-38 senescence dataset & 14,457 cells; six libraries; three conditions \\
Circulating immune-cell atlas & 19,727 labelled profiles across seven species \\
Spermatogenesis atlas & 96,464 labelled nuclei across eleven species \\
Brassicaceae multi-tissue atlas & 86,015 labelled profiles across four species \\
Brassicaceae root atlas, GSE268881 & 184,951 labelled profiles across five species \\
TotalX collection & 58,467 PBMC profiles; 301,515 developing-brain profiles; author-provided cell-line matrices \\
Xenium Prime 5K mouse brain & 63,173 cells; 5,006 assayed genes \\
Purified PBMC dataset & 105,868 cells; 21,932 genes \\
Perirhinal cortex dataset & 17,535 nuclei; 59,357 genes; two sample groups \\
VASA-seq developmental dataset & Author-annotated developmental and lineage subsets at E6.5, E7.5 and E8.5 \\
Human neural-induction dataset, GSE150062 & 3,423 verified non-targeting control cells \\
\bottomrule
\end{longtable}

\normalsize

Repositories and accession identifiers are given in the dataset descriptions above. Counts refer to the adopted source matrices or labelled atlas cohorts, rather than the size of every source publication. Smaller analysis subsets and training, validation and test memberships are specified in the corresponding methods.

\small
\setlength{\LTleft}{0.05\textwidth}
\setlength{\LTright}{0pt}
\begin{longtable}{@{}p{0.45\textwidth}p{0.20\textwidth}p{0.18\textwidth}@{}}
\caption{Species composition of the adopted multispecies atlases.}\label{tab:methods-s3}\\
\toprule
Dataset & Species & Retained labelled profiles \\
\midrule
\endfirsthead
\toprule
Dataset & Species & Retained labelled profiles \\
\midrule
\endhead
Brassicaceae roots (GSE268881) & Arabidopsis thaliana & 47,033 \\
Brassicaceae roots (GSE268881) & Camelina sativa & 27,485 \\
Brassicaceae roots (GSE268881) & Eutrema salsugineum & 33,975 \\
Brassicaceae roots (GSE268881) & Schrenkiella parvula & 36,021 \\
Brassicaceae roots (GSE268881) & Sisymbrium irio & 40,437 \\
Circulating immune cells & Whitespotted bamboo shark & 1,580 \\
Circulating immune cells & Ciona intestinalis & 2,016 \\
Circulating immune cells & Half-smooth tongue sole & 6,567 \\
Circulating immune cells & Human & 3,005 \\
Circulating immune cells & Pacific white shrimp & 4,625 \\
Circulating immune cells & Rhesus macaque & 984 \\
Circulating immune cells & Chinese softshell turtle & 950 \\
Testis & Common marmoset & 6,384 \\
Testis & Chicken & 8,413 \\
Testis & Gorilla & 7,323 \\
Testis & Human & 16,011 \\
Testis & Rhesus macaque & 7,618 \\
Testis & Opossum & 9,926 \\
Testis & Mouse & 5,264 \\
Testis & Gibbon & 6,572 \\
Testis & Platypus & 6,751 \\
Testis & Bonobo & 9,873 \\
Testis & Chimpanzee & 12,329 \\
Brassicaceae multi-tissue & Arabidopsis lyrata & 13,495 \\
Brassicaceae multi-tissue & Arabidopsis thaliana & 42,108 \\
Brassicaceae multi-tissue & Brassica oleracea & 7,682 \\
Brassicaceae multi-tissue & Capsella rubella & 22,730 \\
\bottomrule
\end{longtable}

\normalsize

\subsection{Third-party comparison methods}\label{sm:comparators}

\textbf{Model releases and task scope.} Released pretrained models and their original inference procedures supplied frozen cell representations or native expression predictions. Gene mapping followed each model's supported identities and species conventions. Unsupported species or genes had unavailable predictions rather than zero scores. Geneformer V2, scGPT, scLong, scPRINT-2, TranscriptFormer and UCE formed the six cellular comparators in Fig. 2. Expression comparisons in Fig. 3 additionally included CellFM. The model subset varied with the task and native output support.

\textbf{Geneformer V2.} The V2-104M release~\citesupp{geneformer} used Ensembl-mapped expression and the authors' rank-value tokenization. Its frozen 768-dimensional CLS representation used the official second-to-last-layer extraction setting. Downstream labels trained only the classifier described in \ref{sm:cellular-evaluation}.

\textbf{scGPT.} The whole-human release~\cite{scgpt} used its supported vocabulary and official binning. Cell analysis used 512-dimensional CLS features and maximum sequence length 1,200. Completion used 51 expression bins and at most 600 nonzero visible-context genes. Hidden expression was excluded from bin construction. Native predictions were inverse-binned through visible-expression boundaries to log1p counts per 10,000 without query-truth calibration.

\textbf{scLong.} The released model~\cite{sclong} retained the fixed 27,000-gene axis used for these analyses. Cell features mean-pooled merged gene decodings, excluding the terminal padding position. Completion retained the frozen native expression-regression head. Missing expression on the fixed gene axis followed the authors' input convention; target measurements were excluded from visible-context normalization and input.

\textbf{scPRINT-2.} The small-v2 release~\citesupp{scprint2} used official gene encoding, positional encoding, expression preprocessing and generative inference. Cell analysis used the 256-dimensional base summary representation, at most 1,200 context genes and the released expression-dependent sampling procedure. Expression inference used visible raw counts and explicit gene queries. Human comparisons jointly normalized generated target counts and observed counts. Current multispecies comparisons normalized predictions to observed-input library size only, with depth multiplier one. With native intensity $\mu$ and zero-inflation logit $z$, complete output before normalization was $\mu[1-\operatorname{sigmoid}(z)]$ and the intensity endpoint used $\mu$. The latter was not interpreted as the strict mean conditional on a positive observed count.

Gorilla, platypus and legacy chicken identities were added through the official frozen gene-encoding and positional-encoding extension without updating pretrained weights. Bonobo assay identifiers followed the chimpanzee namespace used by the comparison. Exact gene support fixed multispecies target membership before reading new prediction scores; identifier conventions did not change the biological species of the cells.

\textbf{TranscriptFormer.} TF-Metazoa~\cite{transcriptformer} used raw counts, supported protein-sequence-derived identities and official inference. Its 2,048-dimensional cell representations supplied annotation, integration and reference mapping without task-specific representation fitting. Species and gene support were retained for each atlas.

\textbf{UCE.} The released 33-layer model~\cite{uce} used official raw-expression preprocessing and protein-sequence-derived identities. Inference sampled 1,024 gene tokens with padded length 1,536 and returned 1,280-dimensional cell representations. Additional cell filtering was disabled to preserve the common cohort; gene eligibility followed native vocabulary support. Features supplied classification, integration and reference mapping.

\textbf{CellFM.} Completion used the released 800-million-parameter model~\citesupp{cellfm} and its original scalar cell-wise expression decoder. Weights were frozen, with no new head or query-truth recalibration. Human vocabulary support defined eligible targets; human identities were not substituted for unsupported nonhuman identities. CellFM therefore entered human completion comparisons without fabricated nonhuman scores.

\textbf{Classification and mapping.} Third-party cell representations used training-fold standardization and logistic regression with $C=1$, L-BFGS, at most 2,000 iterations, no class weighting. CellNet used automatically regularized LDA, as in \ref{sm:cellular-evaluation}. Reference mapping used a common cosine kNN classifier for CellNet, scPRINT-2, TranscriptFormer and UCE, with identical partitions, class-frequency correction and validation-selected settings.

\textbf{Reference-supervised expression.} Ridge used at most 512 high-variance visible genes selected from reference cells, reference-fitted standardization and $\alpha=100$, with the target mean as intercept. Cosine kNN used the same feature selection and ten neighbours. Expression predictions were weighted averages of measured reference expression, with weights proportional to nonnegative cosine similarity, normalized to sum to one; zero total similarity gave uniform weights. Predictions were constrained to nonnegative log-expression values. Reference membership matched the adapted decoder for each budget. These methods received target-expression supervision from reference cells and were distinguished from frozen pretrained prediction.

\textbf{Molecular associations.} Expression comparison used absolute Pearson correlation over the same selected cells, candidate genes and measured zeros. Static identity comparison used cosine similarity between learned 64-dimensional gene-identity vectors. Published reference sets and full eligible candidate rankings were shared. These comparisons measured co-expression or context-independent identity similarity, respectively.

\subsection{Evaluation metrics and statistical summaries}\label{sm:metrics}
\renewcommand{\theequation}{S.\arabic{equation}}
\setcounter{equation}{0}

\textbf{Classification.} Accuracy was the fraction of correctly labelled query cells. For class $k$, precision was $P_k=\mathrm{TP}_k/(\mathrm{TP}_k+\mathrm{FP}_k)$ and recall was $R_k=\mathrm{TP}_k/(\mathrm{TP}_k+\mathrm{FN}_k)$. Class F1 was $2P_kR_k/(P_k+R_k)$, with undefined class contributions assigned zero. Macro-F1 averaged class F1 equally. Tabula Sapiens tissue-conditioned evaluation used biological labels present in each true test fold, first averaging donor folds within tissue and then weighting eighteen tissues equally. Global all-class scores used their predefined union of true and predicted labels; seen-class results were reported separately. Balanced accuracy averaged recall over true test classes. Experimental-state classification reported macro-F1 and balanced accuracy averaged equally over held-out donor/library folds; per-state scores, predicted probabilities and confusion matrices were retained.

Rare-cell F1 used labels with training prevalence at most 1\%, at least twenty cohort cells across at least three donors, and positive training and test support in the fold. Eligible labels could differ by fold. Eligible-class F1 was averaged within each fold, then across the three donor-held-out folds equally rather than pooling predictions. The separately reported 5\% minority threshold did not define the primary rare-cell endpoint. Cell counts and donor- or library-held-out scores described distinct sampling and evaluation units.

\textbf{Reference mapping.} Within-species F1 used predefined fine labels; cross-species F1 used shared coarse labels. Coverage was the fraction of all target-test cells with a true label present in the source reference. F1 was evaluated on this eligible subset. Empty subsets retained coverage but undefined F1. Within-species family means weighted species equally; cross-species means weighted eligible ordered directions equally. Shared cross-species kNN settings were selected by validation macro-F1, weighting the four comparison methods and five development task families equally. SD across five overlapping splits described partition sensitivity rather than independent biological replication.

\textbf{Integration.} Metrics followed scIB definitions~\citesupp{scib}. NMI normalized mutual information between graph clusters and cell labels by the arithmetic mean of their entropies. ARI corrected pairwise cluster–label agreement for chance. Cell-label silhouette width was rescaled as $(s+1)/2$. The equal-weight mean of NMI, ARI and rescaled cell-type silhouette defined AvgBIO. Batch silhouette used $1-|s_{\mathrm{batch}}|$ within labels, averaged within each label and then across eligible labels. Graph connectivity averaged each label's fraction of cells in its largest connected neighbourhood component. Their arithmetic mean defined batch performance. Higher values indicated better conservation or mixing. Aggregate scores weighted eligible tissues equally. Metrics used high-dimensional representations and graphs; UMAP supplied visualization~\cite{mcinnes2018umap}.

\textbf{Expression reconstruction.} Complete reconstruction used the expected-expression output $p_{cg}M_{cg}$ across all query cells, including measured zeros. Conditional intensity used decoder magnitude $M_{cg}$ only for measured-positive cells. For gene $g$, complete and positive-intensity PCC were

\begin{equation}
\begin{aligned}
\rho_g^{\mathrm{complete}}&=\operatorname{corr}_{c\in\mathcal C}
\left(y_{cg},\widehat y_{cg}^{\mathrm{complete}}\right),\\
\rho_g^{+}&=\operatorname{corr}_{c\in\mathcal C_g^{+}}
\left(y_{cg},\widehat y_{cg}^{\mathrm{intensity}}\right),\\
\mathcal C_g^{+}&=\{c:n_{cg}>0\}.
\end{aligned}
\label{eq:expression-pcc}
\end{equation}

Conditional-intensity PCC required at least 20 measured-positive cells per target, with the positive-cell mask applied only during scoring and held identical across methods. Detection $d_g=N^{-1}\sum_c\mathbf1(n_{cg}>0)$ defined cumulative gene-eligibility thresholds of 0, 1, 2, 5, 10, 20, 30, 50 and 70\%. Thresholding occurred after inference and did not expose target-positive labels to the model; complete PCC at the 20\% threshold still used all cells. Primary summaries averaged finite gene-wise PCC values equally over the specified target sets. Reference-adaptation duration was selected using mean all-cell ncRNA PCC on a separate validation cohort.

Undefined correlations were not replaced by zero. Human comparisons used common supported, finite target sets defined separately for each model combination. Updated multispecies comparisons used identity-supported, assay-qualified sets fixed before reading new scores, without score-dependent exclusions. Output-component controls used common finite target intersections where specified, including targets with at least 20 measured-positive cells. Coverage, eligibility and defined correlation were reported separately. On $K$ matched cell–gene entries, error metrics were

\begin{equation}
\operatorname{MAE}=\frac1K\sum_{(c,g)}|\widehat y_{cg}-y_{cg}|,
\qquad
\operatorname{RMSE}=\sqrt{\frac1K\sum_{(c,g)}(\widehat y_{cg}-y_{cg})^2}.
\label{eq:prediction-errors}
\end{equation}

Errors used matched cells, targets and expression scales. Reference-mean predictions had applicable errors but undefined gene-wise PCC. Cell-wise correlations, where reported, compared each cell's matched target genes. Heatmaps standardized expression separately per gene for display; scoring preceded display transformation.
\textbf{Molecular retrieval and network scores.} For a candidate gene $g$ and fixed seed set $\mathcal S$, module-retrieval score was the mean centred cosine similarity

\begin{equation}
R_k(g)=\frac{1}{|\mathcal S|}\sum_{s\in\mathcal S}A_k(g,s),
\qquad g\notin\mathcal S.
\label{eq:module-score}
\end{equation}

Average precision integrated precision over recall increments along the complete candidate ranking:

\begin{equation}
\operatorname{AP}
=\sum_t\bigl(\operatorname{Recall}_t-\operatorname{Recall}_{t-1}\bigr)
\operatorname{Precision}_t.
\label{eq:average-precision}
\end{equation}

Seeds and query genes were excluded where appropriate. PBMC module AP was calculated separately in B Naive, T Naive and monocytes and averaged equally. Top-20 recall was the fraction of evaluable reference partners among the retrieved top twenty after excluding the query. External recurrence counted discovery top-twenty members retained in the second cohort's top twenty, with discovery lists fixed before validation. Coding retrieval candidates required model support and detection in at least $\max(5,\lceil0.05N\rceil)$ of $N$ context cells; queries required at least five measured-positive cells. Published reference sets and eligible candidate rankings were shared between methods.

Programme coherence was mean pairwise within-programme association:

\begin{equation}
C_k(\mathcal P)
=\frac{2}{|\mathcal P|(|\mathcal P|-1)}
\sum_{\substack{g,j\in\mathcal P\\g<j}}A_k(g,j).
\label{eq:programme-coherence}
\end{equation}

Expression comparisons used absolute Pearson correlation over the same selected cells, candidates and measured zeros. Background z-scores used 200 size-matched random programmes. Network pathway enrichment used upper-tail hypergeometric tests against the analysed gene panel; pathways with 5--1,000 background genes were eligible. Benjamini--Hochberg correction was applied jointly across 10,720 community--pathway tests, and community annotations required at least three overlapping genes and adjusted $P<0.05$. For community expression scores, variable genes were standardized across 504 protocol-matched PBMC cells, averaged over community genes per cell and then within cell type.

Network sensitivity used RNA Pearson correlations across the 126 monocytes used to estimate representations; links with $r\ge0.30$ were highlighted. Descriptive non-edge controls matched each model link to one random non-edge partner for the same source gene, excluding self-pairs and existing links, and compared finite correlations on paired eligible links. Representation sensitivity used two disjoint 63-cell halves and a within-cell expression-value permutation; Spearman correlation between half-cell cosine scores and mean permutation-associated score change was evaluated on the original edge set. Genes, links and pathway tests were not treated as biological replicates.

\textbf{Prediction attribution and input removal (Fig. 4).} The attribution cohort comprised 24 TotalX SSTX cells, six each from monocyte, T, B and NK populations. CST3 and NEAT1 were jointly absent within a 7,302-gene hidden partition, leaving 29,197 observed genes. Before attribution calculation, six cells per population were sampled across the interval between the 55th and 90th percentiles of summed original predictions for the two targets. This selected cohort was distinct from the observed-context network cohort in \ref{sm:associations}. Integrated gradients~\citesupp{sundararajan2017} explained hidden-target complete log-expression prediction $f_g$. For observed expression $\boldsymbol x$, zero-expression reference $\boldsymbol x_0$ and coordinate $j$,

\begin{equation}
\operatorname{IG}_{j,g}=(x_j-x_{0j})\int_0^1
\frac{\partial f_g(\boldsymbol x_0+t(\boldsymbol x-\boldsymbol x_0))}{\partial x_j}\,dt.
\label{eq:integrated-gradients}
\end{equation}

The integral used 24-step trapezoidal approximation with fixed gene identity, measurement and detection information. Signed gene-pixel contributions were summed within native 16-by-16 patches. Attribution completeness compared summed contributions with the actual-minus-reference prediction; it was a numerical diagnostic rather than reconstruction accuracy.

Removal tests zeroed selected expression without renormalizing remaining inputs or changing identity, measurement or detection information. Decrease was original minus modified prediction. Each cell's 32 highest positive-contribution detected inputs were compared with the other target's selected inputs and eight expression-matched random sets. Random trials were averaged within cell. Named-set controls matched expressed-input count and expression mass. Extra decrease was the selected set's decrease minus the same cell's mean random decrease. The observation unit was the cell. These computational changes assessed dependence of the specified prediction on its input.

\textbf{Cellular interpretation (Fig. 4 supplements).} CD4 attribution used the donor-held-out LDA score difference between Th17 and Th2. Gradient multiplied by input gave signed local contributions to this log-posterior-odds contrast. Ninety-six prespecified cells, twelve per donor/state, supplied summaries. Removal of the 10\% strongest supporting patches was compared with six equal-patch-count random trials per cell. Only trials matching removed expression mass and detected-gene count within 10\% were averaged. The strictly matched subset contained 38 Th17 cells. Retrospectively selected illustrations were distinguished from cohort summaries.

Six-component sparse NMF decomposed frozen final-eight-layer cellular features. Standardization used donors D1–D3; values were clipped to $[-4,4]$ and split into positive/negative channels. Fitting used 300 cells per donor/state, giving 2,700 cells, nonnegative double singular-value initialization, factor regularization 0.001, L1 ratio 0.8, tolerance $3\times10^{-4}$, at most 1,000 iterations. The fitted transformation was applied to donor D4 without refitting: 712 Th0, 655 Th17 and 1,012 Th2 cells. Gene associations correlated log1p counts per 10,000 with activations after removing donor-specific means. Eligible genes were detected in at least 488 cells. Ranking used D1–D3, with selected associations evaluated in D4. Functional names were annotations based on associated genes. Binary ROC AUC compared training-defined highest- and lowest-activation states in D4, measuring the probability that a positive-state cell had a larger activation than a negative-state cell, with half weight for ties; it was not three-class classification accuracy.

\textbf{Uncertainty.} Expression means and paired differences used 2,000 shared target-gene bootstrap draws and percentile 95\% intervals. Complementary gene partitions reused query cells and were not independent cell cohorts. Spatial intervals resampled 130 nonempty physical blocks of $500\times500\,\mu\mathrm m$ with 2,000 shared draws, recomputing PCC; they describe within-section variation. Selected PBMC context comparisons used 1,000 within-protocol cell draws, shared across conditions. Attribution trials were averaged within cells, and paired cell-bootstrap summaries described the selected cohort. Output-component comparisons used shared target-gene bootstrap indices where specified. Spatial marker permutation controls preserved the prediction distribution while permuting cell assignments. Tissue, donor, library, species, gene and cell aggregation units were retained according to the endpoint. Descriptive SD and conditional bootstrap intervals were distinguished from independent-donor confidence intervals.

\subsection{Expression processing and sampling}\label{sm:expression}

Species-qualified identifiers were retained throughout gene matching and image generation. The principal pretraining collection contained 231,190,557 cell profiles, with 1,015,163 genes represented by sequence-derived identities.

For count inputs, expression was normalized to 10,000 counts per cell and transformed using log1p. With $n_{cg}$ denoting the input count and $L_c$ its measured library size, the image value was

\begin{equation}
x_{cg}
=\frac{1}{6}\operatorname{clip}\!\left(
\log\!\left(1+10^4\frac{n_{cg}}{L_c}\right),0,6
\right).
\label{eq:input-scaling}
\end{equation}

The preprocessing policy preferred an available raw expression source and otherwise used the supplied expression matrix. Inputs identified as already log-transformed retained that scale before clipping and image rescaling. Dataset-level measurement masks recorded the gene coverage independently of expression values. No additional filtering by cell or gene counts was applied during construction of the principal pretraining collection.

The principal pretraining collection was used without a separate validation or test partition. Sampling probabilities were proportional to $n_r^{0.4}$, where $n_r$ is the number of profiles in sampling group $r$, moderating differences in group abundance.

\subsection{Sequence features, image mapping and augmentation}\label{sm:sequence}

Gene-associated DNA sequences were matched to species-aware gene identifiers before feature extraction. Sequence features were extracted from the 1-billion-parameter Evo 2 base model, using the output of the third linear projection in the multilayer perceptron of block 20. Features at valid nucleotide positions were mean-pooled; long sequences were encoded as overlapping chunks, whose pooled features were then averaged to obtain one 1,920-dimensional vector per gene. A frozen incremental principal-component transform reduced these vectors to 256 dimensions. Independent learned projections generated the 64-dimensional mean-identity and expression-modulated identity features in equation~\eqref{eq:main-patch-token}.

For image mapping, DNA-derived gene vectors were first reduced to 64 dimensions by PCA and then projected to two dimensions by t-SNE, using cosine distance, perplexity 30, PCA initialization and an automatic learning rate~\cite{maaten2008visualizing}. The coordinates were scaled to the image grid. Balanced greedy k-d tree assignment placed genes at nearby available pixels while favouring less occupied patches, producing a fixed, species-specific mapping with at most one gene per pixel. For each cell, normalized expression and the assay measurement indicator were written at each gene's assigned pixel. Thus, measured zero-expression genes remained distinguishable from genes not measured by the assay.

The expression branch used two channels, expression and visibility. The expression-modulated branch used 64 identity channels, with $\delta=0.02$. Each branch projected to 768 dimensions. During training, adjacent horizontal positions could be swapped within a patch. Patches were selected with probability $0.5$ and neighbouring pairs with probability $0.3$. Gene identities and their expression values moved together. The permutations were disabled at inference.

Each cell generated two global views with patch-mask fractions sampled from $[0,0.1]$, and six local views with fractions from $[0.35,0.8]$. Additional global/local augmentation parameters are summarized in Table~\ref{tab:methods-augmentation}.

\small
\setlength{\LTleft}{0.05\textwidth}
\setlength{\LTright}{0pt}
\begin{longtable}{@{}p{0.45\textwidth}p{0.20\textwidth}p{0.18\textwidth}@{}}
\caption{Global and local view augmentation parameters.}\label{tab:methods-augmentation}\\
\toprule
Augmentation & Global views & Local views \\
\midrule
\endfirsthead
\toprule
Augmentation & Global views & Local views \\
\midrule
\endhead
Expression noise scale & 0.005 & 0.01 \\
Expression dropout base & 0.005 & 0.02 \\
Expression dropout amplitude & 0.03 & 0.12 \\
Gamma range & 0.97–1.03 & 0.93–1.07 \\
Patch bias standard deviation & 0.03 & 0.08 \\
\bottomrule
\end{longtable}

\normalsize

The additional gene-mask ratio was $0.01$. Patch masking propagated to the visibility channel. Shape augmentation, cropping, global layout augmentation and image mixing were disabled.

\subsection{Projector and Gaussian regularization}\label{sm:projector}

The cellular projector used hidden width 1,024 and output dimension 256. SIGReg pooled the view and cell axes before evaluating the projected distribution. For random unit direction $\mathbf d_r$, its empirical characteristic function was

\begin{equation}
\widehat{\varphi}_r(t)
=\frac{1}{BV}\sum_{c=1}^{B}\sum_{v=1}^{V}
\exp\!\left(it\,\mathbf d_r^{\mathsf T}\mathbf z_c^{(v)}\right).
\label{eq:sig-characteristic}
\end{equation}

The regularizer compared this function with the standard Gaussian characteristic function:

\begin{equation}
\mathcal L_{\mathrm{SIGReg}}
=\frac{BV}{R}\sum_{r=1}^{R}
\int_{0}^{3}
2e^{-t^2/2}
\left|\widehat{\varphi}_r(t)-e^{-t^2/2}\right|^2\,dt.
\label{eq:sigreg-objective}
\end{equation}

We used $R=1,024$ directions and trapezoidal integration at 17 equally spaced points. Characteristic-function moments were aggregated across all training devices, pooling views and cells for the regularization objective.

\subsection{Decoder, likelihood and expression weighting}\label{sm:decoder}

The decoder used hidden dimension 512, eight attention heads and three cross-attention blocks. Patch context was read from the final four encoder layers and pooled into 32 spatial bins. The cellular summary was appended as an additional context token. Empty context bins were excluded from attention.

With fixed context keys and values, decoder block $\ell$ updated its contextual state as

\begin{equation}
\begin{aligned}
\mathbf u_{cg}^{(0)}&=\mathbf 0,\\
\mathbf b_{cg}^{(\ell)}
&=\mathbf u_{cg}^{(\ell-1)}
+\operatorname{MHA}_{\ell}
\left(\mathbf q_g+\mathbf u_{cg}^{(\ell-1)},\mathbf K_c,\mathbf V_c\right),\\
\mathbf u_{cg}^{(\ell)}
&=\mathbf b_{cg}^{(\ell)}
+\operatorname{FFN}_{\ell}
\left(\operatorname{LN}_{\ell}(\mathbf b_{cg}^{(\ell)})\right).
\end{aligned}
\label{eq:identity-branches}
\end{equation}

The final state was normalized before the interaction projection in equation~\eqref{eq:main-contextual-gene}. Separate expression-output branches produced a raw magnitude $a_{cg}$ and a rate parameter $b_{cg}$. For the compound Poisson–exponential model, parameters were obtained using

\begin{equation}
\begin{aligned}
M_{cg}&=\operatorname{softplus}_{50}(a_{cg})+\epsilon,\\
\lambda_{cg}&=\operatorname{softplus}(b_{cg})+\epsilon,\\
p_{cg}&=1-e^{-\lambda_{cg}},\\
\theta_{cg}&=\frac{p_{cg}M_{cg}}{\lambda_{cg}},
\qquad
\widehat{x}_{cg}=p_{cg}M_{cg},
\end{aligned}
\label{eq:decoder-outputs}
\end{equation}

where $\epsilon=10^{-6}$ and $\operatorname{softplus}_{50}$ denotes softplus with sharpness parameter 50. $M_{cg}$ is the conditional positive-expression mean, and $p_{cg}$ is the probability of positive expression.

The compound distribution has mass $e^{-\lambda}$ at zero. For positive $y$, its density is

\begin{equation}
f(y\mid\lambda,\theta)
=e^{-\lambda-y/\theta}
\sqrt{\frac{\lambda}{\theta y}}
I_1\!\left(2\sqrt{\frac{\lambda y}{\theta}}\right),
\qquad y>0,
\label{eq:compound-density}
\end{equation}

where $I_1$ denotes the modified Bessel function of the first kind. The likelihood was evaluated using numerically stabilized Bessel functions.

Targets were drawn from measured genes hidden in each encoder view. Up to 256 targets with scaled expression greater than $0.03$ and 256 measured-zero targets were sampled per cell. The likelihood used their retained expression values. For sampling stratum $s$, let $N_{cs}$ be the candidate count and $K_{cs}$ the sampled count. The corrected likelihood was

\begin{equation}
\mathcal L_{\mathrm{CPE}}
=-\frac{1}{B_{\mathrm{valid}}}
\sum_{c\in\mathcal B_{\mathrm{valid}}}
\frac{1}{\sum_s\overline N_{cs}}
\sum_s\sum_{g\in\mathcal T_{cs}}
\frac{\overline N_{cs}}{\overline K_{cs}}
\log f(x_{cg}\mid\lambda_{cg},\theta_{cg}),
\label{eq:expression-likelihood}
\end{equation}

with $\overline N_{cs}=\max(N_{cs},1)$ and $\overline K_{cs}=\max(K_{cs},1)$, matching the numerical treatment of empty strata. Losses were averaged equally over valid cells and views.

Positive-expression protection used the raw magnitude output $a_{cg}$ and frozen source-conditioned gene statistics. For positive target set $\mathcal T_c^+$, a reliable standardized deviation was clipped and squared:

\begin{equation}
\begin{aligned}
d_{cg}
&=\left[
\operatorname{clip}\!\left(
\frac{|x_{cg}-\mu_{sg}|}{|\sigma_{sg}|+\epsilon},0,1
\right)\right]^2,\\
\omega_{cg}
&\propto\frac{\rho}{|\mathcal T_c^+|}
+(1-\rho)\frac{d_{cg}}{\sum_{j\in\mathcal T_c^+}d_{cj}+\epsilon},\\
\mathcal L_{\mathrm{dev}}
&=\frac{1}{B_+}\sum_{c\in\mathcal B_+}
\sum_{g\in\mathcal T_c^+}\omega_{cg}(a_{cg}-x_{cg})^2.
\end{aligned}
\label{eq:deviation-loss}
\end{equation}

Here, $s$ indexes the source dataset, $\rho=0.25$, and target weights were normalized to sum to one within each cell. Unreliable statistics received no deviation contribution; cells without reliable deviation signal used uniform weights. Source-specific gene statistics were used when available; otherwise, gene-level estimates were used. These statistics entered loss weighting only.

\subsection{Species-balanced cell-type prototypes}\label{sm:prototypes}

Fine cell labels were mapped to 25 shared coarse classes. For class $k$ and species $s$, an exponential moving average maintained a prototype $\boldsymbol\mu_{ks}$ from normalized cellular features. Class anchors averaged the contributing species prototypes and were subsequently unit-normalized.

A class entered the contrastive objective after at least two species had each contributed 16 cells. Per-cell features averaged the encoder cellular summaries over views before normalization. For eligible cell $c$ with class $k_c$, the objective was

\begin{equation}
\mathcal L_{\mathrm{proto}}
=\frac{1}{|\mathcal E|}\sum_{c\in\mathcal E}
\eta_c
\left[-\log
\frac{\exp(\mathbf z_c^{\mathsf T}\boldsymbol\pi_{k_c}/\tau)}
{\sum_{k\in\mathcal K_{\mathrm{eligible}}}
\exp(\mathbf z_c^{\mathsf T}\boldsymbol\pi_k/\tau)}
\right].
\label{eq:prototype-loss}
\end{equation}

$\boldsymbol\pi_k$ denotes the normalized class anchor, $\tau=0.05$, and $\eta_c$ is inversely proportional to the cumulative count of the corresponding class–species group. These weights were rescaled to mean one over the eligible cells. Prototype updates used momentum $0.05$. The separate species-centred regularizer and adversarial species objective were disabled.

\subsection{Architecture and optimization}\label{sm:architecture}

The selected model used a base transformer with 768-dimensional hidden states. Its gene identity table and principal-component basis were frozen. The learned input projections, encoder, projector and expression decoder were optimized jointly. Architecture and optimization settings are summarized in Table~\ref{tab:methods-training}.

\small
\setlength{\LTleft}{0.05\textwidth}
\setlength{\LTright}{0pt}
\begin{longtable}{@{}p{0.60\textwidth}p{0.25\textwidth}@{}}
\caption{Model architecture and optimization settings.}\label{tab:methods-training}\\
\toprule
Component or setting & Value \\
\midrule
\endfirsthead
\toprule
Component or setting & Value \\
\midrule
\endhead
Transcriptome image & $384\times384$ \\
Patch size & $16\times16$ \\
Encoder hidden dimension & 768 \\
Encoder depth / attention heads & 12 / 12 \\
Mean and modulated identity dimensions & 64 each \\
Projector hidden / output dimensions & 1,024 / 256 \\
Decoder hidden dimension / blocks / heads & 512 / 3 / 8 \\
Pooled patch context tokens & 32 \\
Encoder layers supplying decoder patch context & Final 4 \\
Stochastic depth & 0.1 \\
Devices & 8 GPUs \\
Cells per device & 74 \\
Effective batch size & 592 \\
Gradient accumulation & 1 \\
Precision & bfloat16 mixed precision \\
Optimizer & AdamW~\cite{loschchilov2019decoupled} \\
Peak learning rate & $2\times10^{-5}$ \\
Minimum learning rate & $5\times10^{-7}$ \\
Weight decay & 0.05 \\
Gradient norm clipping & 1 \\
\bottomrule
\end{longtable}

\normalsize

Training ran for 14 days with a cosine learning-rate schedule, including learning-rate warm-up and delayed decoder activation with a separate decoder loss-weight warm-up. The final saved checkpoint was used for all analyses.

\subsection{Cellular representation evaluation}\label{sm:cellular-evaluation}

\textbf{Representations and classification heads.} Annotation, rare-population recognition and experimental-state classification used the 768-dimensional feature

\begin{equation}
\begin{aligned}
\mathbf r_c&=\operatorname{LN}\!\left(\sum_{\ell=L-7}^{L}\mathbf c_c^{(\ell)}\right),\\
\mathbf u_c&=\operatorname{Standardize}_{\mathrm{train}}(\mathbf r_c),\\
\widehat y_c&=\arg\max_k\Biggl[
\mathbf u_c^{\mathsf T}\widehat\Sigma^{-1}\widehat\mu_k
-\tfrac12\widehat\mu_k^{\mathsf T}\widehat\Sigma^{-1}\widehat\mu_k\\
&\qquad+\log\widehat\pi_k
\Biggr].
\end{aligned}
\label{eq:lda-readout}
\end{equation}

Here, $L$ is the final encoder block, $\mathbf c_c^{(\ell)}$ is its raw CLS state, and $\widehat\mu_k$ and $\widehat\pi_k$ are training-estimated class means and empirical priors. $\widehat\Sigma$ is the shared within-class covariance estimated with automatic shrinkage~\cite{ledoitwolf2004}. LDA used a least-squares solution with automatically estimated covariance shrinkage. The Gaussian discriminant readout was motivated by CellNet's distributional pretraining objective; its covariance was estimated from the downstream features rather than fixed to an identity matrix.

Geneformer V2, scGPT, scLong, scPRINT-2, TranscriptFormer and UCE supplied their respective frozen cell representations. Their classification heads used logistic regression with inverse regularization strength $C=1$, L-BFGS optimization, a maximum of 2,000 iterations, no class weighting. Feature means and standard deviations were estimated independently within every training fold. External annotation and state recognition retained the classifier settings established for Tabula Sapiens. scLong representations were obtained by mean-pooling its merged gene decodings over its 27,000-gene axis, excluding the terminal padding position.

\textbf{Tabula Sapiens annotation.} The complete Tabula Sapiens input contained 549,018 labelled cells after excluding nine cells without valid labels. Tissue-conditioned leave-one-donor-out evaluation scored 276,385 cells across 18 tissues and 49 eligible tissue–donor folds. Each fold fitted its classifier on the remaining donors in that tissue; score definitions and tissue-level aggregation are given in \ref{sm:metrics}.

\textbf{External cohorts and rare populations.} The kidney cohort comprised 107,344 cells, 45 donors and 57 author-defined labels. The lung cohort comprised 69,650 cells, three donors and 59 labels. Each cohort used three fixed, label-blind GroupKFold partitions over donors, with every cell tested once. Classification used the same CellNet CLS aggregation and LDA head described above; comparison models retained logistic regression.

The union of eligible labels contained 34 kidney and 17 lung populations. Rare-label criteria and fold-level score aggregation are specified in \ref{sm:metrics}.

\textbf{Experimental cellular states.} CD4 polarization used 9,777 cells from four donors, restricted to the author-defined Naive starting population, 10X technical batch 1 and Th0, Th2 or Th17 conditions. Four leave-one-donor-out folds evaluated the original experimental condition labels. WI-38 evaluation used 14,457 cells from six culture libraries spanning RS, IR and ETO induction. Source quality control required at least 300 detected genes, 1,000–120,000 counts and a mitochondrial fraction below 12\%. Two folds held out one complete library per condition, swapping the Rep1 and Rep2 groups. The WI-38 replication unit was a culture/library from the same cell line.

Both tasks used the frozen final-eight-layer representation, training-fitted standardization and LDA for CellNet, with logistic regression for the six comparators. State-classification metrics and fold summaries are specified in \ref{sm:metrics}. An exploratory kidney analysis evaluated frozen features from 3,262 author-defined PT-S1/S2 cells from 26 donors, comparing AKI and CKD across three state-stratified donor folds. It used the same heads and was reported separately from the primary experimental-state cohorts.

\textbf{Integration.} The full Tabula Sapiens visualization index comprised 692,102 cells. Metrics used 682,741 labelled cells from 26 eligible multi-donor tissues. scImagene supplied the raw CLS state from the sixth block counted from the output, without additional feature postprocessing. The graph used 90 Euclidean neighbours, Leiden resolution $0.05$ and random seed 0. All comparison models used their respective frozen features with the same graph settings.

Biological conservation averaged NMI, ARI and scaled cell-label silhouette width, with tissues weighted equally. Batch performance averaged the label-conditioned batch silhouette score and graph connectivity. The batch silhouette score entered this average directly. UMAP visualization used 50 principal components, 15 Euclidean neighbours, minimum distance $0.05$, 200 epochs and random seed 42. Kidney and lung integration used the same feature and graph settings; external metric computation used at most 100,000 cells per cohort.

\textbf{Reference mapping.} The reported mapping comparison included CellNet, UCE, TranscriptFormer and scPRINT-2. CellNet used the raw CLS state from the eighth block counted from the output. All four methods shared cosine distance and inverse-distance weighting; within-species identification and cross-species transfer used the distinct protocols below.

Within-species evaluation assigned approximately 80\% of retained cells to reference and 20\% to query. Shared settings were 10 neighbours and class-frequency correction exponent $\gamma=0.25$, selected by reference-internal cross-validation on CellNet and then fixed for all four methods. Evaluation was repeated five times with different random seeds, using matched reference and query splits across all four methods. Queries were sampled by stratification by author-provided labels from the previously defined test partition; prior parameter-selection validation cells were excluded from these query pools. Reference and query cell identifiers were disjoint. Standardization and class frequencies were estimated from the new reference partition. The fine-label summaries included three immune, eleven germline and four plant species whose author-provided labels refined the corresponding coarse labels. Author-provided labels at coarse resolution were reported separately in the species-level analysis. The repeated splits overlapped in cells, and their SD described split sensitivity. Root-atlas and GSE268881 analyses of the same biological cells used identical reference and query membership.

Cross-species mapping used three mutually exclusive partitions within each species, stratified by harmonized coarse cell labels: approximately 50\% reference, 25\% validation and 25\% test cells. The reference partition contained at least one cell of each retained class; class-wise integer rounding produced small deviations from the nominal fractions. For each ordered source–target pair, only source-reference cells supplied kNN neighbours and labels. A separate standardization transform was fitted for each species using that species' reference features. Thus, target-reference features informed target-domain normalization without using their labels, while target validation and test cells did not contribute to the fitted normalization parameters.

Shared kNN settings were selected from 10, 30 or 100 neighbours, uniform or inverse-distance weights, and class-frequency correction exponents of 0 or 0.5. Selection followed the validation macro-F1 criterion summarized in \ref{sm:metrics}. With these settings fixed, CellNet's final-eight-layer representation and single raw CLS states from the sixth and eighth blocks counted from the output were compared using the validation cells. The selected configuration used the eighth-from-last raw CLS state, 30 neighbours, inverse-distance weighting and $\gamma=0.5$. For class $k$, the neighbour-vote score was multiplied by $\pi_k^{-\gamma}$, where $\pi_k$ was its frequency in the source reference partition. Validation cells were not added to the reference set for final testing. Additional plant tasks retained this selected configuration without further parameter selection; their validation partitions were reserved and did not enter the reported test scores.

Final cross-species eligibility, coverage and score aggregation followed \ref{sm:metrics}. These analyses evaluate transfer with access to unlabelled target-reference features. The cells had been examined in earlier analyses, so the results represent internal cell-level holdout evaluation rather than an independent external test. Ordered directions share species and were not treated as independent biological replicates.

Table~\ref{tab:methods-s1} gives the actual retained reference, validation and test counts for each cross-species mapping cohort. Membership was mutually exclusive and exhaustive within each row; the nominal 50:25:25 proportions reflect class-wise stratification and integer rounding. Root-atlas entries under the atlas name and GSE268881 share the same cells and partitions and are represented once.

\small
\setlength{\LTleft}{0.05\textwidth}
\setlength{\LTright}{0pt}
\begin{longtable}{@{}p{0.17\textwidth}p{0.17\textwidth}p{0.105\textwidth}p{0.105\textwidth}p{0.105\textwidth}p{0.105\textwidth}@{}}
\caption{Reference, validation and test partitions for cross-species mapping.}\label{tab:methods-s1}\\
\toprule
Atlas & Species & Retained profiles & Reference & Validation & Test \\
\midrule
\endfirsthead
\toprule
Atlas & Species & Retained profiles & Reference & Validation & Test \\
\midrule
\endhead
Brassicaceae roots (GSE268881) & Arabidopsis thaliana & 47,033 & 23,514 & 11,757 & 11,762 \\
Brassicaceae roots (GSE268881) & Camelina sativa & 27,485 & 13,740 & 6,871 & 6,874 \\
Brassicaceae roots (GSE268881) & Eutrema salsugineum & 33,975 & 16,984 & 8,492 & 8,499 \\
Brassicaceae roots (GSE268881) & Schrenkiella parvula & 36,021 & 18,008 & 9,005 & 9,008 \\
Brassicaceae roots (GSE268881) & Sisymbrium irio & 40,437 & 20,216 & 10,109 & 10,112 \\
Circulating immune cells & Whitespotted bamboo shark & 1,580 & 789 & 394 & 397 \\
Circulating immune cells & Ciona intestinalis & 2,016 & 1,008 & 503 & 505 \\
Circulating immune cells & Half-smooth tongue sole & 6,567 & 3,282 & 1,640 & 1,645 \\
Circulating immune cells & Human & 3,005 & 1,501 & 749 & 755 \\
Circulating immune cells & Pacific white shrimp & 4,625 & 2,312 & 1,156 & 1,157 \\
Circulating immune cells & Rhesus macaque & 984 & 491 & 246 & 247 \\
Circulating immune cells & Chinese softshell turtle & 950 & 472 & 238 & 240 \\
Testis & Common marmoset & 6,384 & 3,191 & 1,596 & 1,597 \\
Testis & Chicken & 8,413 & 4,205 & 2,103 & 2,105 \\
Testis & Gorilla & 7,323 & 3,661 & 1,830 & 1,832 \\
Testis & Human & 16,011 & 8,004 & 4,002 & 4,005 \\
Testis & Rhesus macaque & 7,618 & 3,808 & 1,903 & 1,907 \\
Testis & Opossum & 9,926 & 4,962 & 2,481 & 2,483 \\
Testis & Mouse & 5,264 & 2,630 & 1,315 & 1,319 \\
Testis & Gibbon & 6,572 & 3,285 & 1,642 & 1,645 \\
Testis & Platypus & 6,751 & 3,375 & 1,687 & 1,689 \\
Testis & Bonobo & 9,873 & 4,936 & 2,467 & 2,470 \\
Testis & Chimpanzee & 12,329 & 6,164 & 3,081 & 3,084 \\
Brassicaceae multi-tissue & Arabidopsis lyrata & 13,495 & 6,744 & 3,372 & 3,379 \\
Brassicaceae multi-tissue & Arabidopsis thaliana & 42,108 & 21,049 & 10,524 & 10,535 \\
Brassicaceae multi-tissue & Brassica oleracea & 7,682 & 3,837 & 1,919 & 1,926 \\
Brassicaceae multi-tissue & Capsella rubella & 22,730 & 11,360 & 5,679 & 5,691 \\
\bottomrule
\end{longtable}

\normalsize

\textbf{Expression-input diagnostics and visualization.} Kidney input diagnostics retained the model weights, complete gene identities, measurement masks, pixel layout, cell identifiers and donor folds. Four settings used all expression, zeroed non-protein-coding expression, zeroed LINC01187 and LINC00996, or zeroed the matched coding controls PHLDA2 and BDNF. Coding controls were selected from the nearest 50 genes by mean expression and detection rate. These experiments refitted the same training-fold LDA head on the resulting frozen features. State and rare-cell UMAP overlays used the evaluated CLS features; displayed RNA values came from the original measurements.

\subsection{Withheld-expression evaluation}\label{sm:expression-evaluation}

\textbf{Visible inputs and target truth.} Frozen gene-completion analyses used the pretrained expression decoder; reference-based adaptation was evaluated separately as described below. For each evaluation, visible genes $\mathcal O_c$ and hidden targets $\mathcal T_c$ were fixed before inference. Targets were absent from encoder expression, gene identifiers and measurement masks, while their DNA-derived identities were supplied to decoder queries. Input normalization used only counts that actually entered the encoder:

\begin{equation}
\begin{aligned}
L_c^{\mathrm{obs}}&=\sum_{g\in\mathcal O_c}n_{cg},\\
x_{cg}^{\mathrm{input}}&=\frac{1}{6}\operatorname{clip}\!\left[
\log\!\left(1+10^4\frac{n_{cg}}{L_c^{\mathrm{obs}}}\right),0,6\right],\\
y_{cg}&=\log\!\left(1+10^4\frac{n_{cg}}{L_c^{\mathrm{assay}}}\right).
\end{aligned}
\label{eq:completion-input}
\end{equation}

$\mathcal O_c$ contains supported, measured, visible encoder genes, including measured zeros. $L_c^{\mathrm{assay}}$ is the full measured library used only to construct the separate scoring truth $y_{cg}$. Predictions were returned to the log1p CP10k scale using the fixed training range. Frozen CellNet predictions used no task-specific fitting or truth-based calibration.

\textbf{Human total-RNA completion.} TotalX PBMC and developing human brain each supplied 5,000 fixed query cells. Coding-panel and full-assay evaluations used five complementary gene partitions, with approximately four fifths of genes visible and one fifth withheld in each pass. Each gene was scored from the pass in which it was hidden; the query cells were shared across passes. A separate coding-to-non-protein-coding setting retained all measured protein-coding context and simultaneously withheld all non-protein-coding targets.

Non-coding RNA summaries included annotated lncRNAs and the explicitly defined other-RNA biotypes. Other non-protein-coding features, including pseudogene categories outside this RNA scope, were tracked separately. Native molecular coverage was assessed using assay annotations and supported identities before score-based filtering.

\textbf{Output-component controls.} Within the full-assay evaluation, gene-wise scores for conditional magnitude, complete prediction and a one-UMI control were aligned by dataset, target and gene partition. The control was $u_c=\min\{\log(1+10^4/L_c^{\mathrm{obs}}),6\}$. Metric eligibility and uncertainty were handled as described in \ref{sm:metrics}. These full-assay diagnostics were reported separately from the common-coding-panel model comparison; no additional model inference or target-based calibration was performed.

\textbf{Comparison methods and output scale.} scGPT supplied frozen expression predictions using 51-bin visible-expression input and at most 600 nonzero context genes. Its inverse-bin predictions were already on the log1p CP10k scale. scPRINT-2 supplied model-generated counts from visible raw-count context. Its prediction library combined observed counts and generated target counts before conversion to log1p CP10k. These human comparisons used the same measured truth and target identifiers as CellNet within supported vocabularies. scLong retained its released expression-regression head, and CellFM retained the frozen 800-million-parameter model's scalar expression decoder without target-based recalibration. Multispecies scPRINT-2 used observed-input-depth normalization, as specified in \ref{sm:comparators}.

Ridge and kNN used target-expression supervision from a reference partition disjoint from the query cells. Up to 10,000 reference cells were sampled without replacement. Feature selection retained at most 512 high-variance visible genes. Ridge used training-fitted standardization and $\alpha=100$; kNN used cosine similarity and $k=10$, with nonnegative similarities normalized to sum to one and uniform weights when their sum was zero. These methods transferred reference target information to the same held-out queries. A constant reference-mean predictor was retained as a baseline.

\textbf{Reference-based decoder adaptation.} Adaptation started independently from the pretrained decoder for each cohort and expression-hiding protocol. The encoder, DNA-derived gene identities, image layout and context pooling remained fixed. Decoder parameters were updated using the expression objective described in \ref{sm:decoder}, with target expression supplied only by non-test reference cells. Hidden targets remained absent from the corresponding encoder inputs. Ridge and kNN used the same reference-cell membership as the adapted decoder.

For the primary human coding-to-non-protein-coding comparison, 500 cells were used for decoder training. A separate set of 2,000 non-test validation cells per cohort was excluded from training and used to select the number of epochs under the expression-reconstruction criterion in \ref{sm:metrics}; the selected duration was 30 epochs. Thus, 500 denotes the training-reference budget, with an additional labelled validation set used for model selection. Decoder optimization used a learning rate of $2\times10^{-5}$, batches of 256 cells. The complete predefined test partition was excluded from development, and final evaluation used the original 5,000 query cells per cohort. Multispecies adaptation used 500 non-test reference cells with the fixed 30-epoch duration. Under gene-axis five-fold evaluation, each reference cell was assigned one expression-independent hidden fold, and all folds trained a shared cohort decoder. A gene contributed supervision only when hidden from that cell's encoder input. Receiving target-expression labels during adaptation was distinguished from frozen prediction of genes lacking expression-pretraining exposure.

Reference-size sensitivity analysis used 500, 2,000, 10,000 or all available non-test reference cells. Smaller human reference budgets retained the separate 2,000-cell validation cohort for epoch selection. After selection was fixed, full-reference fits used all 39,970 PBMC or 225,765 brain non-test cells. These are completion-reference counts; annotation training retained its separate partitions.

\textbf{Multispecies completion and exposure status.} Five gene partitions were evaluated separately in chimpanzee (1,849 cells), bonobo (1,481), gorilla (1,098), macaque (1,143), mouse (790), platypus (1,012) and chicken (1,262). Multispecies scoring followed the all-cell and measured-positive criteria in \ref{sm:metrics}. Each target was matched to its actual decoder gene identity; a species-independent identifier suffix was accepted only when it uniquely identified a supported gene.

Expression-pretraining exposure was assigned by matching the actual decoder gene identity to genes with measured expression exposure during pretraining. Genes represented by sequence identity but lacking measured expression exposure were distinguished from identities absent from the pretraining gene collection. Bonobo targets decoded through previously active chimpanzee identities were assigned the exposure status of those actual identities. The current scPRINT-2 comparisons used a common target set fixed by exact gene-identity support before examining prediction scores. Support included pretrained identities and species identities added through the official frozen gene encoder. The common expression-unseen subsets comprised 2,108 bonobo, 12 mouse and 533 chicken genes. This exposure definition refers specifically to CellNet pretraining.

\textbf{Direct spatial validation.} One Xenium Prime 5K mouse-brain coronal section supplied 63,173 cells. A fixed uniform sample of 12,000 cells was drawn without replacement. Twenty prespecified lineage markers were simultaneously withheld from encoder expression, identities and measurement masks. The remaining 4,986 supported measured genes supplied observed context under equation~\eqref{eq:completion-input}.

Complete predictions were evaluated against direct measurements of the same markers in the same 12,000 cells. Original micrometre coordinates were used for paired expression maps and uncertainty estimation. Model input consisted of expression and gene identity. Slc17a7, Gad1, Aqp4 and Mog were the prespecified main display markers. A smaller-panel sensitivity analysis used the same cells and targets with 1,000 randomly sampled visible non-target genes.

Reference-based spatial prediction used the remaining 51,173 non-query cells from the same section. Decoder adaptation, ridge and kNN therefore received same-section reference measurements, with all 12,000 query identities excluded. Coordinates supplied visualization and uncertainty estimation rather than model input.

\subsection{Gene associations, molecular networks and non-coding module recovery}\label{sm:associations}

Associations used the frozen 512-dimensional final interaction state in equation~\eqref{eq:main-contextual-gene}. For each cell population, gene features were averaged across cells, centred across the complete decoder panel and L2-normalized before cosine scoring, as in equation~\eqref{eq:main-association}. Static learned identity and expression-based comparators were evaluated as specified in \ref{sm:metrics}.

The principal association cohorts included TotalX PBMC (1,671 selected cells across 17 types), TotalX brain (3,700 across 37 types), PurifiedPBMC (1,000 across ten types) and perirhinal cortex (720 across ten types). At most 100 cells were sampled per annotated type. Measured genes with supported molecular identities supplied the encoder context. Already log-transformed sources retained their supplied values before clipping and image rescaling; raw-count sources used library normalization and log1p transformation.

\textbf{Hidden-panel module recovery.} The fixed PBMC decoder panel contained 1,020 genes, including 373 non-coding candidates. Expression values and measurement masks were set to zero at every panel target location; the fixed pixel identity layout was retained. Thirty members of the published TotalX PBMC module were randomly selected as seeds. Removing seeds left 343 candidates, comprising 117 held-out module members and 226 background candidates.

\textbf{Observed-context neighbourhoods.} Pair and programme association analyses retained measured query expression. Decoder panels combined reference-programme genes, fixed queries, the established non-coding pool, up to 2,048 frequently detected genes and up to 512 non-coding genes. Eligibility required detection in at least five cells of the analysed type. Each query was excluded from its own ranking. All eligible cells contributed to the population mean and expression comparator, including cells with zero query expression.

Cross-dataset comparisons mapped discovery genes into the second cohort and repeated the same feature aggregation and cosine scoring. For HLA-DRA and CD3D, the discovery top-20 neighbours were fixed before external recurrence evaluation. Supplementary complex retrieval queried each of the eight CCT members separately in HEK293T and untreated Huh7. Top-20 recall followed the common definition in \ref{sm:metrics}. Additional complex retrieval analyses included the MCM2–7 and RNA-exosome reference screens.

\textbf{Gene-network construction.} The molecular network used 126 TotalX CD14-positive monocytes, matched by assay protocol to the primary PBMC comparisons. The encoder received all 36,499 measured genes supported by the model, including expression of the queried genes. Source log1p expression was clipped to $[0,6]$ and divided by 6. The 3,279-gene decoder panel was the deduplicated union of the 2,048 protein-coding genes and 512 genes of other annotated biotypes with the highest detection rates across 1,026 selected PBMC cells, all 1,625 supported protein-coding genes detected in at least seven of the 126 platelets, and supported platelet-module reference genes and prespecified queries or markers, including SMANTIS, LINC01750, SMILR, GP9 and ITGA2B. Genes required a supported molecular identity and an unambiguous gene-symbol and pixel assignment. The decoder panel defined network nodes; it did not restrict the genes observed by the encoder.

For each gene, the final 512-dimensional interaction states were averaged across the 126 monocytes, centred across all 3,279 panel genes and L2-normalized, following equation~\eqref{eq:main-association}. Let $A(g,j)$ denote the resulting cosine similarity, and let $\mathcal N_{15}(g)$ contain the 15 highest-scoring other genes. An undirected edge was retained when its cosine was positive and either gene selected the other:

\begin{equation}
w_{gj}=
\begin{cases}
A(g,j), & A(g,j)>0\ \text{and}\ \bigl[j\in\mathcal N_{15}(g)\ \text{or}\ g\in\mathcal N_{15}(j)\bigr],\\
0, & \text{otherwise},
\end{cases}
\qquad g\ne j.
\label{eq:network-edge}
\end{equation}

Duplicate undirected edges were merged, yielding 36,638 links. Communities were identified by weighted Leiden optimization~\citesupp{leiden2019} of the Reichardt–Bornholdt configuration-model objective~\citesupp{reichardt2006}, using cosine edge weights, resolution 1.0, with iterations continued to convergence. This yielded eight communities. Network display coordinates used a weighted Fruchterman–Reingold layout~\citesupp{fruchterman1991} with 500 iterations. Expression correlations and pathway annotations did not determine edges, community assignments or graph coordinates.

\textbf{Community enrichment and expression scores.} Community annotations and expression-based scores followed the common definitions and statistical procedures in \ref{sm:metrics}.

\textbf{Network sensitivity analyses.} Sensitivity analyses used the same 126 monocytes and two disjoint 63-cell halves, with one within-cell expression-value permutation. Correlation, matched non-edge and half-cell comparison statistics followed \ref{sm:metrics}. These analyses characterize association scores within the sampled population; they do not estimate independent biological replication, de novo graph recovery or causal regulation.

\textbf{Exploratory PBMC cell-context cases.} The frozen model encoded 126 platelets and 300 cells each from B Naive, T Naive and CD14-positive monocyte populations. Primary comparisons matched each population to 126 cells with 8 10X-dUTSO, 74 SSTX and 44 SSTX-pA+ profiles. All 36,499 supported measured genes supplied encoder context, including query expression; the decoder panel contained 3,279 genes. A common pool of 2,736 protein-coding candidates required measured-positive expression in at least seven cells of any primary population. Cosine similarities between the learned 64-dimensional gene-identity vectors provided a context-independent comparator.

The ncRNA panel was screened post hoc for population-preferred neighbours. Prioritization used top-20 model ranks, query/target detection, contrasts with the other populations, ranks relative to expression and identity, and preservation in cell halves and the two larger protocol strata. Four representative cases were displayed in Fig. 5g,h: MALAT1–PF4, MALAT1–LYZ, MALAT1–AFF3 and LINC00486–ITGA2B. Partner programme assignments were supported by external primary literature~\citesupp{sun2023malatplatelet,tsukumo2022aff3,liu2021megakaryocyte}. A single within-cell expression-value-shuffled input realization was re-encoded with the same frozen model. Uncertainty used the within-protocol cell-bootstrap procedure in \ref{sm:metrics}. Technical analyses used disjoint 63-cell halves, protocol subsets and expanded B/T/monocyte populations of 300 cells; platelets remained at 126. The candidate pool remained fixed.

\subsection{Host-cell states and programme coherence}\label{sm:host-states}

The host-cell analysis used 500 untreated and 500 DENV-exposed Huh7 profiles, with host RNA providing the expression context. Frozen CLS features supplied cellular representations for UMAP visualization. Programme scores were constructed by standardizing each measured host gene across the 1,000 selected cells and averaging the relevant signature members; the quiescent-lncRNA score subtracted the mean down-signature score from the mean up-signature score.

Within the 500 exposed profiles, the five scores for active UPR stress, antiviral inflammation, quiescent lncRNA, MAPK signalling and chromatin regulation were standardized across cells and clustered using k-means with two clusters, 10 initializations. The cluster with the higher active-UPR-stress score was labelled active-like; the other was labelled quiescent-like, yielding 267 and 233 cells, respectively. Because the host-only matrix lacked viral-load measurements, these groups approximate the published infection states using host expression signatures. Separation was measured in the original high-dimensional CLS space, with UMAP used for display. This analysis did not fit the LDA state-recognition head described in \ref{sm:cellular-evaluation}.

Frozen final interaction features were extracted separately for each cell group. Programme coherence, its expression comparator and background calibration followed \ref{sm:metrics}.

\subsection{Developmental and neural programme retrieval}\label{sm:programme-retrieval}

Published programme members provided reference sets for query-based retrieval. CellNet supplied frozen final interaction states, averaged within each biological context and centred across the entire decoder panel. All mapped measured RNA contributed to encoder context, including the non-coding query. Decoder panels defined the queried and ranked genes; they did not restrict the full encoder input to those panel genes. Candidate eligibility and retrieval scoring followed \ref{sm:metrics}.

\textbf{Mouse developmental programmes (Fig. 5j).} VASA-seq analysis used a 2,695-gene decoder panel. The main query contexts comprised Foxl2os in 100 E8.5 mesenchymal cells, C130021I20Rik in 100 E8.5 neural/ectodermal cells, and Gm50321 in 122 early vascular-lineage cells. The vascular context contained three E6.5 and 119 E7.5 cells.

Programme references came from the original VASA-seq Supplementary Table 5. Coding markers were filtered by adjusted $P<0.01$ and log fold change greater than 1, followed by the top 60 genes in the published marker-score ranking. Candidate coverage retained 60/2,149 members/candidates for Foxl2os, 60/2,152 for C130021I20Rik and 46/2,117 for Gm50321. The VASA source contributed to pretraining; this analysis assessed programme retrieval in the specified biological contexts.

For within-cohort stability, sorted cell indices were assigned alternately to two disjoint halves. Foxl2os and C130021I20Rik used 50/50 cells; Gm50321 used 61/61. Each half's gene features were averaged and centred independently, and expression correlations were recomputed on the identical half-cell set. Candidate and reference sets remained fixed to the full-cohort definitions.

\textbf{Human neural programme (Fig. 5k).} GSE150062 analysis used all 3,423 author-verified non-targeting control cells and a 2,615-gene decoder panel. SERTAD4-AS1 was measured in 13 cells; all 3,423 cells contributed to gene representations and expression correlations. Input expression was already log-transformed and retained its original normalization before clipping to $[0,6]$ and dividing by 6.

The published NSC density-state-12 marker table defined the programme reference. Markers were filtered by adjusted $P<0.01$ and average log2 fold change greater than $0.25$, followed by the top 60 in the author-provided fold-change ranking. Coverage retained 59 members among 2,082 eligible coding candidates. The original perturbation study supplied the biological reference; model inference used the non-targeting controls with observed query RNA.

\end{document}